\documentclass{aa}
\usepackage{graphicx}
\usepackage{txfonts}
\usepackage{amsbsy} 
\usepackage{soul}

\makeatletter
\newcommand\footnoteref[1]{\protected@xdef\@thefnmark{\ref{#1}}\@footnotemark}
\makeatother

\begin{document}

\title{Regularized 3D spectroscopy with CubeFit: method and application to the
  Galactic Center Circumnuclear disk\thanks{The data presented herein were obtained at the W. M. Keck Observatory, which is operated as a scientific partnership among the California Institute of Technology, the University of California and the National Aeronautics and Space Administration. The Observatory was made possible by the generous financial support of the W. M. Keck Foundation.}\fnmsep\thanks{Figures 4, 5, 8, 9 and B.1 to B.4 are also available in FITS format
at the CDS via anonymous ftp to cdsarc.u-strasbg.fr (130.79.128.5)
or via \url{http://cdsweb.u-strasbg.fr/cgi-bin/qcat?J/A+A/}.} }

\titlerunning{Regularized 3D spectroscopy with CubeFit: method and application to the GC CND}

\author{Thibaut Paumard\inst{1} \and Anna Ciurlo\inst{2} \and Mark R. Morris\inst{2} \and Tuan Do\inst{2} \and Andrea M. Ghez\inst{2}}

\institute{LESIA, Observatoire de Paris, Université PSL, Sorbonne Université, Université Paris Cité, CNRS, 5 place Jules Janssen, 92195 Meudon, France\\
  \email{thibaut.paumard@observatoiredeparis.psl.eu}
  \and
  Department of Physics and Astronomy, University of California Los Angeles, 430 Portola Plaza, Los Angeles, CA 90095, USA\\
  \email{ciurlo@astro.ucla.edu}
}

\abstract
    {The Galactic Center black hole and the nuclear star cluster are surrounded
      by a clumpy ring of gas and dust (the circumnuclear disk, CND) that rotates
      about them at a standoff distance of $\simeq1.5$~pc. The mass and density of
      individual clumps in the CND are disputed.} 
    {We seek to use H$_2$ to characterize the clump size distribution and to investigate the morphology and dynamics of the interface between the ionized interior layer of the CND and the molecular reservoir lying further out (corresponding to the inner rim 
      of the CND, illuminated in ultraviolet light by the central star cluster).}
    {We have observed two fields of approximately $20"\times20"$ in the
      CND at near-infrared wavelengths with the OSIRIS spectro-imager at the Keck
      Observatory. These two fields, located at the approaching and receding nodes of the CND,  best display this interface. Our data cover two H$_2$ lines
      as well as the Br$\gamma$ line (tracing \ion{H}{ii}). We have
      developed the tool CubeFit, an original method to extract maps of continuous physical
      parameters (such as velocity field and velocity dispersion) from
      integral-field spectroscopy data, using regularization to largely preserve spatial resolution in regions of low signal-to-noise ratio.}
    {This original method enables us to isolate compact, bright features
      in the interstellar medium of the CND. Several clumps in the southwestern
      field assume the appearance of filaments, many of which are parallel
      to each other. We conclude that these clumps cannot be
      self-gravitating.}
    {}
    
\maketitle

\section{Introduction}

The circumnuclear disk (\object{CND}) is a well-defined ring of gas and dust orbiting
the Galactic Center black hole and the nuclear cluster of massive young
stars. It has an inner cavity of $\simeq$1.5~pc radius, apparently evacuated by some combination of energetic outbursts caused by the black hole accretion flow and by supernovae from the central cluster of massive stars. The importance
of the CND is that it is a reservoir of gas that
will probably fuel future episodes of star formation and of copious
accretion onto the central black hole, and in a previous phase of its
existence, it may have been responsible for forming the present
cluster of massive young stars \citep{MorrisSerabyn1996,
  MorrisGhezBecklin1999, LuJR+13}.  Consequently, in order to understand the
activity of the central parsec of our Galaxy, especially including the
role of the CND in star formation, it is important to elucidate the evolutionary
path of the CND by understanding its structure and dynamics in as much
detail as possible. Furthermore, because many galaxies with gas-rich
nuclei apparently have nuclear disks surrounding their central
supermassive black holes \citep[e.g.][and references therein]{2015ARA&A..53..365N,2020A&A...634A...1G,2021A&A...652A..65V}, we can gather important but
otherwise unattainable insights from the CND that are applicable to
galactic nuclei in general and even to active galactic nuclei.

The CND is a warm (a few 100 K) molecular medium characterized by
strong turbulence, and clumpiness \citep{GuestenEtal1987, Genzel1989,
  MarrEtal1993, JacksonEtal1993, BradfordEtal2005, Oka+11, Martin+12, Requena-Torres+12, Mills+13, Tsuboi+18_ALMA_CND, Hsieh+21, Dinh+21}.
  The very existence of clumps has led some authors to argue that the clumps are either tidally stable \citep{Shukla+04, ChristopherEtal2005, MonteroCastanoEtal2009}, or that the CND is a transient feature having an age less than a few dynamical times because the clumps have not yet had sufficient time to be tidally sheared out of existence \citep{GuestenEtal1987, Requena-Torres+12}. However, the individual clumps can be transient features produced by instabilities or by large-scale disturbances to the disk \citep{Blank+16, Dinh+21}, even if the disk in which they are produced is itself a long-lived structure.
  
Tidal stability of the clumps requires  densities exceeding $\sim10^7$ cm$^{-3}$.  Such a high density could only apply to a small fraction of the CND volume, because the total mass is constrained by the optically thin far-infrared and submillimeter fluxes from the CND \citep{GEG10,Etxaluze+11}.  
Estimates of the total mass of the CND have varied over a wide range, depending on the densities and volumes inferred for the clumps.
Density estimates, mostly based on the analysis of molecular rotational lines, range up to $\sim10^6$ cm$^{-3}$ \citep{GuestenEtal1987, Oka+11,  Requena-Torres+12, Lau+13, Mills+13, SmithWardle14, Tsuboi+18_ALMA_CND}.  Recently, \citet{Hsieh+21} reported CS observations of a large population of tiny clumps, many having inferred densities in the range from $10^6$ to $10^8$ cm$^{-3}$.  However, almost all of the analyses based on molecular line observations have assumed that the molecular excitation is entirely collisional, although radiative excitation via the rotation-vibration lines is likely to contribute substantially to the excitation in this infrared-bright region \citep{Mills+13}.  Consequently, the inferred densities in most treatments can be regarded as upper limits, and the question of whether any of the clumps in the CND are tidally stable remains open.  The uncertainty in the density determinations and in the density distribution function (or the distribution of both clump sizes and clump densities) underlies the rather uncertain mass estimates, but the mass of the inner ring of the CND, from where most of the molecular lines and infrared continuum emission arises, probably lies in the range from $2$--$10\times10^4 M_{\sun}$.   

\begin{figure*}
  \includegraphics[width=1.0\textwidth]{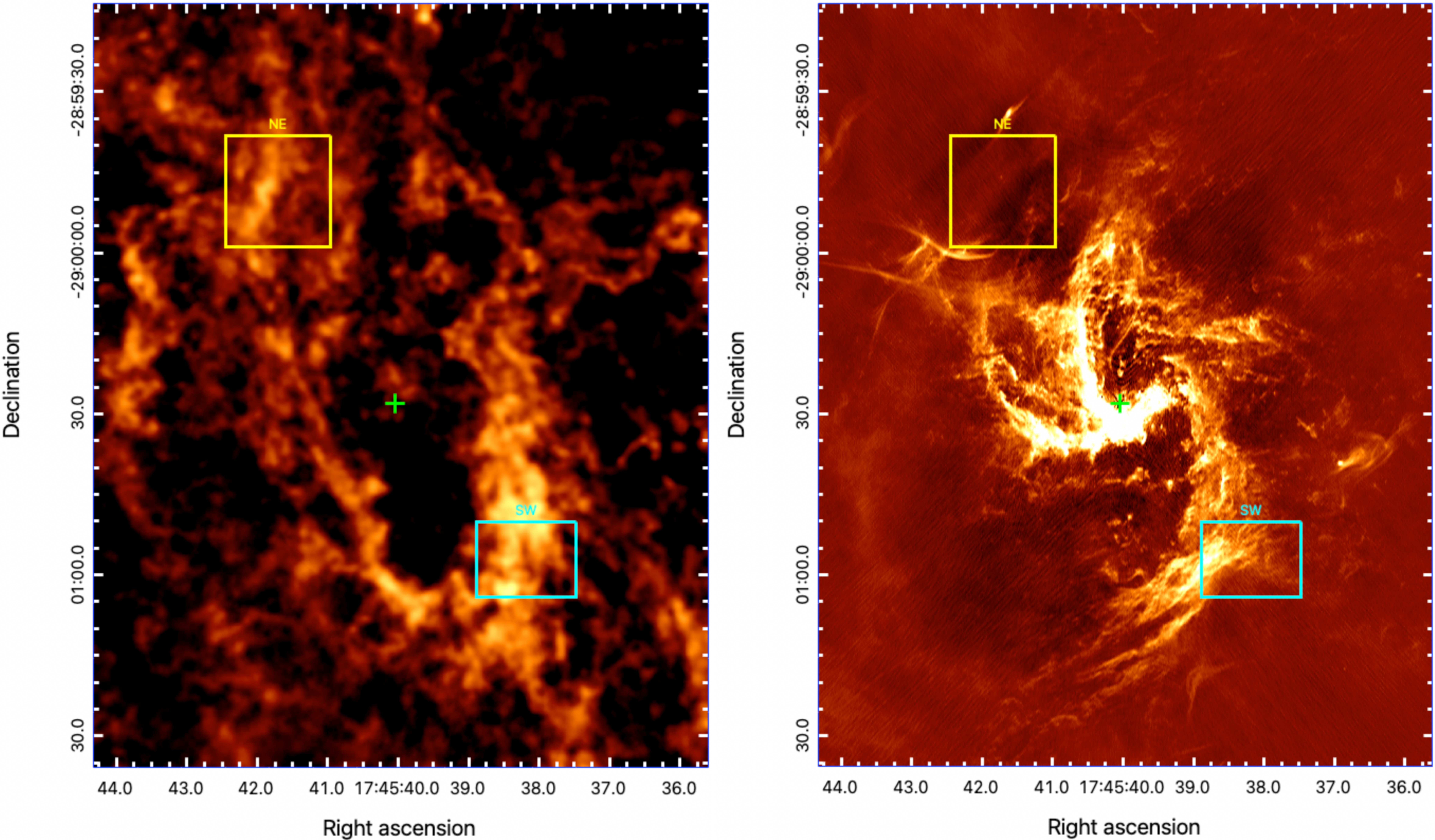}
  \caption{Outline of the two mosaicked regions reported here superimposed on: {\it left} the total intensity map of J=4-3 CS emission, from ALMA \citep{Hsieh+21}, and {\it right} the 6-cm radio continuum image from the VLA \citep{ZMG16, Morris+17}. The location of \object{Sgr~A*} is indicated in each panel with a green cross.
  \label{fig:fieldmaps}}
\end{figure*}

The investigation reported here is therefore motivated in part by the need to elucidate the clump size distribution. 
Clump morphology is another key issue that we propose to
investigate. To date, theoretical treatments of the clump characteristics
all assume spherical clumps, but that assumption is valid only if the clumps are self‐gravitating to the point of being tidally stable, which might apply only to a small fraction of the gas mass.  If the clumps are not tidally stable, tidal shear will have pulled them into elongated streams. Indeed, the H$_2$ morphologies sampled with the Hubble Space Telescope (HST) observations of \citet{YusefZadehEtal2001} suggest that
some fraction of the emission is organized into filamentary features.  The detection of filamentary clumps on even smaller scales would imply an upper limit to their densities that is below the Roche density.

To address these questions, we have observed two fields of the CND using the spectro-imager OSIRIS \citep{LarkinEtal2006} on the Keck II
telescope.  These fields -- the northeastern and southwestern lobes (hereafter NE and SW, respectively) in the nomenclature of \citet{YusefZadehEtal2001} and  \citet{ChristopherEtal2005} -- lie at the opposite nodes of the CND (Fig.~\ref{fig:fieldmaps}).

In order to characterize the size and shape of the smallest
clumps in this dataset, we need to retain the full sampling-limited
spatial resolution (at $0.1''$ per pixel) while reconstructing maps of
the brightness distribution, radial velocity and radial velocity
dispersion of the relatively low signal-to-noise (SNR) interstellar
emission lines contained in the spectral band.  To this end, we
have developed a new method to analyze spectro-imaging data. Inspired by deconvolution and
interferometric image reconstruction, this method reconstructs optimal
parameter maps (typically line flux, radial velocity and velocity
dispersion) under the two constraints that the model needs to be close
to the data in the minimum $\chi^2$ sense and that the parameter maps must be
smooth. This method was also used in another paper on infrared spectroscopy of the Galactic Center
\citep{2016A&A...594A.113C}. Here, we give additional details about the
foundation of the method and provide more thorough testing and validation.

In Sect.~\ref{sect:observations}, we describe our observations
and data reduction procedure. Sect.~\ref{sect:3Dfitting} presents our
technique for building parameter maps as well as first tests of the method using the
strong OH lines present in the data. We apply this method to our data
set and provide more ample validation in Sect.~\ref{sect:application}.
We then discuss our findings in Sect.~\ref{sect:discussion} and offer
concluding remarks in Sect.~\ref{sect:conclusion}.

\section{Observations and data reduction}
\label{sect:observations}
\label{sect:datareduction}

We have used the integral-field spectrograph OSIRIS \citep{LarkinEtal2006}  at the W.M. Keck Observatory fed by the laser guide star adaptive optics system to observe two fields at the inner edge of the CND
near the location of the two nodes of the CND orbit. The observations were carried out on May 13 and 19, 2010 (NE field, 14 frames) and July 24 and 25, 2011 (SW field, 10 frames) and all consist of 900~s exposures taken in the Kn3 band (2.121--2.229~$\mu$m) with a 100~mas plate scale, giving a field of view for each frame of $4.8"\times6.4"$.
The data have been reduced with the OSIRIS pipeline \citep{LockhartEtal2019}, which subtracts darks, assembles the cubes, corrects the sky lines and creates the final mosaics. The mosaics are photometrically calibrated using standard A stars observed the same night as the science observations. The procedure is described in \citet{CiurloEtal2020}. Furthermore each mosaic is astrometrically calibrated using three stars whose absolute position is known through HST observations of the area (Hosek et al., in prep.). The uncertainty on the astrometric calibration is about 0.5''.

The final data products are two mosaics, one for the NE and one for the SW field. The NE field centered at J2000 coordinates (17:45:41.67,-28:59:48.8), near the receding node, is an approximately square field $20"$ on each side
with a small ($<1"$) horizontal gap $\simeq4"$ south of its central
point. The SW mosaic, centered at (17:45:38.17,-29:00:57.469,13.8) covers a rectangular area of $14"\times14"$ near the approaching node. 

In the spectral band covered by our data, we detect two H$_2$ lines
(1-0 S(1) at 2.1217~$\mu$m and 1-0 S(0) at 2.2232~$\mu$m) and the
Br$\gamma$ line of the \ion{H}{i} spectrum (2.1667~$\mu$m), tracing \ion{H}{ii}. The two H$_2$ lines are several orders of magnitude stronger than any other H$_2$ line in the band according to the HITRAN database \citep{HITRAN2016,H2-nu-1-736,H2-S-1-738}. 

The data are dominated by continuum flux from stars. We have estimated
this continuum emission for each spatial pixel as a spline function
going through the average flux in 12--13 featureless regions
(Tables~\ref{tab:free_ranges_SW} and \ref{tab:free_ranges_NE}) of the
spectrum and subtracted it from the data.  Unfortunately the H$_2$
line at $2.12$~$\mu$m is at the edge of the spectral bandpass so that
we have no continuum estimate on the short wavelength side of this
line and must resort to extrapolation there. The presence of the
second line at $2.22$~$\mu$m, to the extent that it can be considered
as sharing the same radial velocity, alleviates this issue. The
SNR per data point ranges from $0$ to
$\approx35$ for H$_2$ from $0$ to $\approx18$ for Br$\gamma$.

\begin{table}[th]
    \centering
    \caption{Wavelength ranges used for continuum estimation for the SW mosaic. For each region, visually selected for its lack of spectral features in the data cube, we list the central wavelength ($\mu$m, observed wavelength in the vacuum), the number of spectral channels and the spectral bandwidth (nm). The bandwidth of a single channel is $0.25$~nm.}
    \label{tab:free_ranges_SW}
    \begin{tabular}{lrc}
    \hline\hline
    $\lambda_\mathrm{center}$ ($\mu$m) & $n_\mathrm{channels}$ & $\Delta\lambda$ (nm)\\
\hline
2.124 & 9 & 2.25\\
2.125875 & 8 & 2.00 \\
2.1295 & 5 & 1.25\\
2.133875 & 8 & 2.00 \\
2.143 & 9 & 2.25 \\
2.160875 & 16 & 4.00\\
2.1685 & 13 & 3.25\\
2.1755 & 21 & 5.25\\
2.1835 & 13 & 3.25\\
2.1915 & 21 & 5.25\\
2.2195 & 13 & 3.25\\
2.227625 & 12 & 3.00\\
\hline\hline
    \end{tabular}
\end{table}

\begin{table}[th]
    \centering
    \caption{Same as Table~\ref{tab:free_ranges_SW} for the NE mosaic.}
    \label{tab:free_ranges_NE}
    \begin{tabular}{lrc}
    \hline\hline
    $\lambda_\mathrm{center}$ ($\mu$m)& $n_\mathrm{channels}$ & $\Delta\lambda$ (nm)\\
    \hline
2.125875 & 8 & 2.00\\
2.13 & 7 & 1.75\\
2.13475 & 7 & 1.75 \\
2.148 & 17 & 4.25 \\
2.157 & 7 & 1.75 \\
2.16975 & 7 & 1.75 \\
2.17475 & 7 & 1.75 \\
2.1785 & 5 & 1.25 \\
2.183375 & 14 & 3.50\\
2.1925 & 13 & 3.25 \\
2.2095 & 17 & 4.25 \\
2.219 & 9 & 2.25 \\
2.228 & 9 & 2.25 \\
\hline
    \end{tabular}
\end{table}

\section{CubeFit: Data modeling with regularized parameter maps}
\label{sect:3Dfitting}
\label{sect:linemaps}

Integral field spectroscopy is a powerful tool that allows the
simultaneous spectroscopic observation of all pixels within a continuous field-of-view,
sometimes with the goal of recording the individual spectra of many
point sources and sometimes to study the spectral properties of
diffuse emission. This paper focuses on the latter. Integral field
spectroscopy data are three-dimensional, with two spatial dimensions
and one spectral dimension: $\mathcal D(l, m, \lambda$) is
some quantity related to the emitted intensity (e.g. flux density)
originating from the direction $(l, m)$ at wavelength
$\lambda$, with uncertainties $\mathcal U(l, m,
\lambda)$. $\mathcal D$ can be interpreted either as a collection of
images recorded in many consecutive wavelength channels or as a
collection of spectra (the \emph{spaxels}) for each pixel of these
images. Individual data points are called \emph{voxels}. Observers
typically want to interpret such data as a set of maps of physical
parameters. Given a 1D spectral model $\mathcal S_{\{a_i\}}(\lambda)$ of $n$ scalar physical parameters $a_0 \dots a_n$, one will try
to construct a 2D map $\boldsymbol a_i$ for each parameter so that the 3D
model
\begin{equation}
  \mathcal M_{\{\boldsymbol a_i\}}(l, m, \lambda) =
  \mathcal S_{\{\boldsymbol a_i(l, m)\}}(\lambda)
\end{equation}
is a good match to the data $\mathcal D$, under a criterion which involves the uncertainties $\mathcal U$.

\subsection{Traditional independent 1D fits}

The most usual approach to this problem is to simply perform a
spectral fit on each individual spaxel spectrum in the cube. In other
terms, one will determine $\boldsymbol a_i(l, m)$ by minimizing for
each point $(l, m)$ in the field-of-view the following quantity:
\begin{equation}
  \chi_{\mathcal S, l, m}^2(\{a_i\}) = \sum\limits_{\lambda} \left(((\mathcal D(l, m, \lambda) - \mathcal S_{\{a_i\}}(\lambda))\cdot \mathcal W(l, m, \lambda))^2\right)
\end{equation}
where $\mathcal W$ is the weight associated with each data point
(usually, $\mathcal W = 1/\mathcal U$). Note that minimizing the 1D $\chi^2$ term for each point $(l,m)$ is equivalent  to globally minimizing the 3D $\chi^2$:
\begin{equation}
  \chi_\mathcal M^2(\{\boldsymbol a_i\}) = \sum\limits_{l, m, \lambda} \left(((\mathcal D - \mathcal M_{\{\boldsymbol a_i\}})\cdot \mathcal W)^2\right) = \sum\limits_{l,m} \chi_{\mathcal S, l, m}^2(\{\boldsymbol a_i(l, m)\})
\end{equation}

This method works very well and is sufficient as long as the SNR in
the emission line is large ($\gtrsim3$--$5$) within the solid angle viewed by each spaxel. However, this condition of high SNR is quite
restrictive. It means that this approach will fail in low SNR areas of
the field, inevitably ending up fitting independent noise spikes.

\subsection{An original regularized 3D fit}

In order to avoid this problem, we propose to add to the $\chi^2$ term
a penalty term that will ensure that the parameter maps are
regular, in the sense of the regularization approach detailed below: we want the maps of observable parameters to be as continuous as possible across adjacent pixels. The measured variations across the field should be representative of physical variations rather than noise, and any discontinuity should be smoothed by the imaging resolution of the instrument.
This treatment inspired by image deconvolution and interferometric
image reconstruction, two problems that have strong similarities with the
one that occupies us; in all these cases, the observer wants to
interpret complex data as a set of regular high-resolution spatial
maps. Our estimator takes the form:
\begin{equation}
  \label{eq:estimator}
  \mathcal E (\{\boldsymbol a_i\}) = \chi^2_\mathcal M(\{\boldsymbol a_i\})
    + \sum\limits_{i=1}^n\mathcal R_i(\boldsymbol a_i)
\end{equation}
where each $\mathcal R_i$ is an appropriate penalty function that
encodes prior knowledge on each parameter map. By globally minimizing
such an estimator over an entire 3D data set, one ensures that the
solution respects a compromise between proximity to the data and this
prior knowledge. Quadratic-linear (or $L_2-L_1$ for short) priors are
often used to smooth small noise gradients while preserving the large
gradients of edges as explained by \citet[and references
  therein]{MugnierFuscoConan2004}. We use the norm from their eq.~9:
\begin{equation}
  \mathcal R_{\mu_i,\delta_i}(\boldsymbol a_i)=\mu_i \delta_i^2 \sum\limits_{l,m} \phi (\nabla \boldsymbol a_i(l,m)/\delta_i) 
\end{equation}
where
\begin{equation}
  \phi(\boldsymbol x) = |\boldsymbol x|-\ln(1+|\boldsymbol x|)
\end{equation}
and
\begin{equation}
  \nabla \boldsymbol a_i(l,m) = [\nabla_l\boldsymbol a_i
    (l,m)^2+\nabla_m\boldsymbol a_i (l,m)^2]^{1/2}\text,
\end{equation}
$\nabla_l \boldsymbol a_i$ and $\nabla_m \boldsymbol a_i$ being the map
finite-difference gradients along $l$ and $m$, respectively. The two
hyperparameters (per parameter map) $\mu_i$ and $\delta_i$ currently have to be set by
hand. $\delta_i$ regulates the transition between the two regimes: small gradients (typical of noise) are penalized whereas strong gradients (more likely physical) are restored. $\mu_i$ allows weighting the
various regularization terms in $\mathcal E$.

We developed CubeFit\footnote{\url{https://github.com/paumard/cubefit}}, a code implementing this method using the Yorick interpreted language\footnote{\url{https://software.llnl.gov/yorick-doc/}}. We borrowed the L2--L1 regularization function from Yoda\footnote{\url{https://github.com/dgratadour/Yoda}} by Damien
Gratadour, a Yorick port of the MISTRAL deconvolution software by
\cite{MugnierFuscoConan2004}. The model-fitting engine is
the conjugate gradient algorithm implemented in the Yorick package
OptimPack\footnote{\url{http://www-obs.univ-lyon1.fr/labo/perso/eric.thiebaut/optimpack.html}}, version 1.3.2. A Python port of CubeFit is under development.

\subsection{Test of the method on OH lines}
\label{sect:OHfit}

The many strong OH lines present in the near infrared spectral band give us
the opportunity to test our method on high signal-to-noise data. In this section, we use CubeFit on versions of the two mosaics which are not sky-subtracted and compare the results with those obtained using a 1D approach. In the process, we determine field-variable corrections to the instrumental spectral resolution and wavelength calibration that we will use in the later sections.

We have used the 8 OH lines between 2.12 and 2.23~$\mu$m in the list from
\citet{1992A&A...254..466O}. For the 1D spectral model $\mathcal
S_{\{a_i\}}(\lambda)$, we use a Doppler-shifted, multi-line Gaussian
profile:
\begin{equation}
  \mathcal G^{\{\lambda_j\}}_{\{I_j\}, v, \sigma}(\lambda)=\sum\limits_j I_j \times \exp\left(-\frac{\left(\frac{\lambda-\lambda_j}{\lambda_j} \times c - v\right)^2}{2 \times \sigma^2}\right)
  \label{eq:G}
\end{equation}
where $c$ is the speed of light. The 10 parameters are the intensities
of the 8 lines $\{I_j\}$, the common Gaussian width $\sigma$ (in the
radial velocity domain) and a common radial velocity $v$. We have
fitted the data with this model using two methods: individual 1D fits
on each spaxel (the maps have been $\sigma$-filtered to remove some
bad fits) and with CubeFit.

\begin{figure*}
  \begin{center}
    \null \hspace{2.6cm} OH linewidth  \hfill Velocity offset \hfill Normalized flux \hspace{2.3cm} \null\\
    \rotatebox{90}{\hspace{1.3cm} 1D fit}\hspace{0.1cm}
    \includegraphics[scale=0.4, viewport=0 50 435 334, clip]{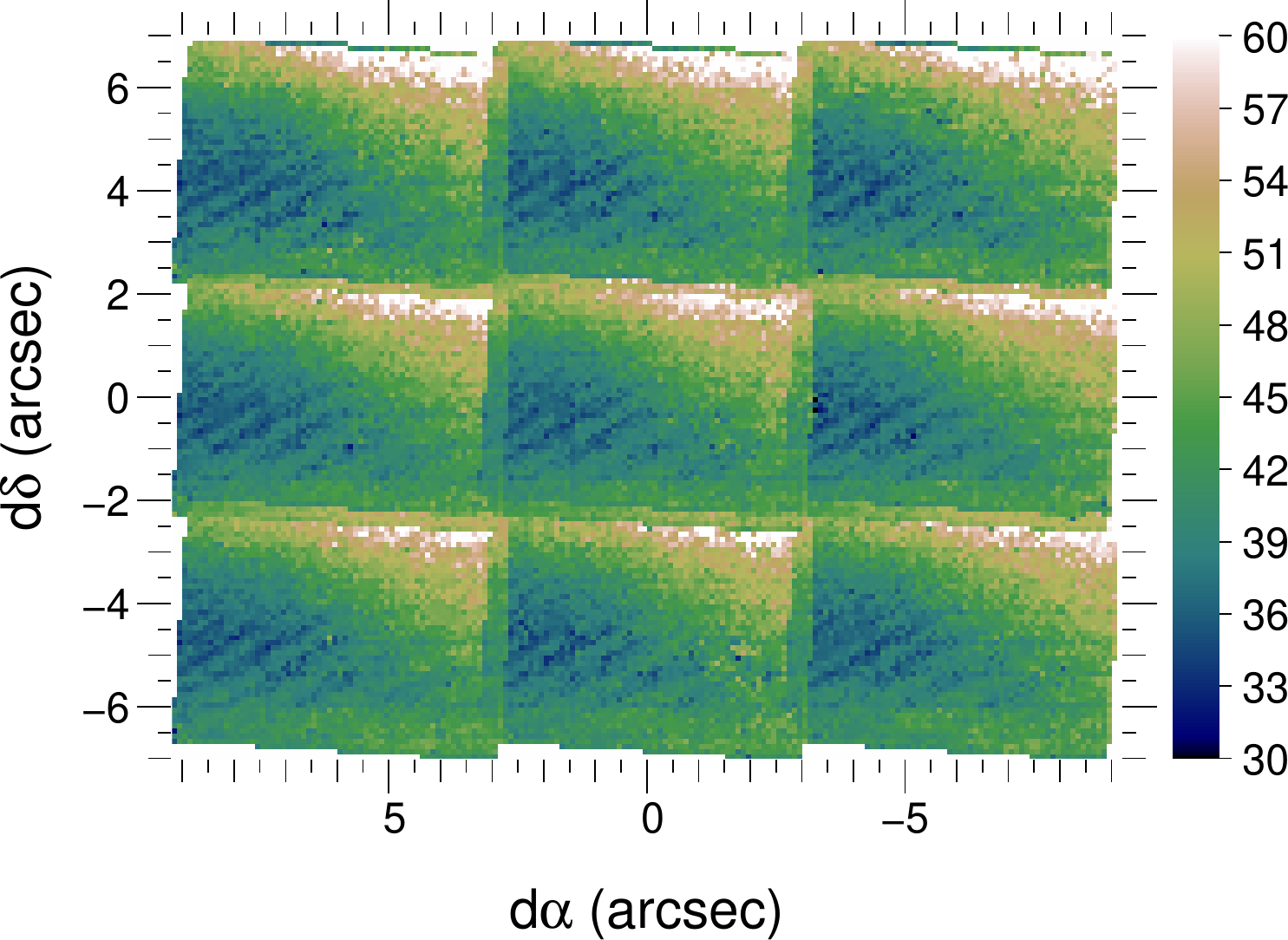}
    \includegraphics[scale=0.4, viewport=44 50 435 334, clip]{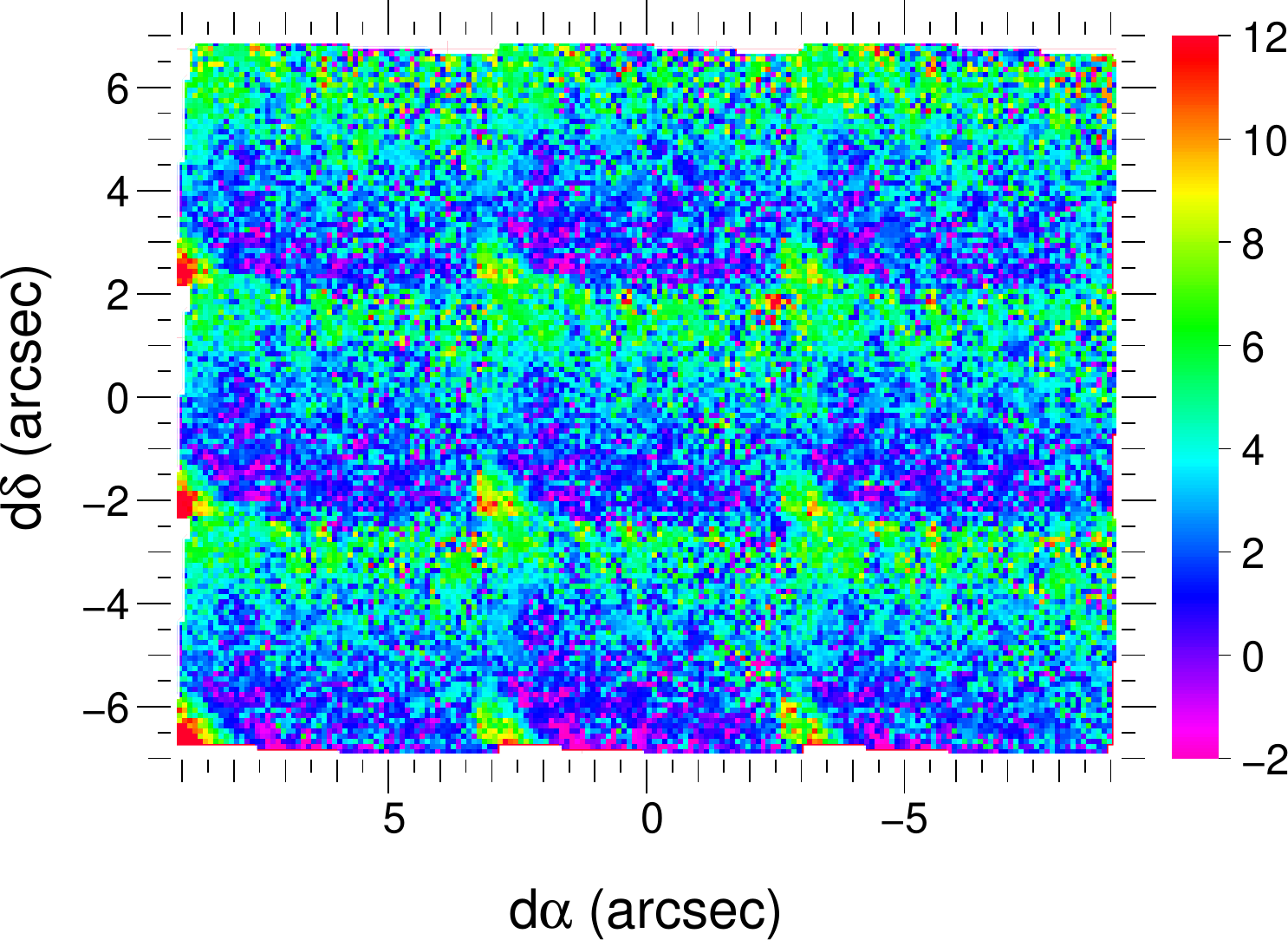}
    \includegraphics[scale=0.4, viewport=44 50 435 334, clip]{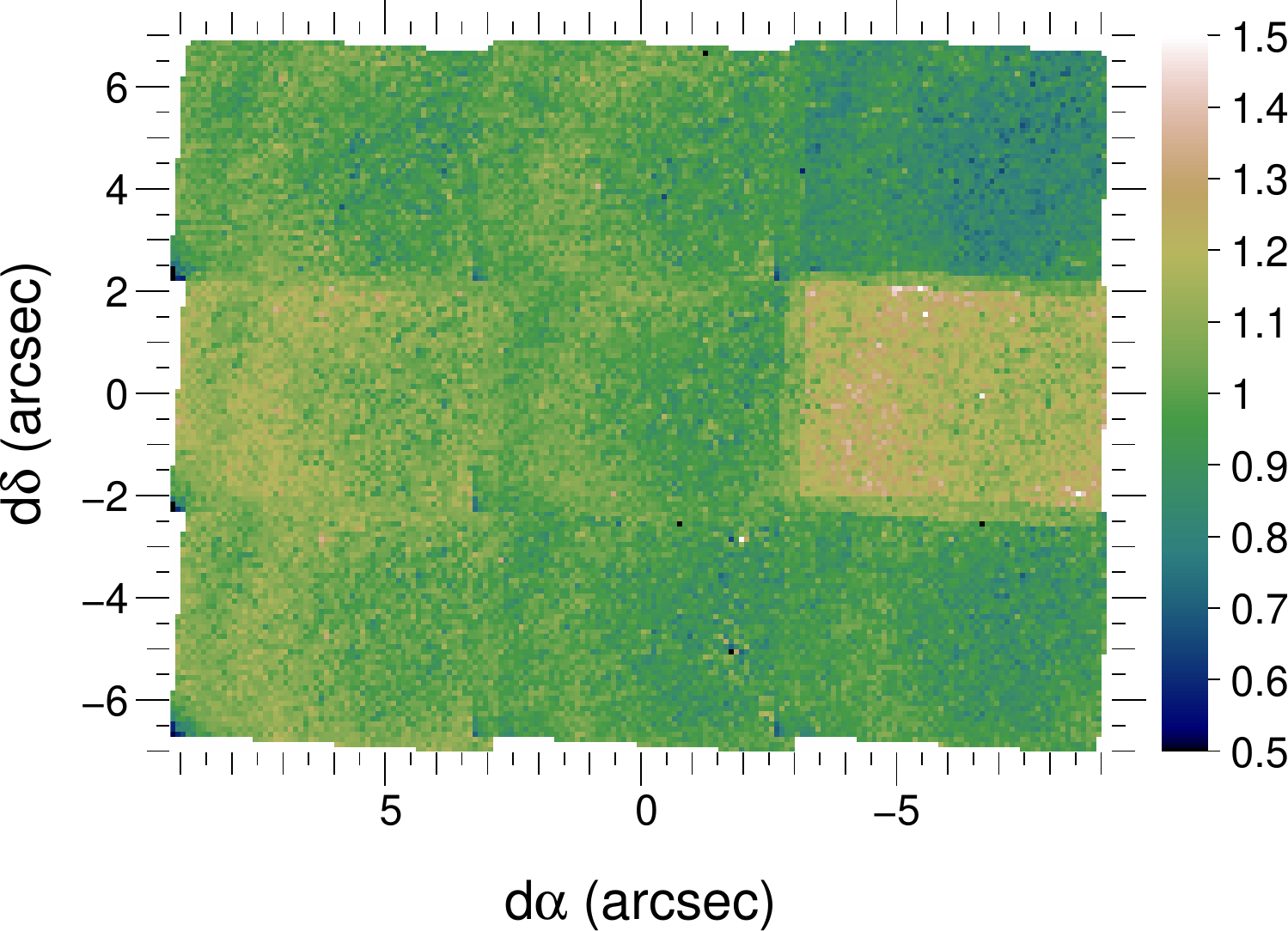}\\
    \rotatebox{90}{\hspace{1.3cm} 3D fit}
    \includegraphics[scale=0.4, viewport=0 50 435 334, clip]{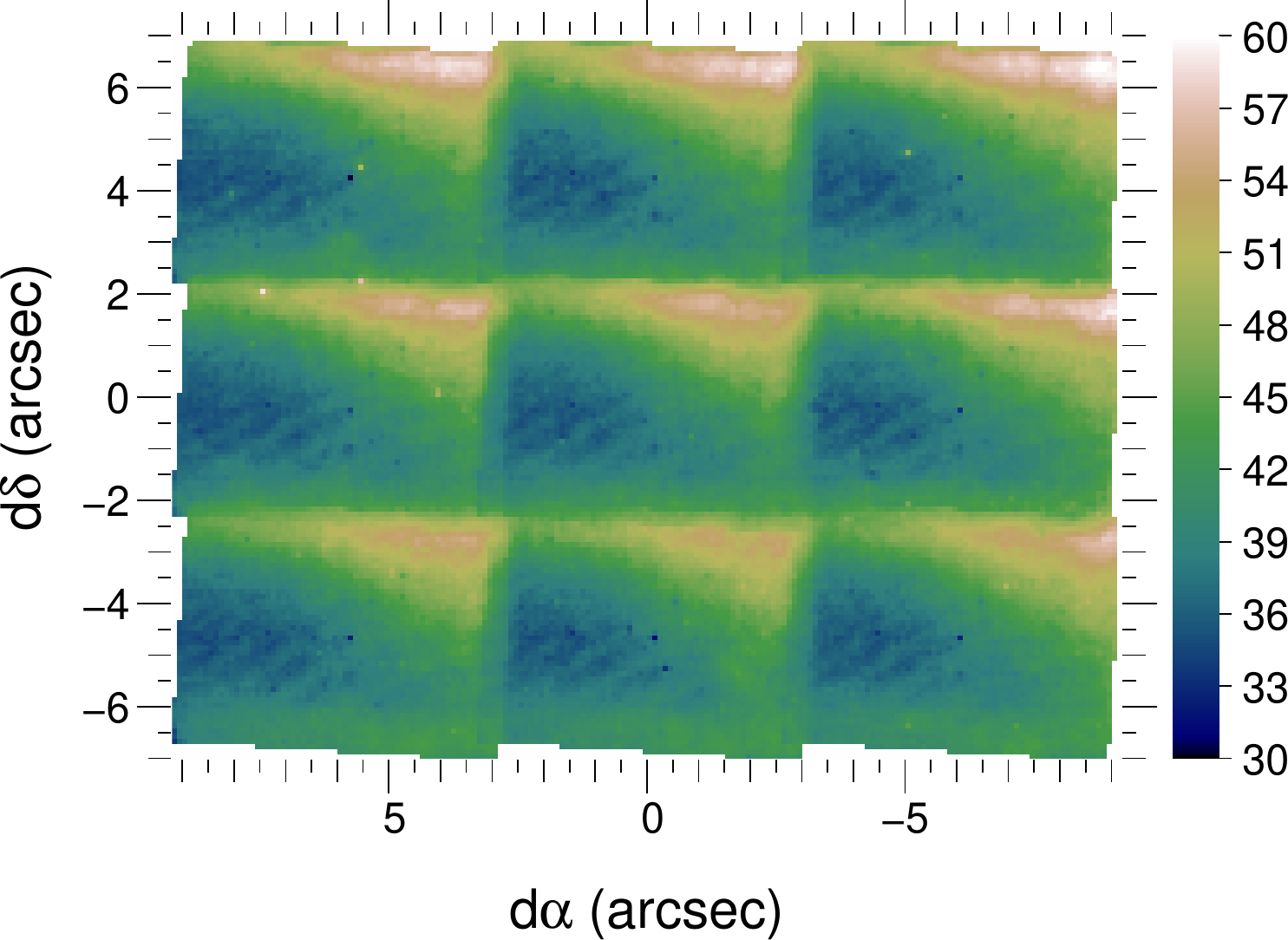}
    \includegraphics[scale=0.4, viewport=44 50 435 334, clip]{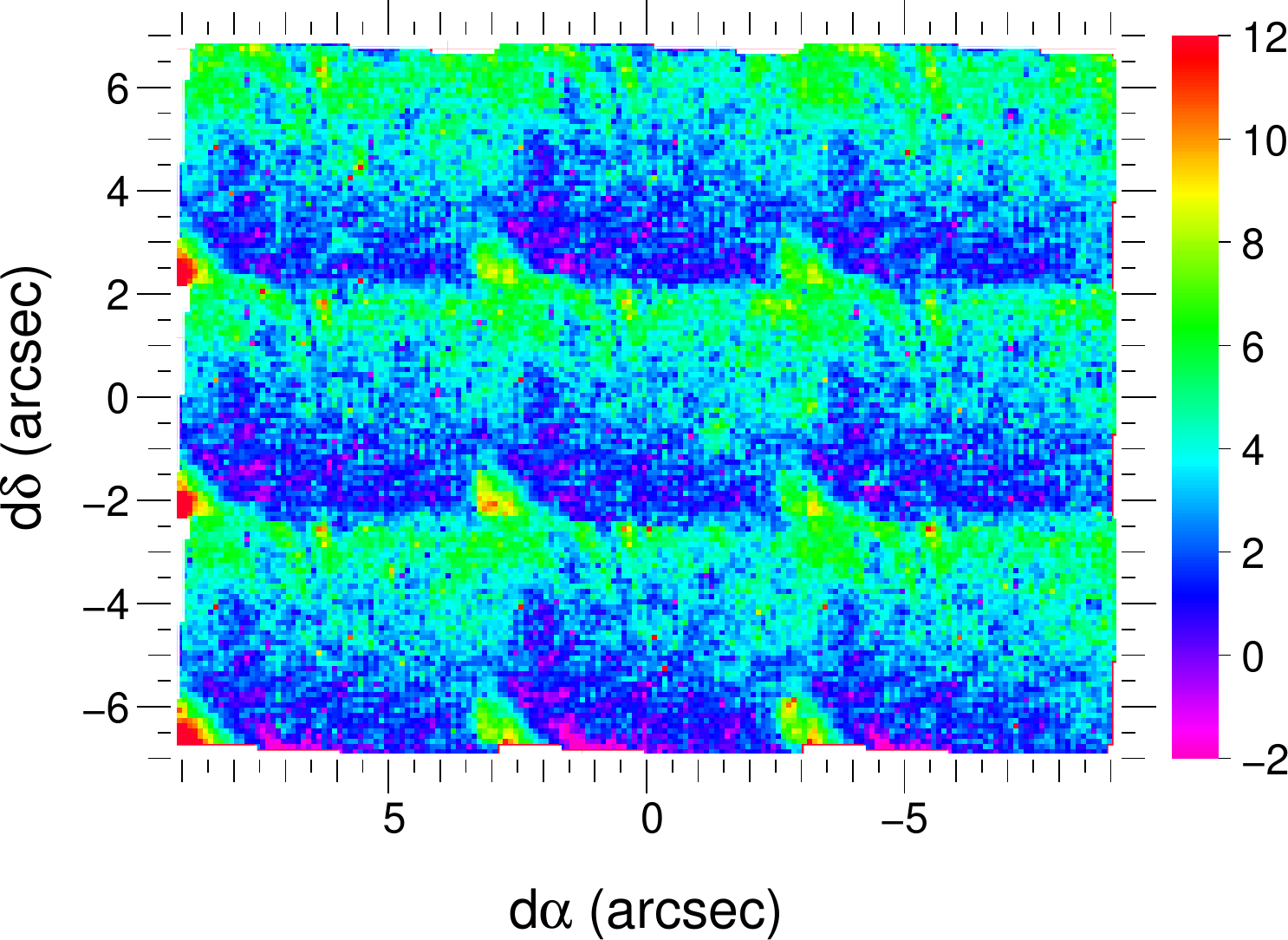}
    \includegraphics[scale=0.4, viewport=44 50 435 334, clip]{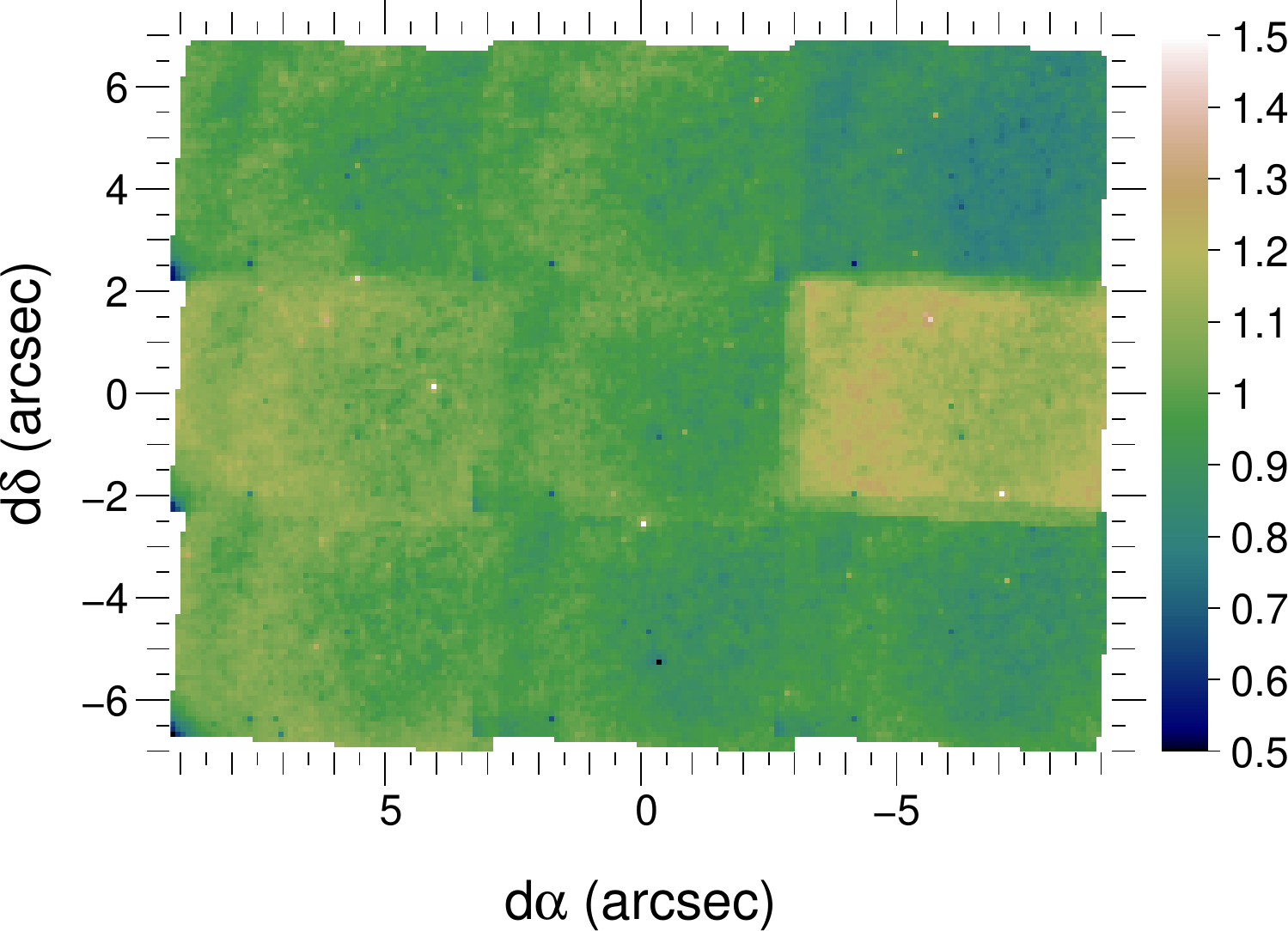}\\
    \rotatebox{90}{\hspace{1.9cm}$\text{3D}-\text{1D}$}
    \includegraphics[scale=0.4, viewport=0 0 435 334, clip]{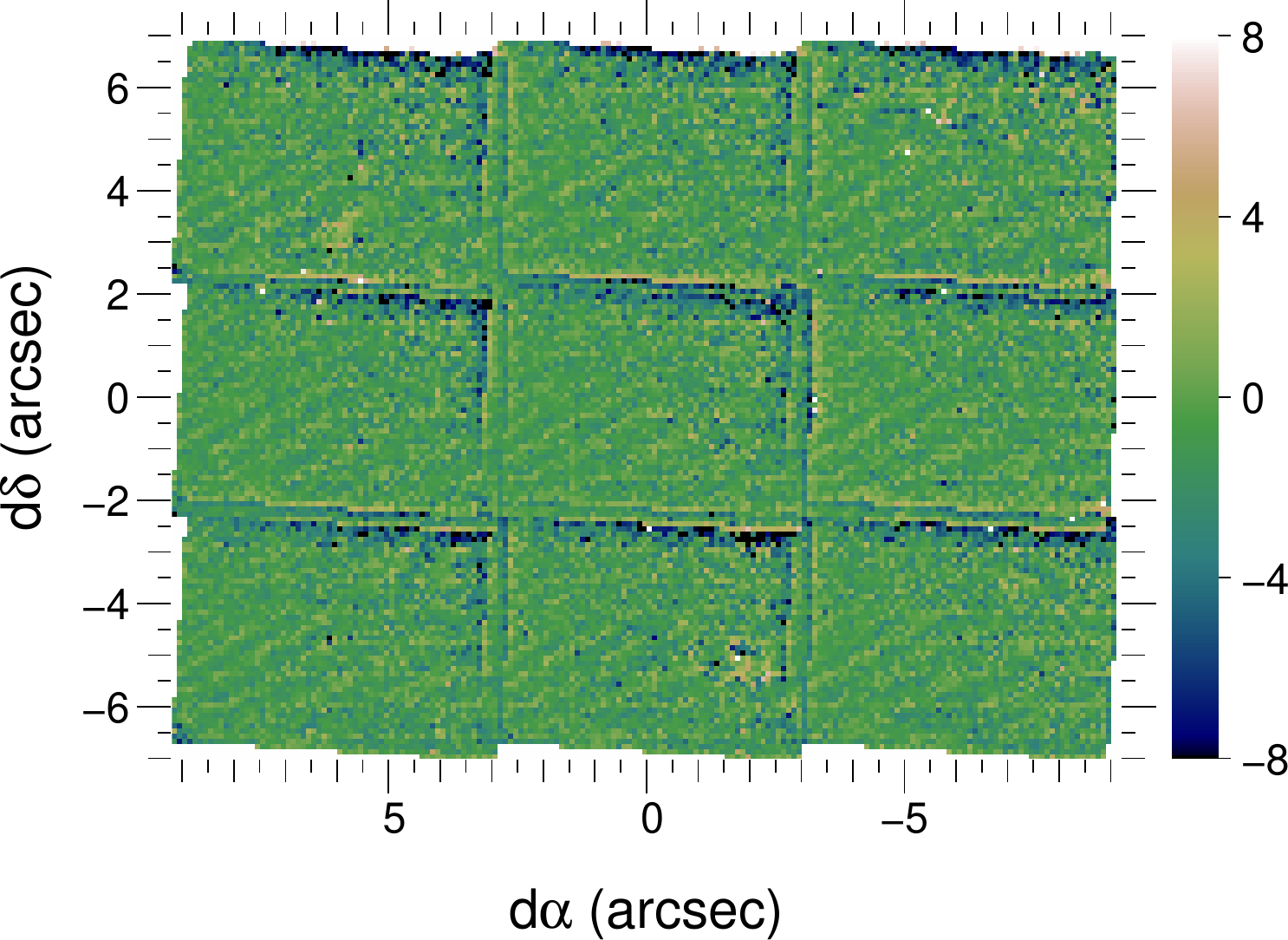}
    \includegraphics[scale=0.4, viewport=44 0 435 334, clip]{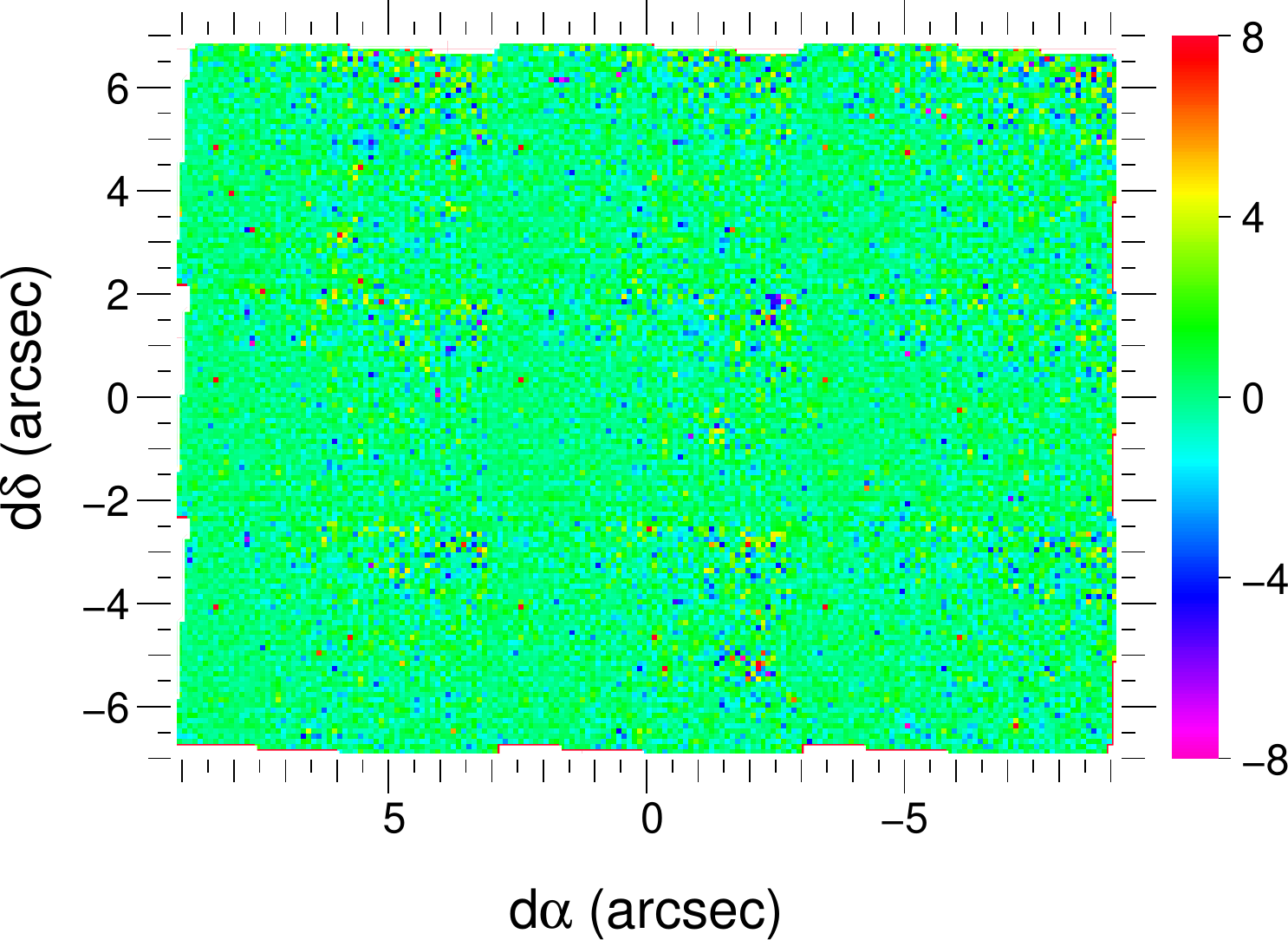}
    \includegraphics[scale=0.4, viewport=44 0 435 334, clip]{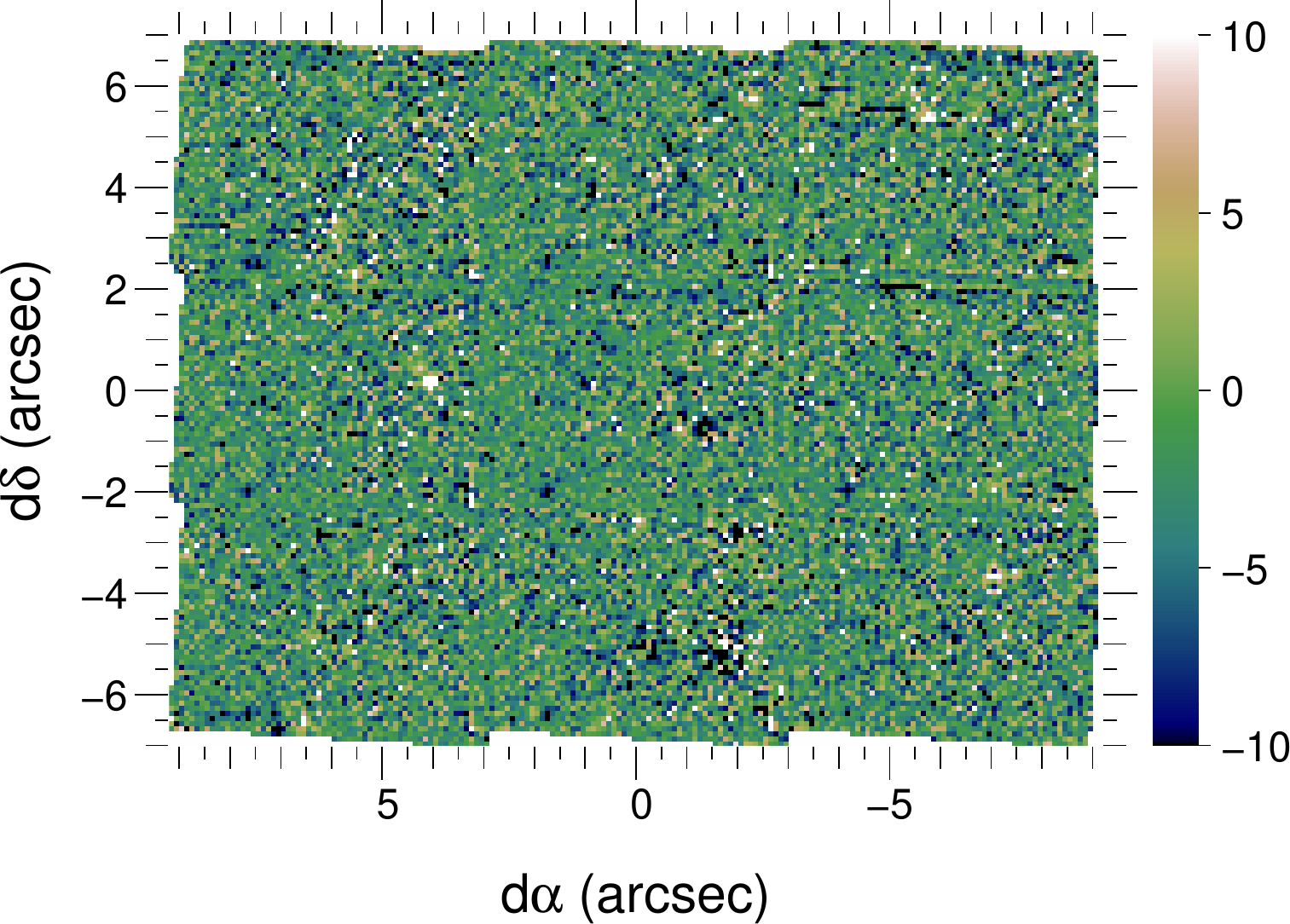}\\
    \vspace{3mm}
    \includegraphics[scale=0.4]{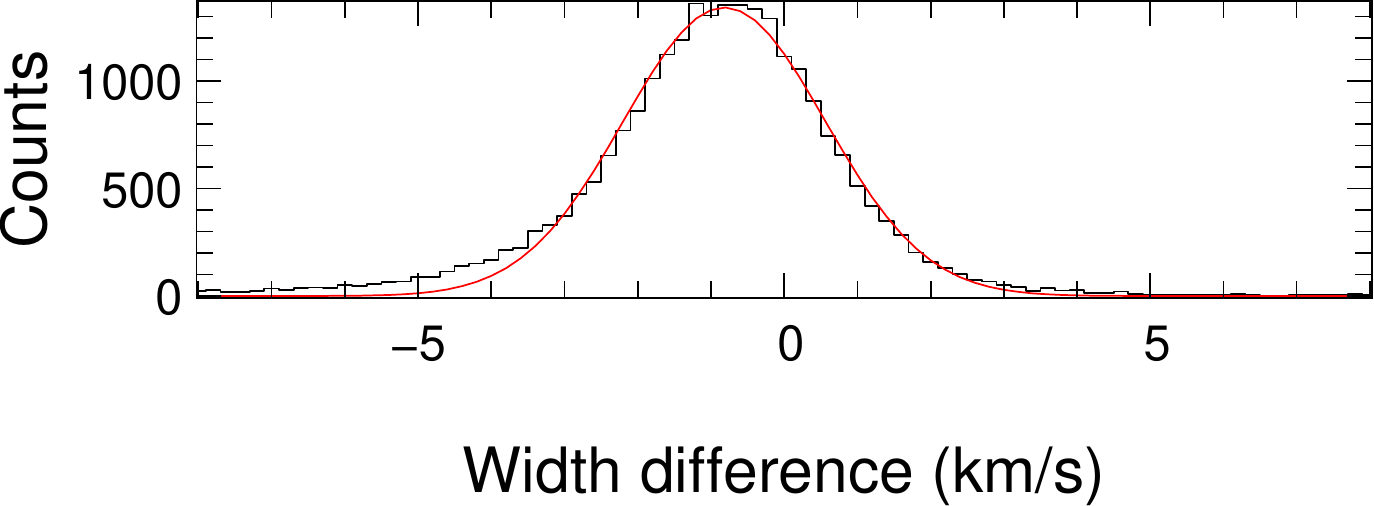}
    \includegraphics[scale=0.4]{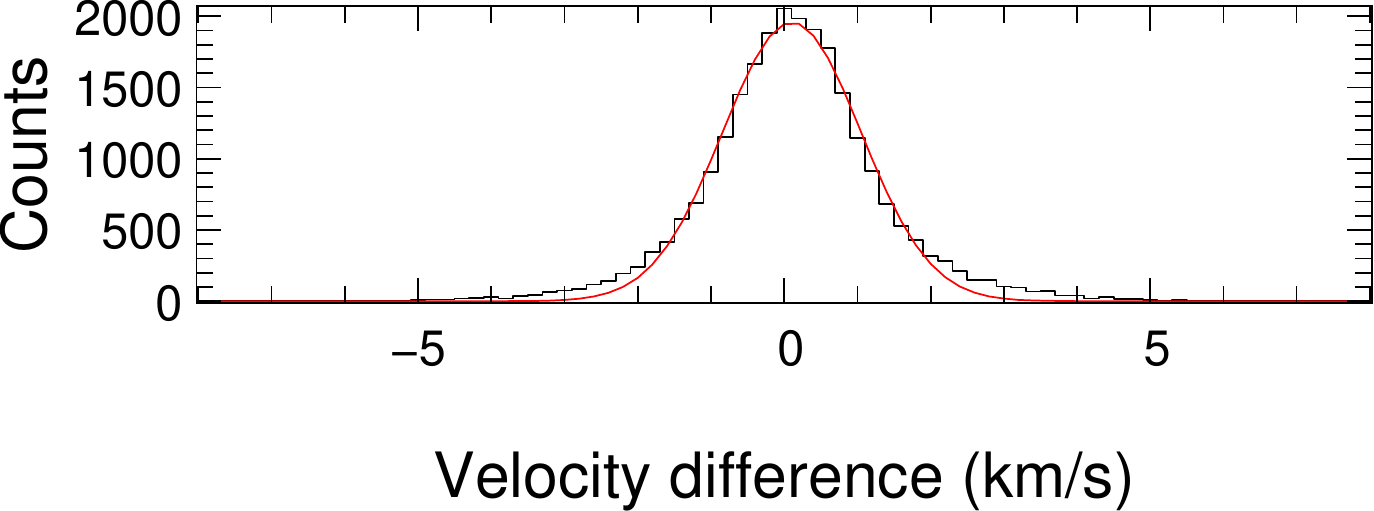}
    \includegraphics[scale=0.4]{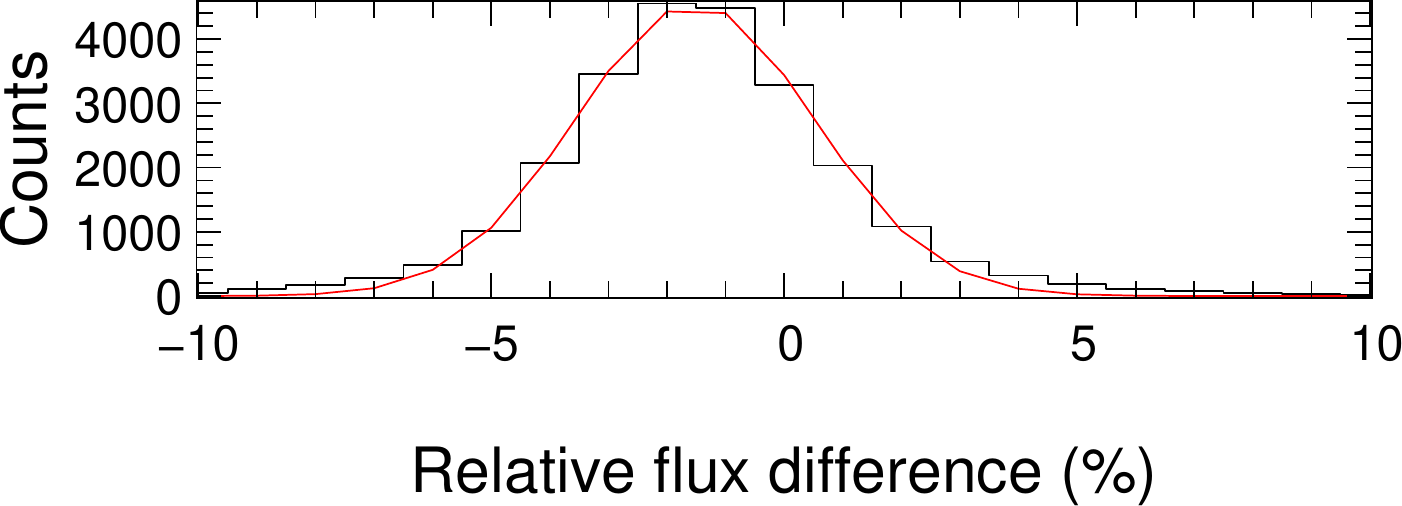}
    \caption{Results of multiple OH line fits for the SW
      mosaic. \emph{Left column:} linewidth (km~s$^{-1}$);
      \emph{center column:} radial velocity offset (km~s$^{-1}$);
      \emph{right column:} flux of the brightest line estimated as
      intensity $\times$ width normalized to the 1D fit median;
      \emph{top row:} median-filtered parameter maps from the 1D fit;
      \emph{second row from top}: regularized maps from the 3D fit;
      \emph{third row from top:} 3D parameter maps minus 3D fit
      (divided by 1D fit for flux); \emph{bottom row:} histograms of
      difference maps (black histograms) and Gaussian fits to those
      histograms (red curves). The moments of the Gaussian fits to the
      histograms are, for the linewidth difference: median
      $\mu=-0.82$~km~s$^{-1}$ and width $\sigma=1.38$~km~s$^{-1}$, for
      the line velocity difference: $\mu=0.10$~km~s$^{-1}$ and
      $\sigma=0.95$~km~s$^{-1}$, for the relative flux estimate
      difference: $\mu=-1.5\%$ and $\sigma=2.0\%$.\label{fig:SWOHfit}}
  \end{center}
\end{figure*}

Figure~\ref{fig:SWOHfit} shows the result of the 1D and 3D fits for
the south-west (SW) mosaic and their differences. The linewidth
(which traces spectral resolution for those intrinsically very thin
telluric lines) varies considerably across the mosaic. In each subfield, the
resolution varies from $\simeq35$~km~s$^{-1}$ in the south-eastern
corner to $\simeq60$~km~s$^{-1}$ in the north-western corner. These
strong variations create sharp edges at the transition between
subfields. In addition, the linewidth map shows some striping at a
much smaller level than this overall gradient. The measured radial
velocity map also shows a trend in the SE--NW direction on the order
of a few kilometers per second. Because of the variations of the
linewidth, it is better to express line strength in terms of flux
($\propto$ intensity $\times$ width) rather than intensity. The
corresponding map shows variations on the order of $5\%$ within
subfields. The differences between subfields are probably due to
actual variations of the airglow spectrum during the night.

The maps produced by the two methods are very similar and exhibit the
same features. As expected, the CubeFit 3D-fit parameter maps are
slightly less noisy, but reproduce the sharpest features only
partially (in particular the sharp edges between subfields and
striping of the linewidth map), which explains the slight bias in the
linewidth distribution (Fig.~\ref{fig:SWOHfit}, bottom-left
panel). Conversely, the CubeFit parameter maps do not need to be
$\sigma$-filtered and some artifacts that can be seen in the 1D-fit
parameter maps are very well corrected by the regularization of
CubeFit.  For instance, artifacts from two very bright stars can be
seen in the 1D flux map at $(d\alpha, d\delta)\simeq(-2'', -3'')$, and
$(-2'', -5'')$. Finally, striping can also be seen in the 3D-fit
velocity map subfields and can hardly be seen in the 1D-fit parameter
map due to the additional noise.

The same analysis on the NE mosaic yields very similar
results. In the rest of this paper, we fit the intrinsic width and
radial velocity of various lines. To this effect, we correct the model
for variable spectral resolution and wavelength calibration residuals
by adding pixel by pixel the OH linewidth
$\boldsymbol\sigma_{\text{OH}}(l,m)$ (in quadrature) and velocity
offset $\boldsymbol v_{\text{OH}}(l,m)$ to the fitted linewidth and
velocity:
\begin{eqnarray}
  \boldsymbol \sigma^{\text{tot}}(l,m) &=& \sqrt{\boldsymbol \sigma(l,m)^2+\boldsymbol \sigma_{\text{OH}}(l,m)^2}\label{eq:sigmatot}\\
  \boldsymbol v^{\text{tot}}(l,m) &=& \boldsymbol v(l,m)+\boldsymbol v_{\text{OH}}(l,m)
\end{eqnarray}
Since the SNR of the OH lines is so high, and in order to fully remove
the sharp edges in the spectral resolution spatial variations, we use
the results of the 1D fit for this purpose.

Further tests of the method are done in the next sections together
with the analysis of the CND data and are summarized in
Sect.~\ref{sect:conclusion}.

\section{Application to the CND data}
\label{sect:application}
To apply CubeFit to the (sky-subtracted) CND data, we chose to use line
flux as a parameter for the Gaussian profile rather than line amplitude. The amplitude
$\boldsymbol I(l,m)$ at any point is linked to the flux
$\boldsymbol F(l,m)$ and the total linewidth
$\boldsymbol\sigma^{\text{tot}}(l,m)$ (eq.~\ref{eq:sigmatot})
by:
\begin{equation}
  \boldsymbol I(l,m) = \frac{\boldsymbol F(l,m)}{\sqrt{2\pi}\boldsymbol\sigma^{\text{tot}}(l,m)}\text{.}
  \label{eq:I}
\end{equation}

The reason for this choice is that line amplitude depends on the instrumental spectral resolution, which in our case varies across the field with sharp edges (Sect.~\ref{sect:OHfit}). With our choice of parameters (line flux, intrinsic width and velocity), we only fit astrophysical quantities, clean of instrumental signatures.

We have also tried fitting the two H$_2$ lines separately as well as
together. In the regions where the two lines have sufficient SNR, the
separate fits did not show a significant difference in radial velocity
or linewidth, with difference histograms compatible with statistical
uncertainties. We therefore chose to fit the two lines together. On
the contrary, Br$\gamma$ shows very different morphology and dynamics
as demonstrated below and is fitted separately from H$_2$. When
fitting two lines together, the quantity of interest is the line ratio
rather than each line flux separately. Dividing one flux map by the
other causes several difficulties. Division by small factors with low
SNR is a classical issue. In addition, the regularization in our
method may lead to two maps with slightly different effective
resolution, which would lead to artifacts in the division. We
therefore use the ratio itself as a parameter rather than the two line
fluxes.

On each mosaic, we have finally performed two independent fits on the
continuum-subtracted data: one on the Br$\gamma$ line alone where the
three parameters are line flux $\boldsymbol F_{\text{Br}\gamma}$,
intrinsic width $\boldsymbol \sigma_{\text{Br}\gamma}$, and intrinsic
radial velocity shift $\boldsymbol v_{\text{Br}\gamma}$, and one on
both H$_2$ lines at once where the four parameters are H$_2$ $\lambda
2.12\mu$m line flux $\boldsymbol F_{\text{H}_2}$, flux ratio between
the two lines $\boldsymbol r_{\text{H}_2}$, common width $\boldsymbol
\sigma_{\text{H}_2}$ and common Doppler shift $\boldsymbol
v_{\text{H}_2}$. The amplitude of the H$_2$ $\lambda 2.12\mu$m line
$\boldsymbol I_{2.12}$ is computed as per eq.~\ref{eq:I} and the
amplitude of the H$_2$ $\lambda 2.12\mu$m line is $\boldsymbol
I_{2.22}=\boldsymbol r \times \boldsymbol I_{2.12}$.

Finally, the 3D model functions $\mathcal M^{\text{Br}\gamma}$ and $\mathcal M^{H_2}$ are
expressed below using the multi-line Gaussian function $ \mathcal
G^{\{\lambda_i\}}$ from eq.~\ref{eq:G}:
\begin{align}
  \mathcal M^{\text{Br}\gamma}_{\boldsymbol F_{\text{Br}\gamma}, \boldsymbol v_{\text{Br}\gamma}, \boldsymbol\sigma_{\text{Br}\gamma}}(l, m, \lambda) &=
  \mathcal G^{\lambda_{\text{Br}\gamma}}_{\boldsymbol I_{\text{Br}\gamma}(l,m), \boldsymbol v_{\text{Br}\gamma}^{\text{tot}}(l,m), \boldsymbol \sigma_{\text{Br}\gamma}^{\text{tot}}(l,m)}(\lambda) \hspace{5mm}\text{and}\\
  \mathcal M^{H_2}_{\boldsymbol F_{\text{H}_2}, \boldsymbol r_{\text{H}_2}, \boldsymbol v_{\text{H}_2}, \boldsymbol\sigma_{\text{H}_2}}(l, m, \lambda) &=
  \mathcal G^{\lambda_{2.12}, \lambda_{2.22}}_{\boldsymbol I_{2.12}(l,m), \boldsymbol I_{2.22}(l,m), \boldsymbol v_{\text{H}_2}^{\text{tot}}(l,m), \boldsymbol \sigma_{\text{H}_2}^{\text{tot}}(l,m)}(\lambda)\text{.}
\end{align}

\begin{figure*}
  \begin{center}
    \includegraphics[width=\textwidth]{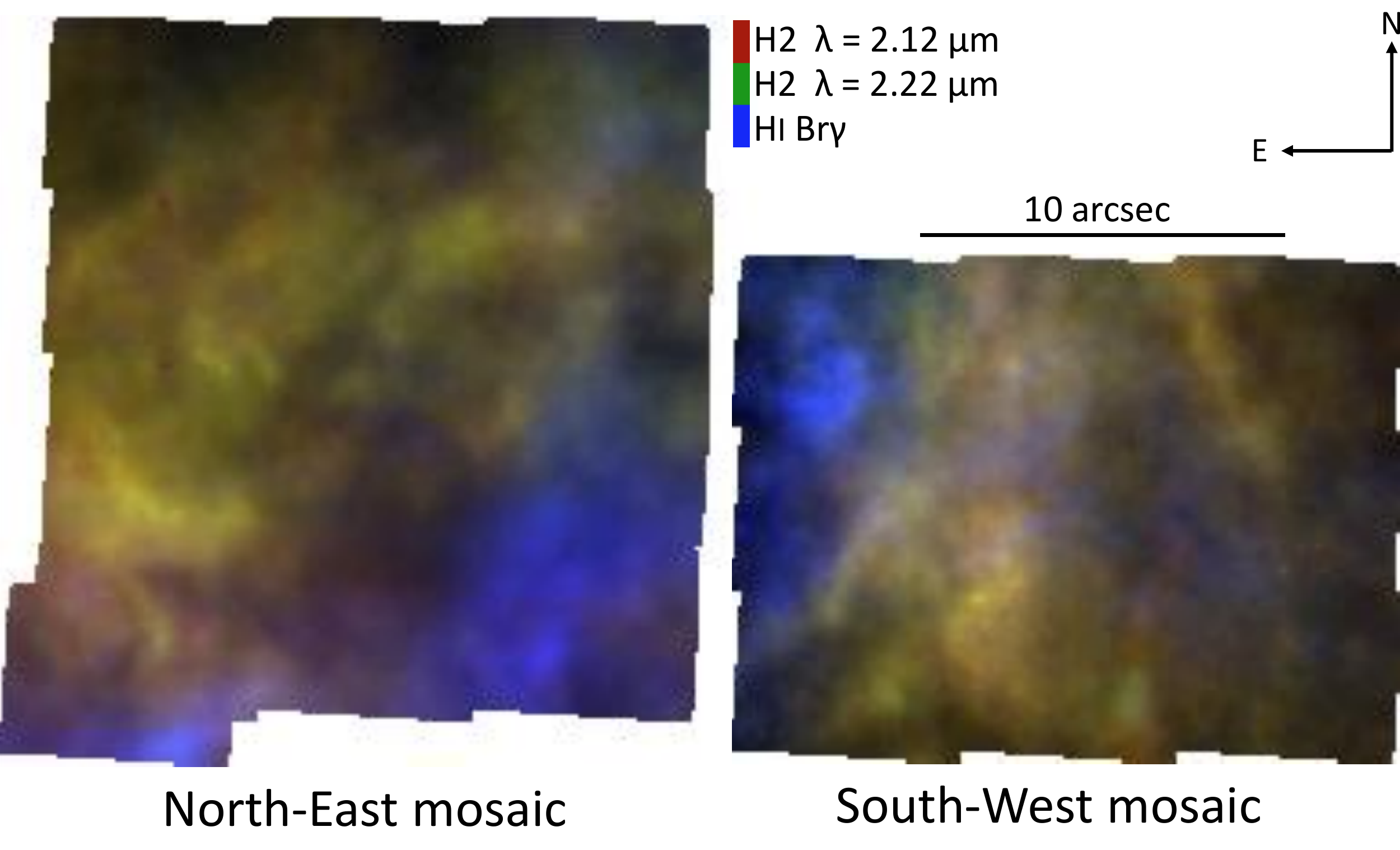}
  \end{center}

  \caption{Composite line maps of the NE (left panel) and
    SW (right panel) mosaics. The flux of each of the three
    lines of interest has been assigned a color (tints represented by
    the color bar at the top). \emph{Red:} H2 ($2.12\,\mu$m),
    \emph{green:} H2 ($2.22\,\mu$m), \emph{blue:} \ion{H}{ii}
    (traced by Br$\gamma$, $2.17\,\mu$m). Line flux is estimated by our fitting
    procedure. Regularization ensures that the maps are smooth and
    provides for extrapolation (in particular, there is no gap anymore
    between the two fields). Actually, it also provides some
    extrapolation but we have cut the field of view to the initial one
    (except for the gap).}
  \label{fig:both-mosaic-3color}
\end{figure*}

The H$_2$ fit on the NE mosaic was performed in two passes. The first
time, only $\boldsymbol F_{\text{H}_2}$ and $\boldsymbol
v_{\text{H}_2}$ were fit. $\boldsymbol r_{\text{H}_2}$ was set to
$0.3$ and $\boldsymbol\sigma_{\text{H}_2}$ to $40\;$km~s$^{-1}$, both
constant across the field. The Gaussian-smoothed and noise-added
result of this fit was then used as the initial guess for a second fit
where all parameters were free. A direct one-pass fit with constant
maps as initial guess for all parameters had led to an erroneous local
minimum. This two-pass treatment was not necessary for the SW
mosaic. The Br$\gamma$ fits were also done in two passes, where
$\boldsymbol\sigma$ was fixed to 20~km~s$^{-1}$ during the first pass.
Figure~\ref{fig:both-mosaic-3color} shows the flux distribution of the
three lines as recovered by CubeFit. The fixed values for the first
pass where chosen as typical values as evaluated in previous tries in
regions where the fit was of good quality.

The resulting model cubes were interactively compared to the data cube
using Cubeview for
Yorick\footnote{\url{https://github.com/paumard/yorick-cubeview}}
\citep{2003PhDT........21P} which was extended for this purpose.  This
inspection revealed imperfect OH subtraction in very low SNR
regions. OH lines very close to Br$\gamma$ and to the H$_2$ lines
could bias our velocity and velocity width measurements. We therefore
have decided to cut the lowest SNR regions out of our final maps. This
threshold is applied on maximum intrinsic line flux per spectral
channel:
\begin{equation}
  \boldsymbol F_\mathrm{chan} = \frac{\boldsymbol F}{\sqrt{2\pi}\boldsymbol\sigma}\times\Delta\lambda
  \label{eq:Fchan}
\end{equation}
where $\Delta\lambda$ is the bandwidth of a spectral channel.  We consider a detection to be significant and parameters to be unbiased on spaxels with $\boldsymbol F_\mathrm{chan}>\boldsymbol F_\mathrm{chan,thresh}$ where $F_\mathrm{chan,thresh}$ is twice the median over the field of
view of the root-mean-square (RMS) of the residuals within each spaxel. In the absence of
calibration residuals such as those caused by the OH lines, our method
would not require a threshold to be set.

The regularization has the effect of smoothing noise without smoothing
significant features in the various maps. It therefore serves our goal
to increase SNR locally without degrading resolution. When data are
missing altogether, the regularization term makes the fitting procedure
act as an interpolation function: the output parameter maps therefore have less
holes than the original data.

For estimating uncertainties, we follow the procedure described in \cite{2016A&A...594A.113C}, which makes use of this property: we generate four independent subsets of the original data by selecting every second row and every second column of spaxels. We then apply CubeFit to each sub-set independently. The RMS of the 4 independent estimates divided by $\sqrt 4 = 2$ is taken to represent the statistical uncertainties of the original data.

Table~\ref{table:errs} lists statistical properties of the maps
and their uncertainties: the flux per channel threshold ($\boldsymbol F_\mathrm{chan,thresh}$) for each map, and within the region where 
$\boldsymbol F_\mathrm{chan}>\boldsymbol F_\mathrm{chan,thresh}$, the minimum and maximum of $\boldsymbol{F}$ and the median of the uncertainties in the four parameters. As a comparison, the median uncertainties using a traditional 1D fitting method (directly estimated by the
Yorick\footnote{\url{https://software.llnl.gov/yorick-doc/}}
Levendberg-Marquadt-based fitting engine
\texttt{lmfit}\footnote{\url{https://github.com/frigaut/yorick-yutils/blob/master/lmfit.i}})
for the NE H$_2$ data are  
$\widetilde{\sigma_F}\approx3.74\times10^{-27}$~W~m$^{-2}$~arcsec$^{-2}$,
$\widetilde{\sigma_r}\approx10$~$\%$,
$\widetilde{\sigma_v}\approx9.4$~km~s$^{-1}$,
$\widetilde{\sigma_\sigma}\approx9.2$~km~s$^{-1}$. The 1D-fit
uncertainties can also be estimated by taking the RMS of each
parameter map over 4 neighboring points, yielding median values of
$\widetilde{\sigma_F}\approx1.24\times10^{-26}$~W~m$^{-2}$~arcsec$^{-2}$,
$\widetilde{\sigma_r}\approx15$~$\%$,
$\widetilde{\sigma_v}\approx13.9$~km~s$^{-1}$ and
$\widetilde{\sigma_\sigma}\approx23.7$~km~s$^{-1}$. The CubeFit
uncertainties are therefore about 10 times smaller than their 1D
counterparts.\label{sect:1Duncertainties}

\begin{table*}[tb]
  \begin{minipage}[t]{\textwidth}
  \caption{Statistical properties of the various
    fits\label{table:errs}. $\boldsymbol{\widetilde{x}}$ denotes the
    median of variable $\boldsymbol{x}$ over the region in which the
    line detection is significant.}  \centering
  \begin{tabular}{cc|ccccccccc}
    \hline\hline
    mosaic&species &
    $\boldsymbol F_\mathrm{chan,thresh}$\footnote{$\times10^{-27}$~W.m$^{-2}$.arcsec$^{-2}$ per channel} &
    $\min(\boldsymbol F)$\footnote{\label{fn:fluxunit}$\times10^{-27}$~W.m$^{-2}$.arcsec$^{-2}$} &
    $\max(\boldsymbol F)$\footnoteref{fn:fluxunit} &
    $\boldsymbol{\widetilde{\sigma_F}}$\footnoteref{fn:fluxunit} &
    $\boldsymbol{\widetilde{\sigma_r}}$\footnote{$\%$} &
    $\boldsymbol{\widetilde{\sigma_v}}$\footnote{\label{fn:vunit}km~s$^{-1}$} &
    $\boldsymbol{\widetilde{\sigma_\sigma}}$\footnoteref{fn:vunit}
    \\
    \hline
    NE&H$_2$ &
    $3.68$ &
    $8.5$ &
    $67.4$ &
    $0.88$ &
    $1.1$ &
    $1.8$ &
    $0.95$
    \\
    NE&Br$\gamma$ &
    $1.72$ &
    $3.2$ &
    $30.2$ &
    $0.46$ &
    -- &
    $1.5$ &
    $2.6$
    \\ 
    SW&H$_2$&
    $3.67$&
    $7.4$ & 
    $75.7$ & 
    $1.11$ & 
    $1.0$ & 
    $1.4$ & 
    $1.5$ 
    \\ 
    SW&Br$\gamma$&
    $1.86$&
    $3.0$ & 
    $53.6$ & 
    $0.42$ & 
    -- & 
    $1.1$ & 
    $1.6$ 
    \\
    \hline\hline
  \end{tabular}
  \end{minipage}
\end{table*}

\begin{figure*}[tbh!]
  \centering
  \includegraphics[scale=0.4, viewport=0 50 460 413, clip]{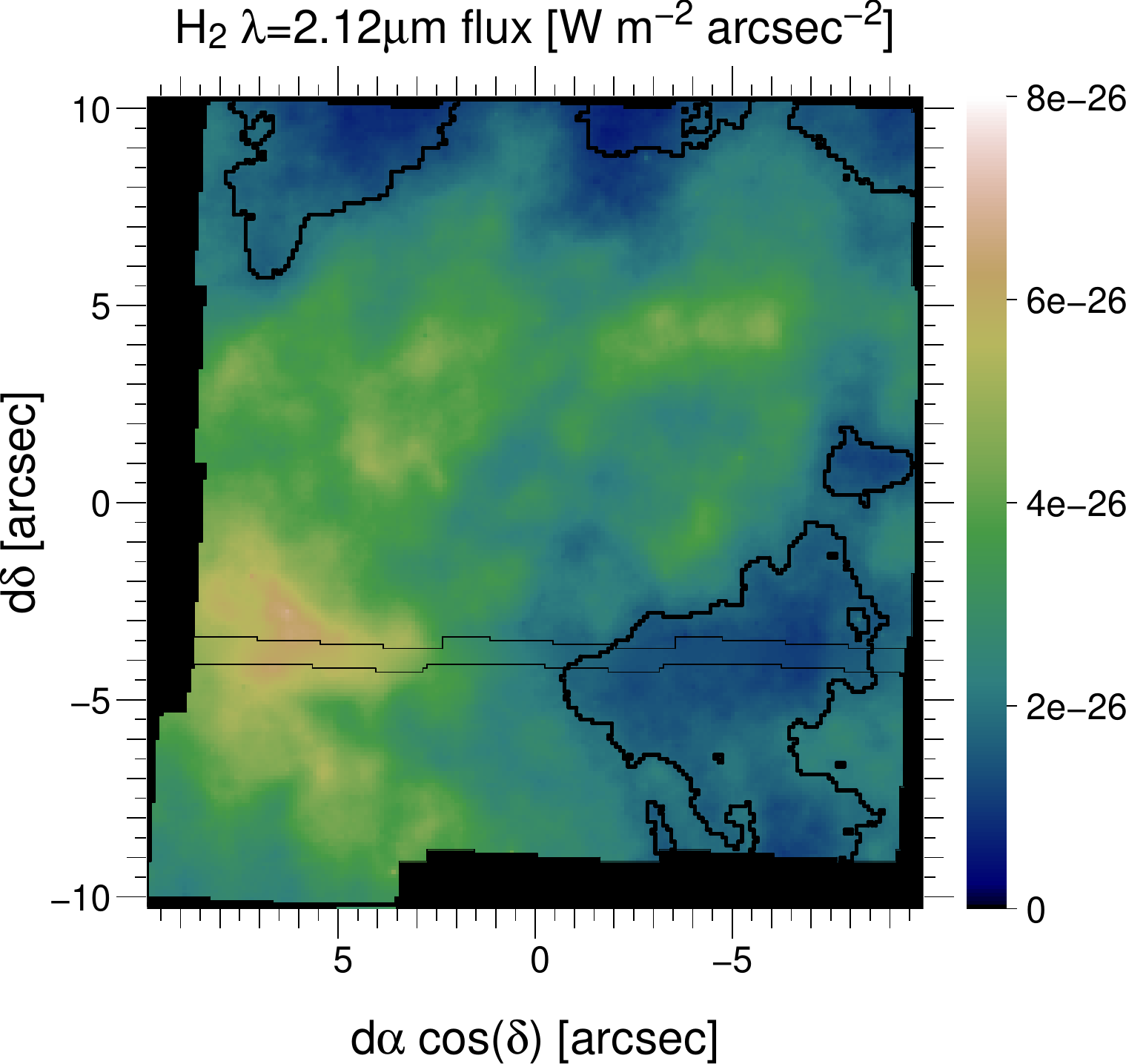}
  \includegraphics[scale=0.4, viewport=44 50 460 413, clip]{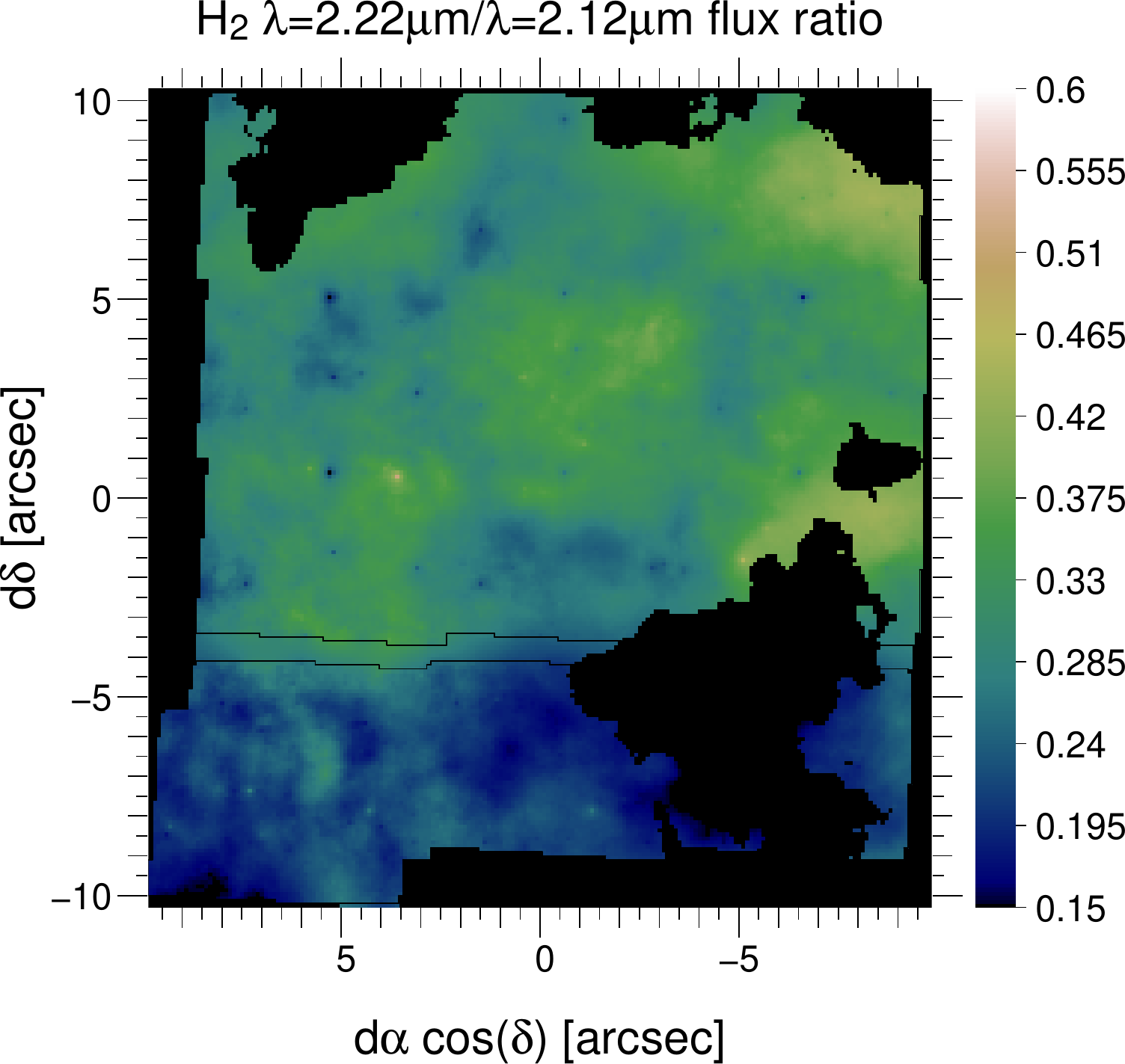}
  \vspace{2mm}\\
  \includegraphics[scale=0.4, viewport=0 0 460 413, clip]{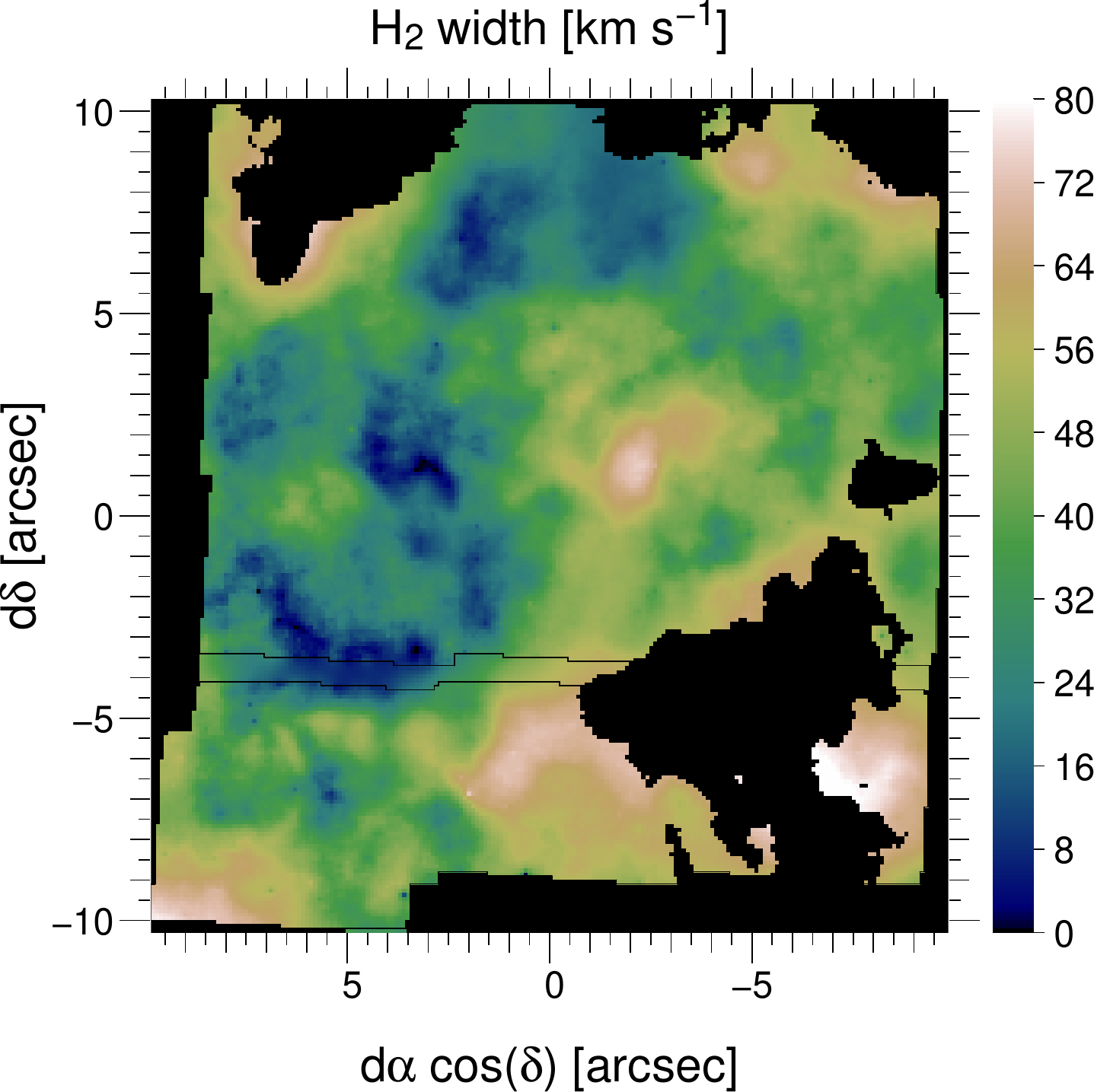}
  \includegraphics[scale=0.4, viewport=44 0 460 413, clip]{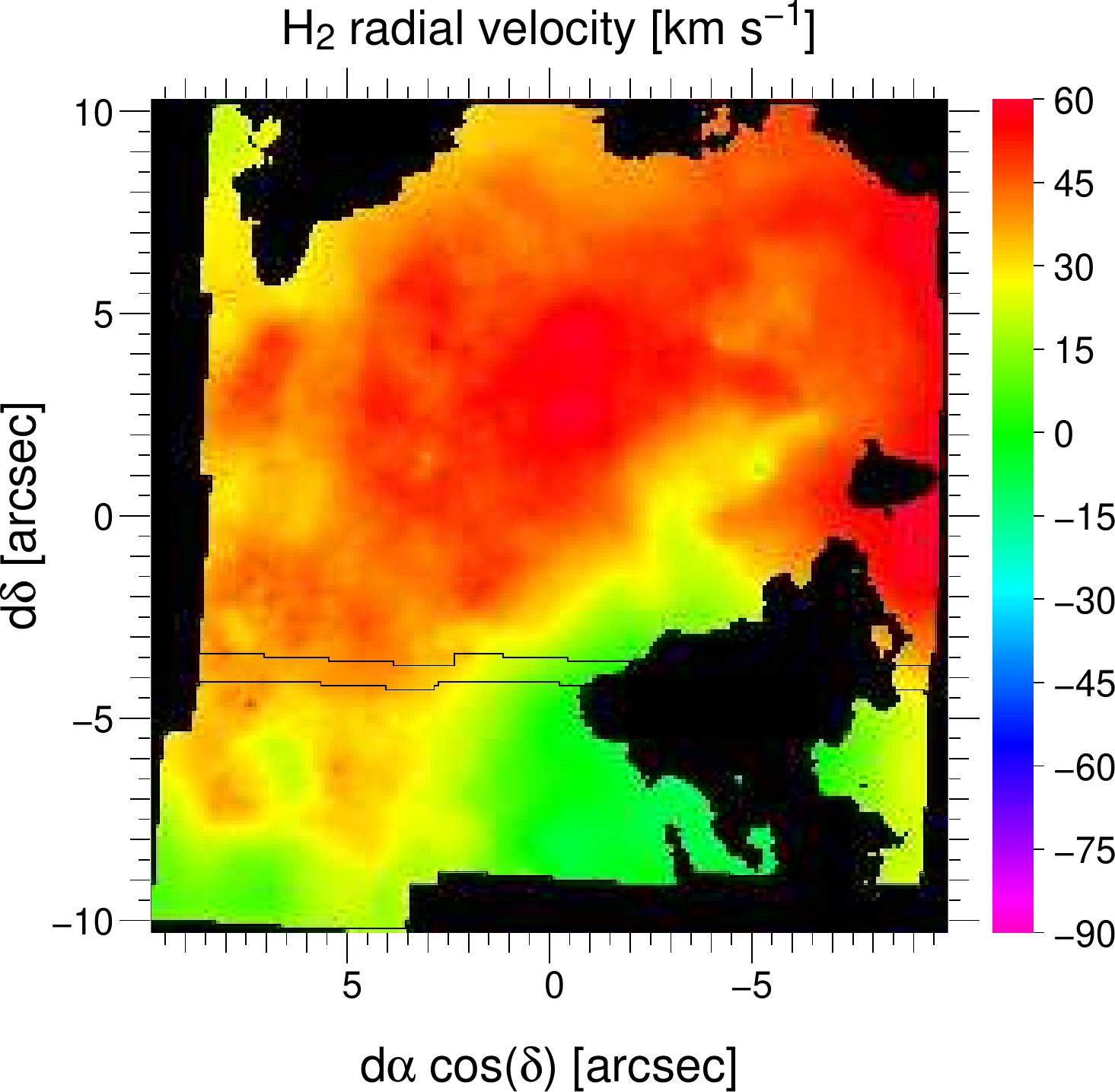}
  \caption{Result of CubeFit on the NE mosaic for both
    H$_2$ lines. A black curve on the flux map delineates the
    amplitude threshold (see text). This region is masked out in the
    other maps. The four maps are, from top left to bottom right:
    H$_2$ $\lambda 2.12\mu$m line flux $\boldsymbol F_{\text{H}_2}$,
    flux ratio between the two lines $\boldsymbol r_{\text{H}_2}$,
    common intrinsic linewidth $\boldsymbol \sigma_{\text{H}_2}$ and
    common intrinsic radial velocity shift $\boldsymbol
    v_{\text{H}_2}$.}
  \label{fig:NE-maps-H2}
\end{figure*}

\begin{figure*}[t]
  \centering \includegraphics[scale=0.4, viewport=0 0 460 413, clip]{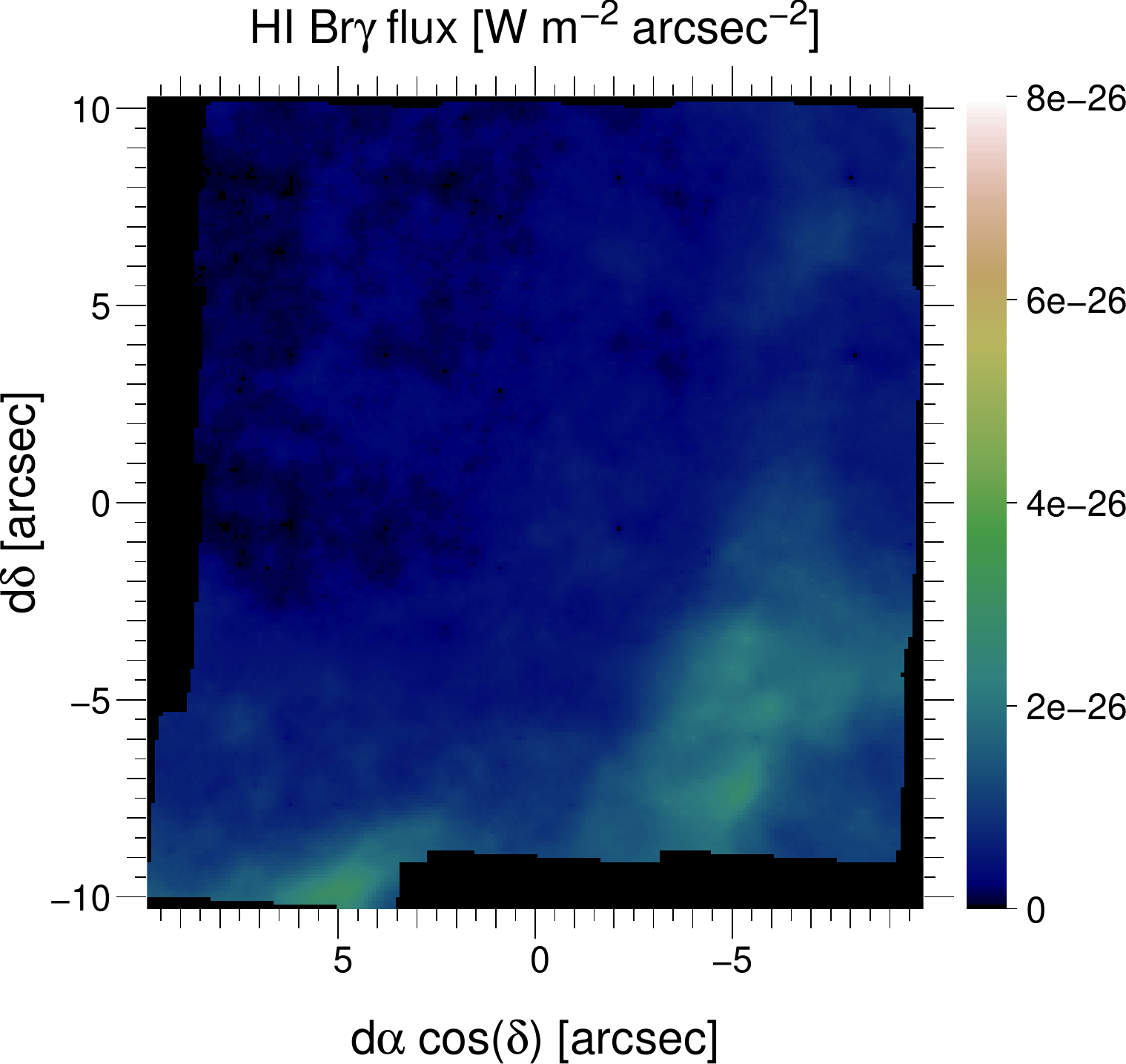}
  \includegraphics[scale=0.4, viewport=44 0 460 413, clip]{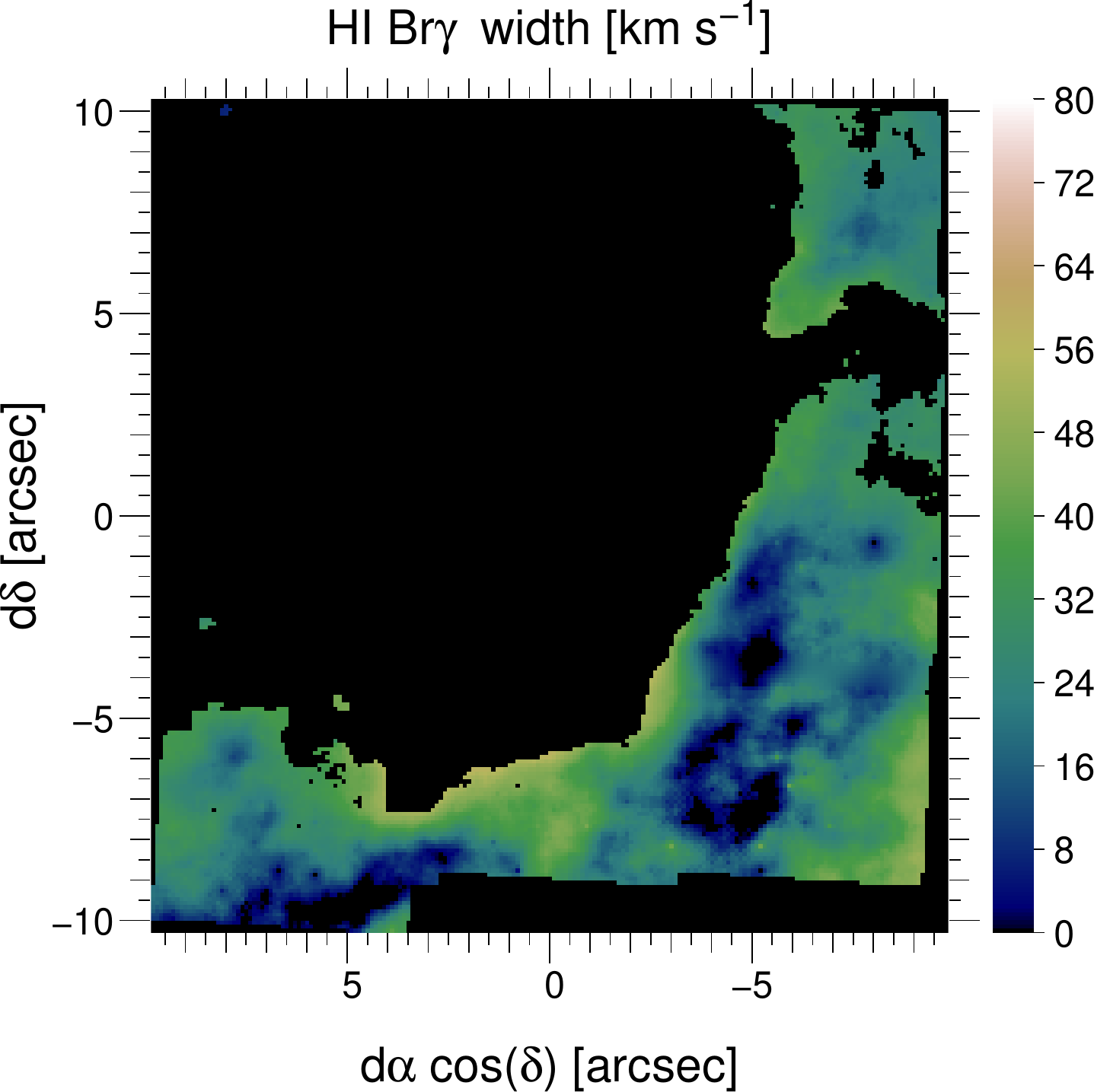}
  \includegraphics[scale=0.4, viewport=44 0 460 413, clip]{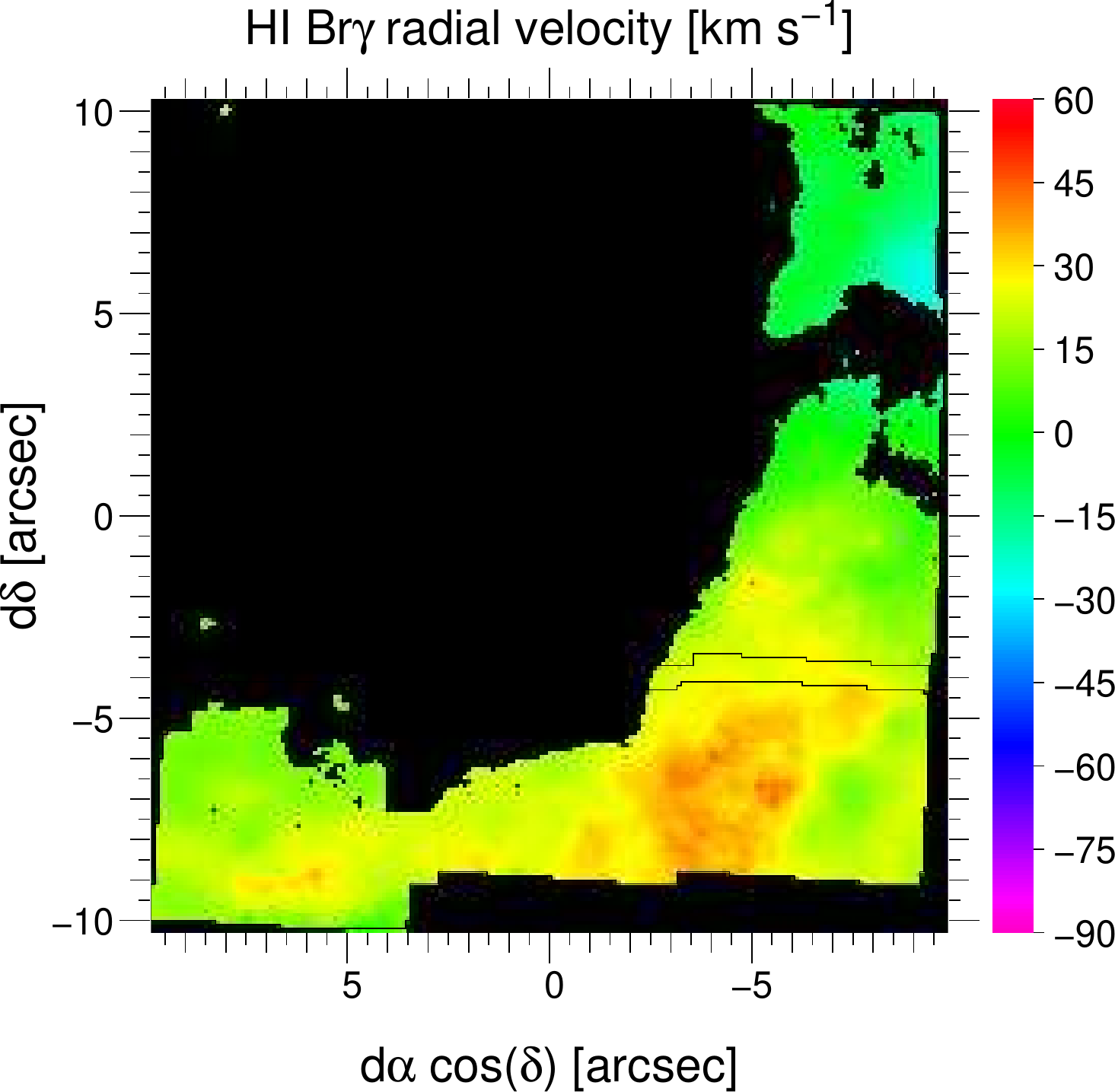}
  \caption{Result of CubeFit on the NE mosaic for the
    Br$\gamma$ line. A black curve on the flux map delineates the
    amplitude threshold (see text). This region is masked out in the
    other maps. The three maps are, from left to right: line flux
    $\boldsymbol F_{\text{Br}\gamma}$, intrinsic width $\boldsymbol
    \sigma_{\text{Br}\gamma}$, and intrinsic radial velocity shift
    $\boldsymbol v_{\text{Br}\gamma}$.}
  \label{fig:NE-maps-HI}
\end{figure*}

\subsection{NE mosaic}

The results for the NE mosaic are presented in
Figs.~\ref{fig:NE-maps-H2} and \ref{fig:NE-maps-HI} for H$_2$ and
Br$\gamma$, respectively. The Br$\gamma$ line arises from neutral hydrogen but it traces the \ion{H}{ii} region. The corresponding uncertainty maps are in
Figs.~\ref{fig:NE-errmaps-H2} and \ref{fig:NE-errmaps-HI} in
Appendix~\ref{appendix:errmaps}. The median uncertainties are listed
in Table~\ref{table:errs}.

H$_2$ is significantly detected almost everywhere in the field,
whereas \ion{H}{ii} is concentrated along the southern and western
edges of the field, i.e. on the area nearest to the central
cavity. The various maps of the two species are very different from
each other, so that H$_2$ and \ion{H}{ii} presumably belong in
distinct volumes of the interstellar medium. In particular the radial
velocity map for H$_2$ shows a rather smooth gradient from
$0$~km~s$^{-1}$ in the southeastern portion of the mosaic to
$60$~km~s$^{-1}$ in the northwestern area, while the velocity measured
in \ion{H}{ii} is near $0$~km~s$^{-1}$ in the northwestern area and
approaches $40$~km~s$^{-1}$ in the southwestern area. This variation
does not appear as a smooth overall gradient. A plausible explanation
is that H$_2$ is dominated by the bulk orbital motion of the CND,
while the ionized gas is more perturbed by its interactions with the
nuclear cluster. The two line flux maps appear clumpy, with
no clear spatial coherence.

\begin{figure}
  \includegraphics[width=\columnwidth]{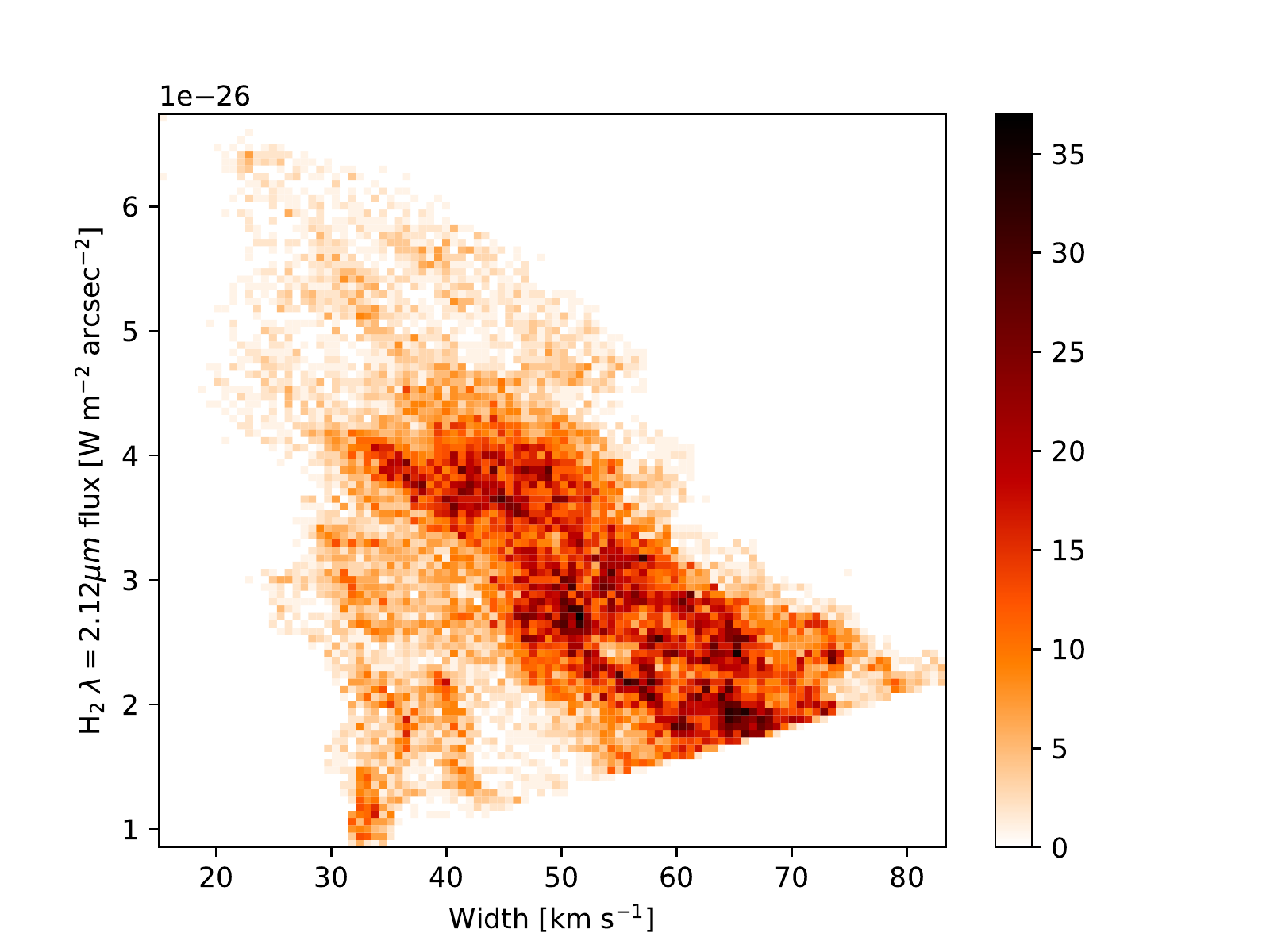}
  \includegraphics[width=\columnwidth]{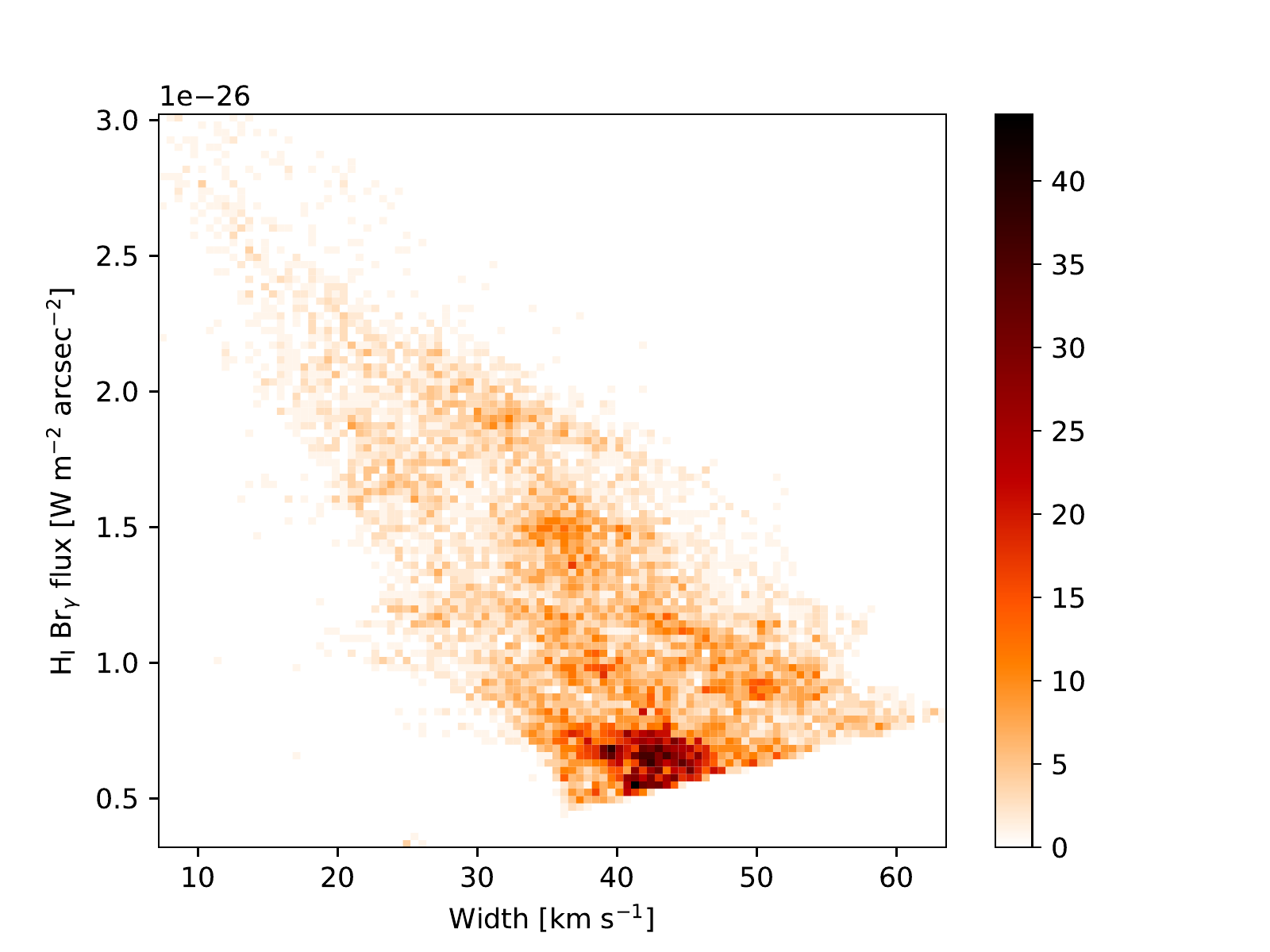}
  \caption{Line flux vs. width density plot (arbitrary units) in the NE
      mosaic. \emph{Top:} H$_2$ ($\lambda=2.12\;\mu$m); \emph{Bottom:}
      \ion{H}{ii}. The threshold imposed on intensity results in a
      tilted lower selection boundary on this plot. For both species, the
      two quantities are anti-correlated. Although remarkable, the
      overall anti-correlation is not very tight. Denser regions in the
      plot could be the trace of individual clumps in the CND
      with a tighter local anti-correlation, modulated by variable foreground
      extinction.\label{fig:NEcor}}
\end{figure}

\begin{figure}
\centering
  \includegraphics[scale=0.4, viewport=0 55 475 380, clip]{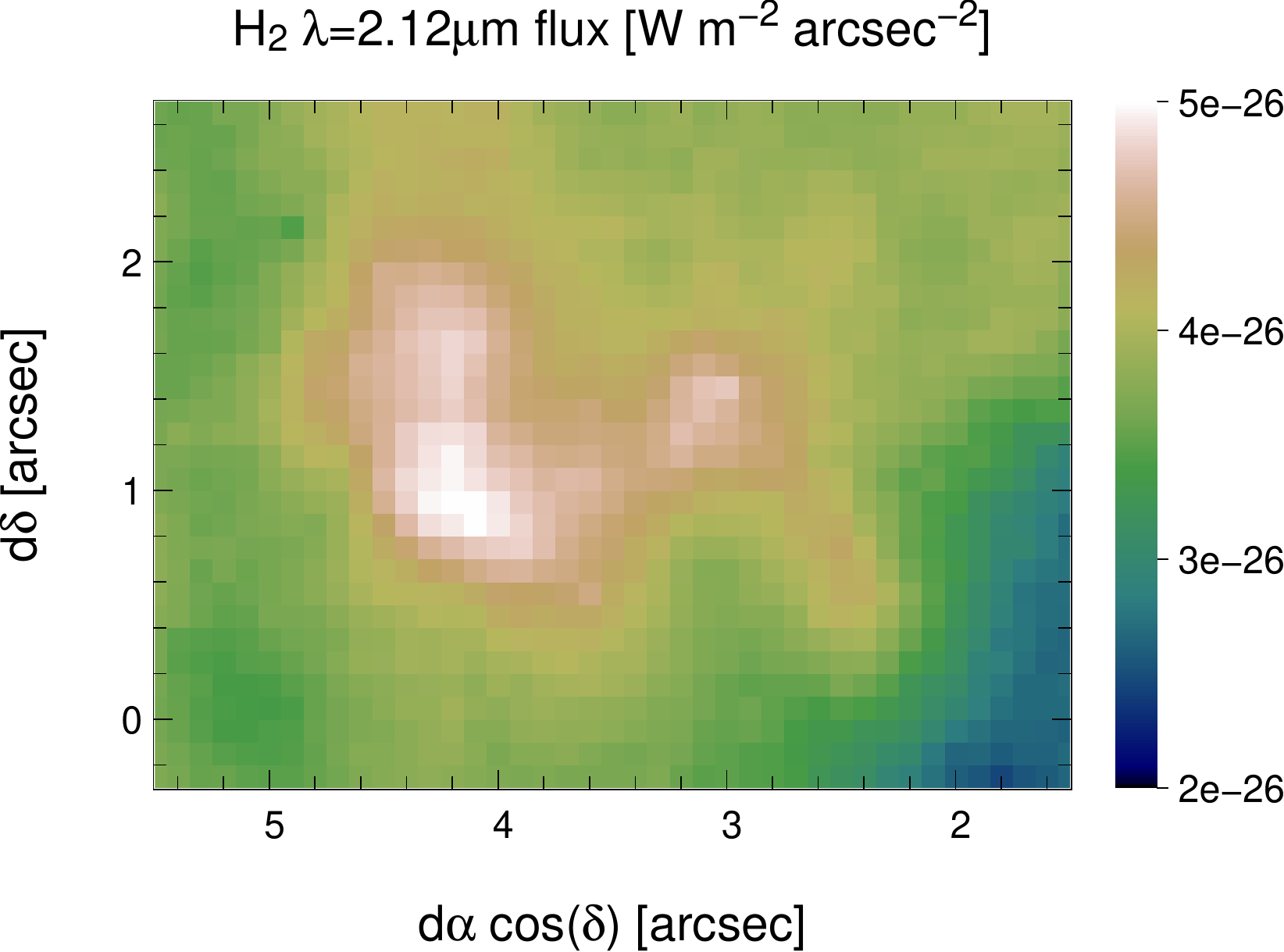}
  \includegraphics[scale=0.4, viewport=0 0 475 380, clip]{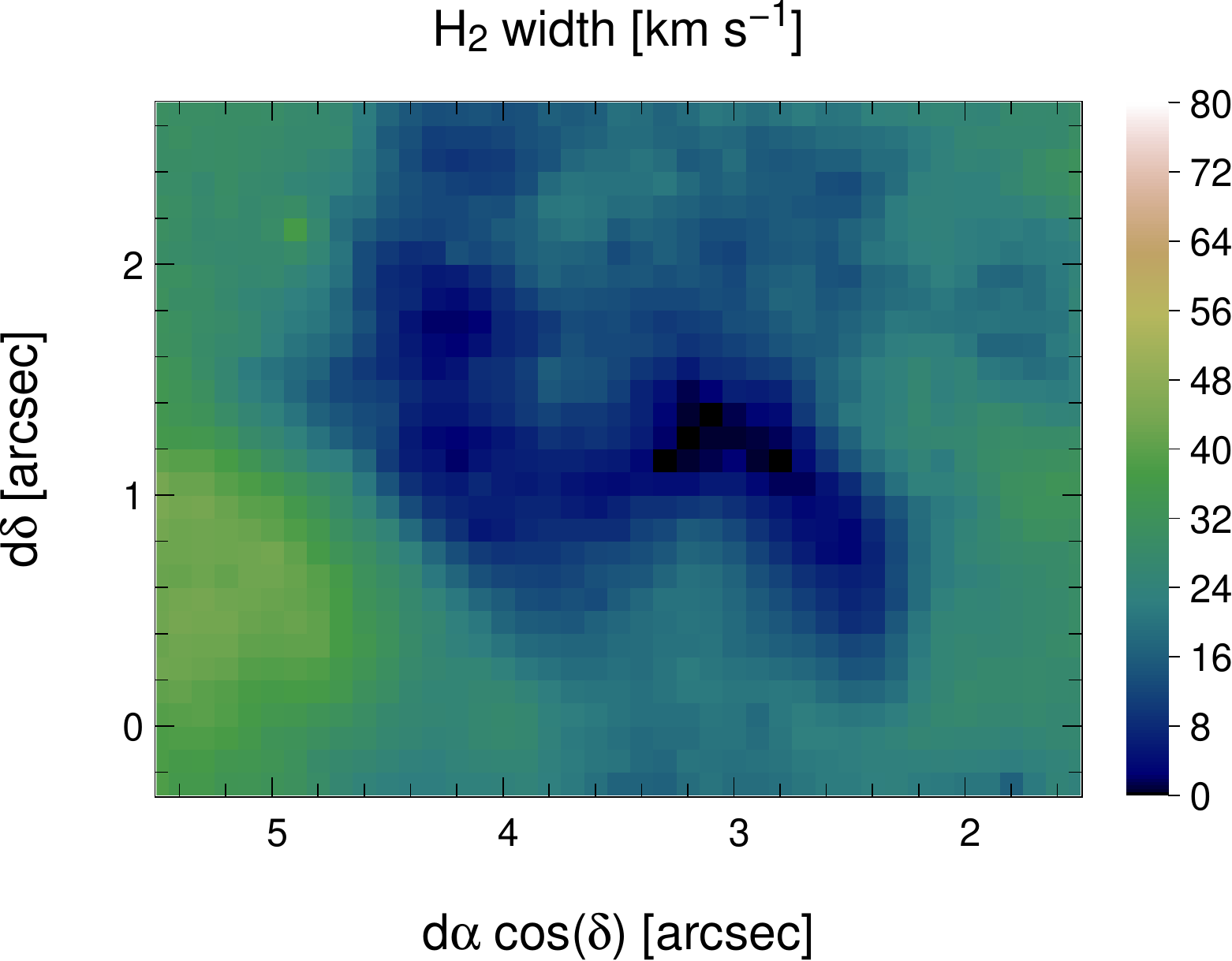}
  \caption{Zoom-in on a feature in the NE H$_2$ mosaic 
      ($\lambda=2.12\;\mu$m) line flux and width maps. The bright
      S-shaped ridge in the flux maps corresponds to a similarly
      shaped valley in the width map.\label{fig:NEzoom}}
\end{figure}

For both H$_2$ and \ion{H}{ii}, the intrinsic linewidth map shows considerable
structure with small, sharp features.  The median of this width over
the spaxels where H$_2$ is reliably detected is
$\widetilde{\boldsymbol{\sigma}_{\text{H}_2}}\approx50$~km~s$^{-1}$,
and the minimum and maximum values are
$\min{\boldsymbol{\sigma}_{\text{H}_2}}\approx15.0\pm0.6$~km~s$^{-1}$
and
$\max{\boldsymbol{\sigma}_{\text{H}_2}}\approx83.4\pm2.4$~km~s$^{-1}$. Similarly,
for \ion{H}{ii},
$\widetilde{\boldsymbol{\sigma}_{\text{Br}\gamma}}\approx40$~km~s$^{-1}$,
$\min{\boldsymbol{\sigma}_{\text{Br}\gamma}}\approx7.21\pm1.48$~km~s$^{-1}$
and
$\max{\boldsymbol{\sigma}_{\text{Br}\gamma}}\approx63.57\pm4.75$~km~s$^{-1}$. Most
interestingly, the linewidth maps appear anti-correlated with the line
flux maps. This anti-correlation can be seen in
  Fig.~\ref{fig:NEcor}, which displays line flux versus linewidth for
  the two species. It is reminiscent of the correlation seen by
  \citet{2019A&A...621A..65C} between dereddened line flux and
  extinction for H$_2$ detected in the central cavity. Like them, we
  can hypothesize that the more densely populated regions in the plot represent individual clumps for which a tight (anti-)correlation exists,
  each one being affected by a different amount of foreground
  extinction. For instance, the H$_2$ width map shows sharp features
in dark colors (i.e. small values) near $(\mathrm d\alpha
\cos(\delta), \mathrm d\delta)\approx (6'', -3'')$ and $(3'', 1'')$
where the flux map shows local maxima of matching
shape. Figure~\ref{fig:NEzoom} shows a zoom-in on the two maps of
  the latter region. Since we are using an original method to measure
line flux and width, we verify in Appendix~\ref{appendix:NE_H2_comp}
that this correlation is not an artifact of this new method.

\subsection{SW mosaic}

\begin{figure*}[!ht]
  \centering
  \includegraphics[scale=0.4, viewport=0 50 460 338, clip]{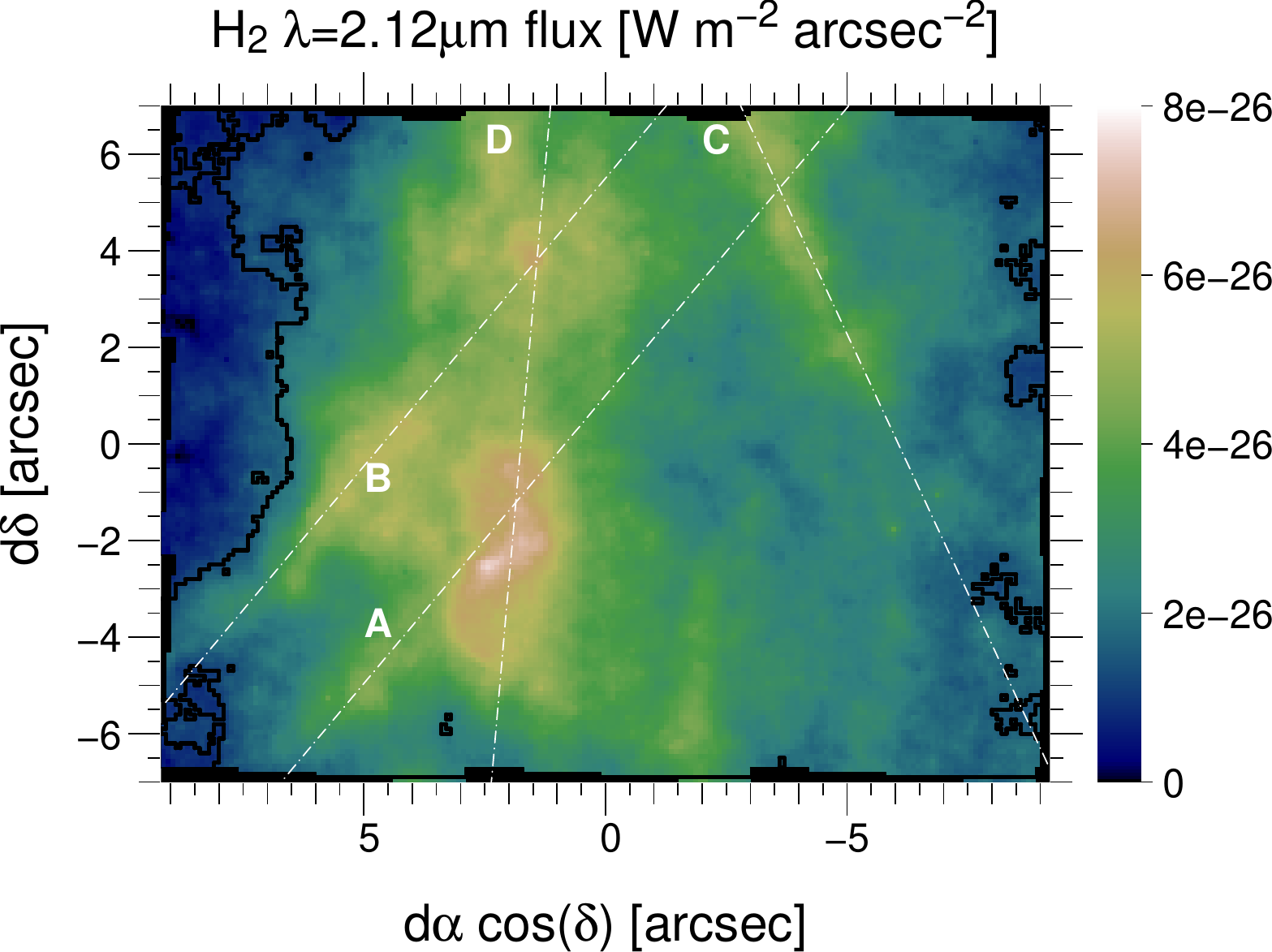}
  \includegraphics[scale=0.4, viewport=44 50 460 338, clip]{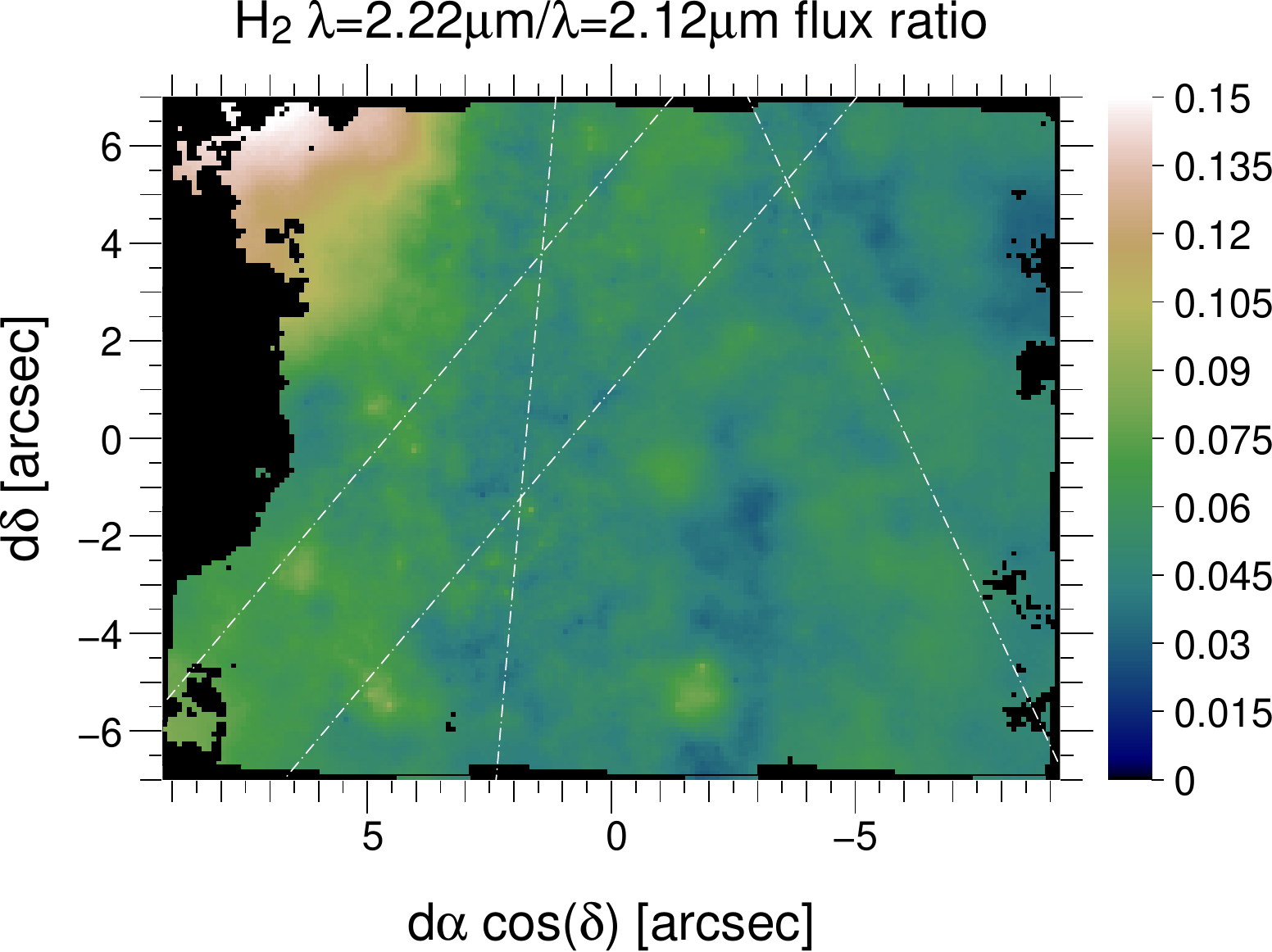}\\
  \vspace{2mm}
  \includegraphics[scale=0.4, viewport=0 0 460 338, clip]{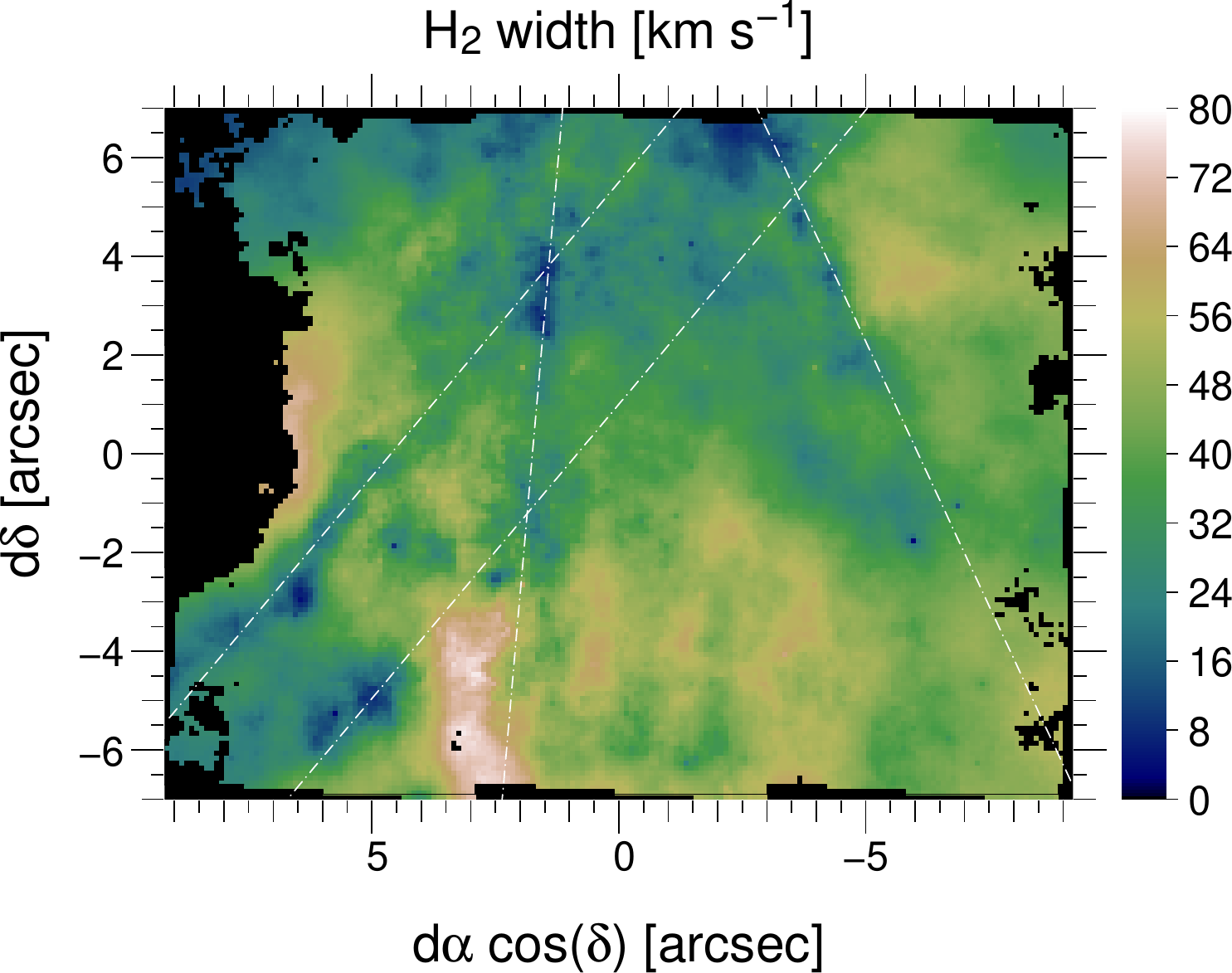}
  \includegraphics[scale=0.4, viewport=44 0 460 338, clip]{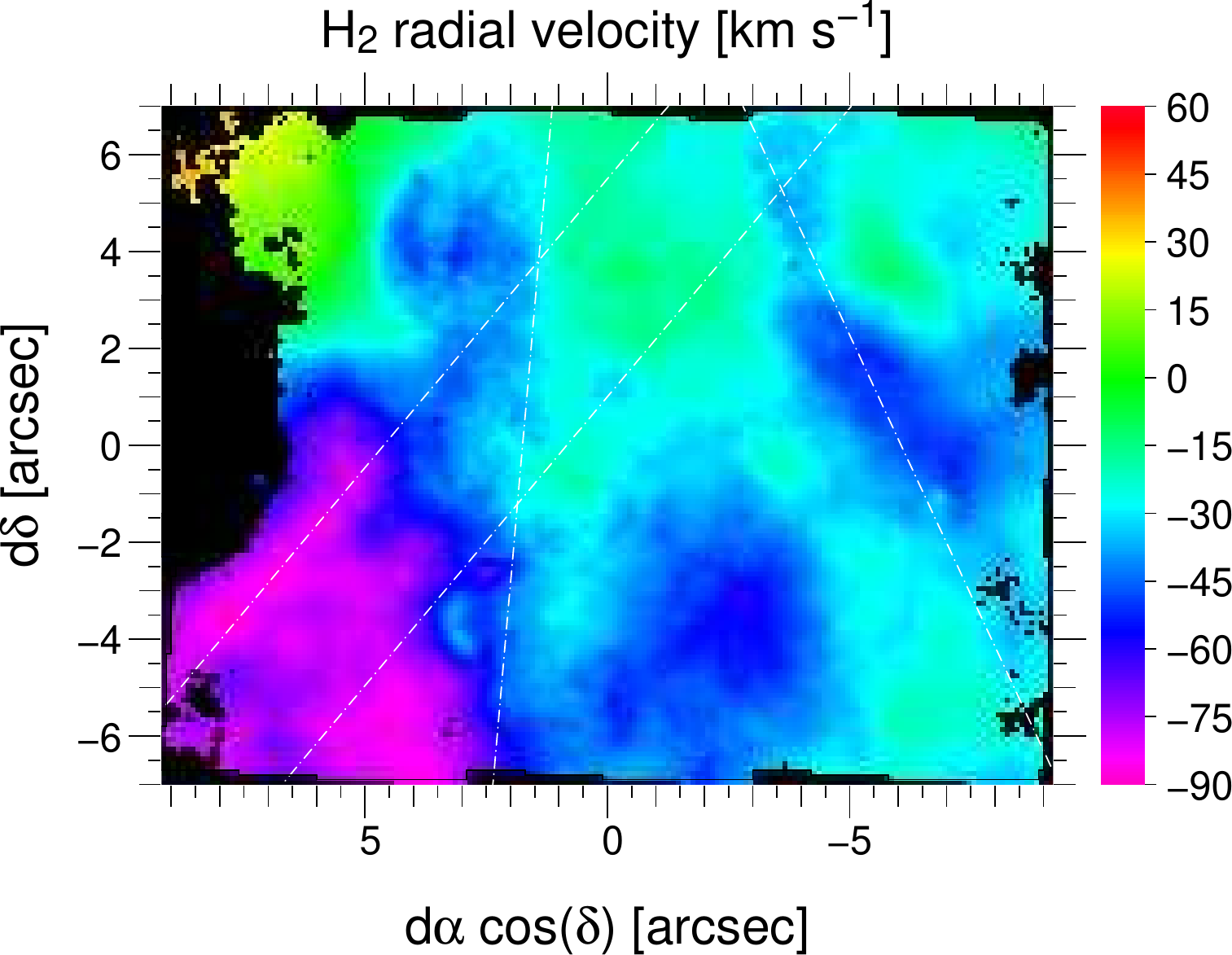}
  \caption{H$_2$ in the SW mosaic: same panels as Fig.~\ref{fig:NE-maps-H2}. White,
    dash-dotted lines indicate the locations of virtual slits used in
    Fig.~\ref{fig:SW-spectrograms} and underline filamentary
    features.}
  \label{fig:SW-maps-H2}
\end{figure*}

\begin{figure*}[!ht]
  \centering
  \includegraphics[scale=0.4, viewport=0 0 460 338, clip]{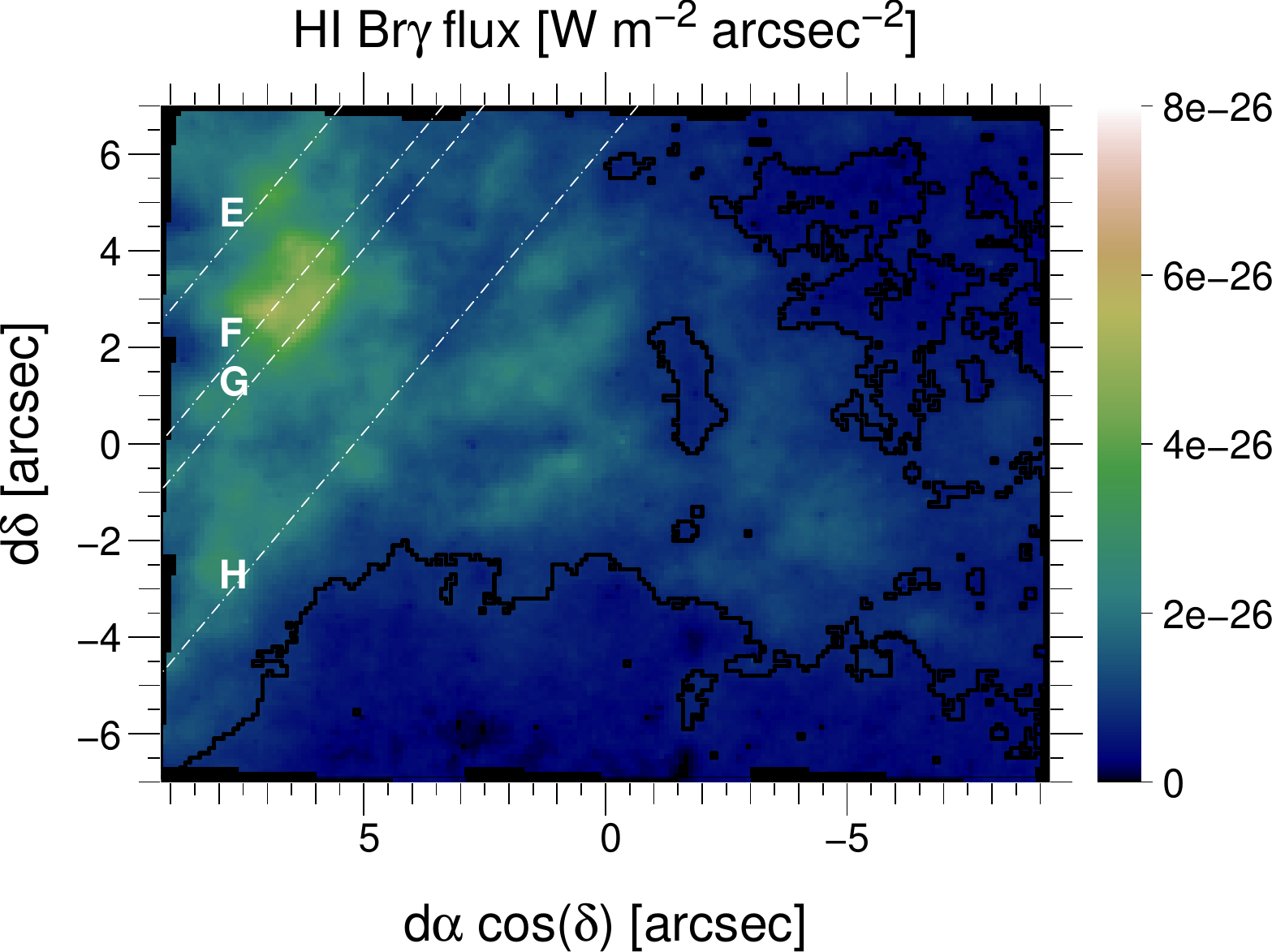}
  \includegraphics[scale=0.4, viewport=44 0 460 338, clip]{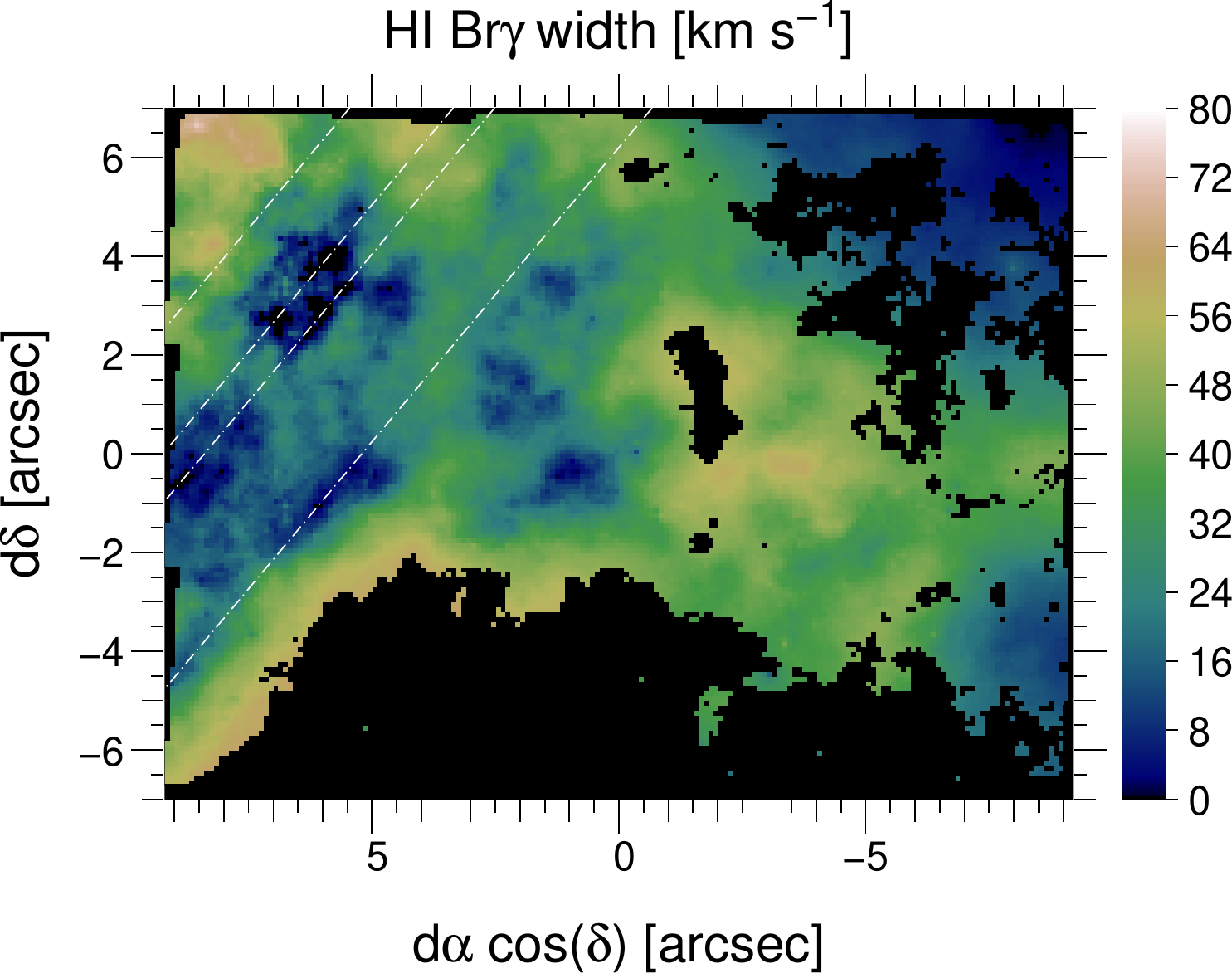}
  \includegraphics[scale=0.4, viewport=44 0 460 338, clip]{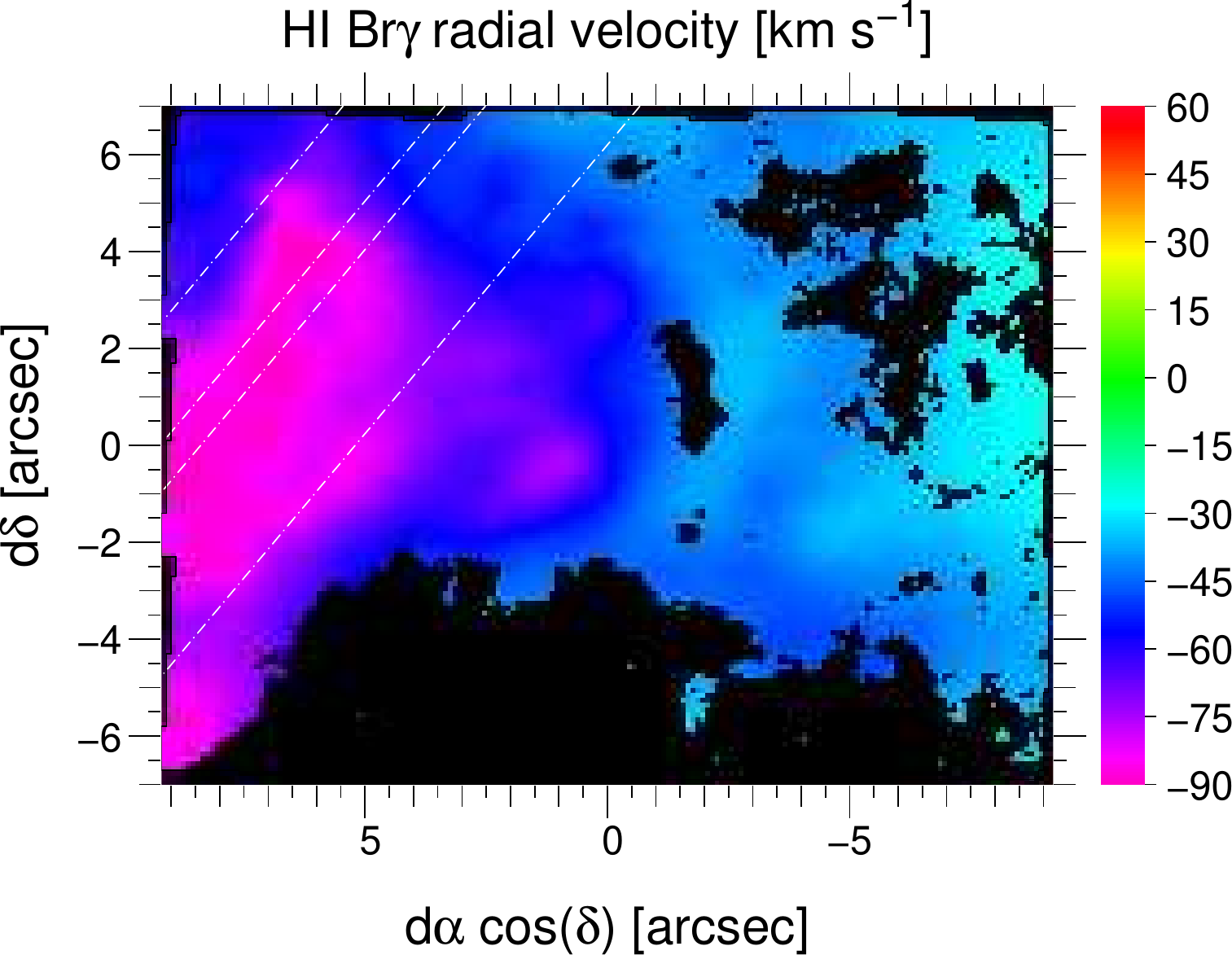}
  \caption{Brackett-$\gamma$ emission in the SW mosaic: same panels as Fig.~\ref{fig:NE-maps-HI}. White,
    dash-dotted lines indicate the location of virtual slits used in
    Fig.~\ref{fig:SW-spectrograms-b} and underline filamentary
    features.}
  \label{fig:SW-maps-HI}
\end{figure*}

\begin{figure*}[!ht]
  \centering
  \includegraphics[width=0.25\textwidth]{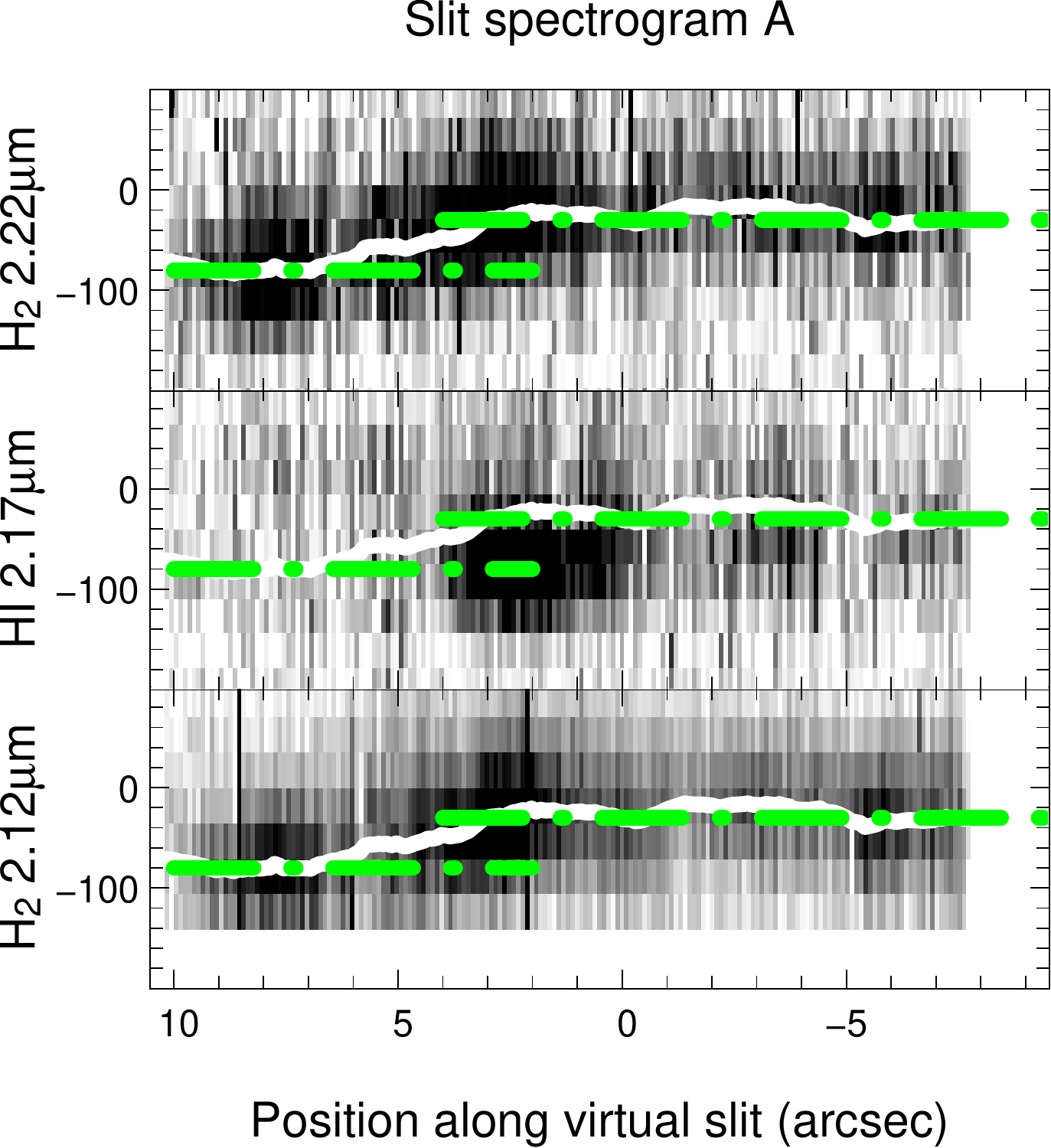}%
  \includegraphics[width=0.25\textwidth]{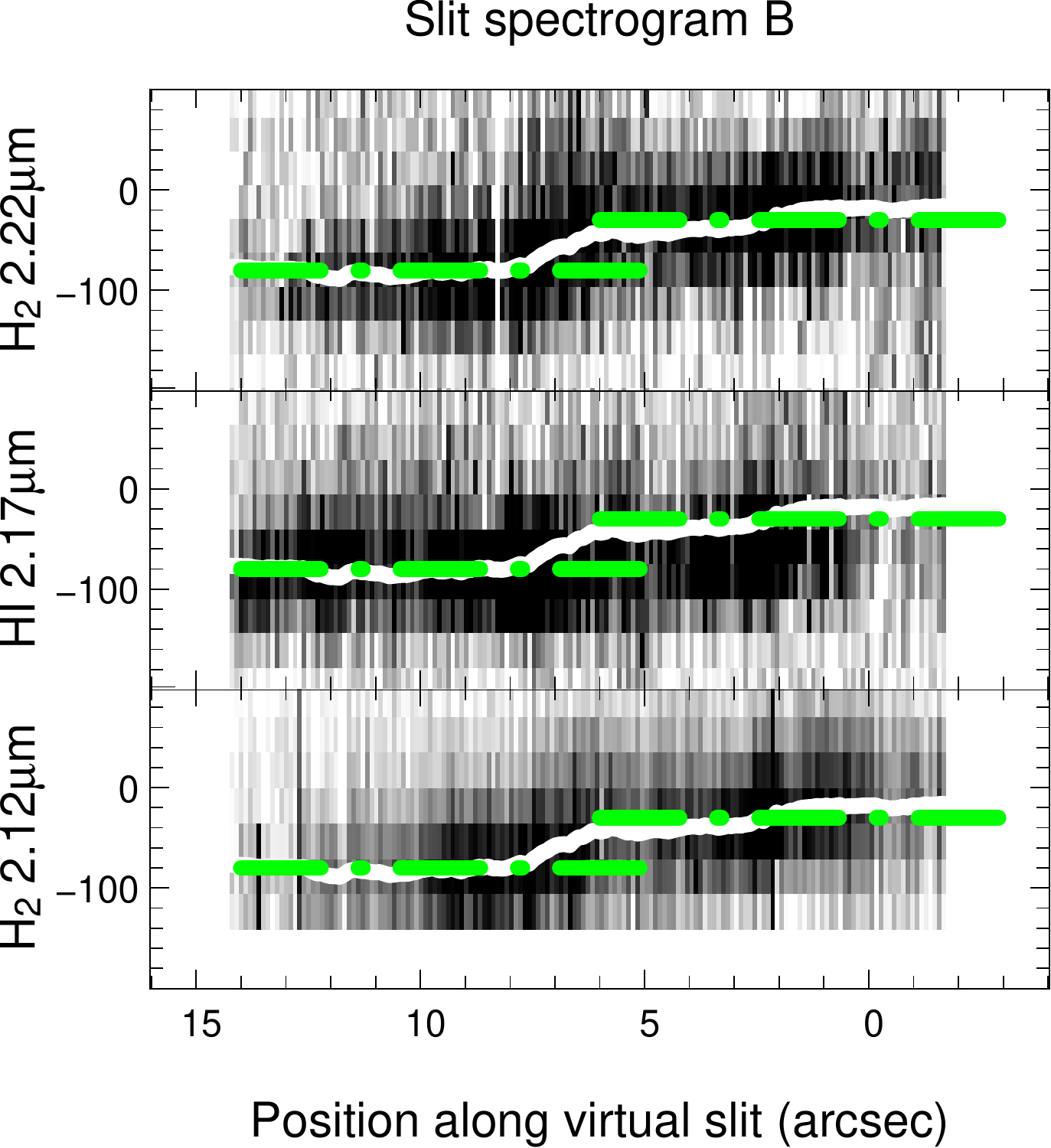}%
  \includegraphics[width=0.25\textwidth]{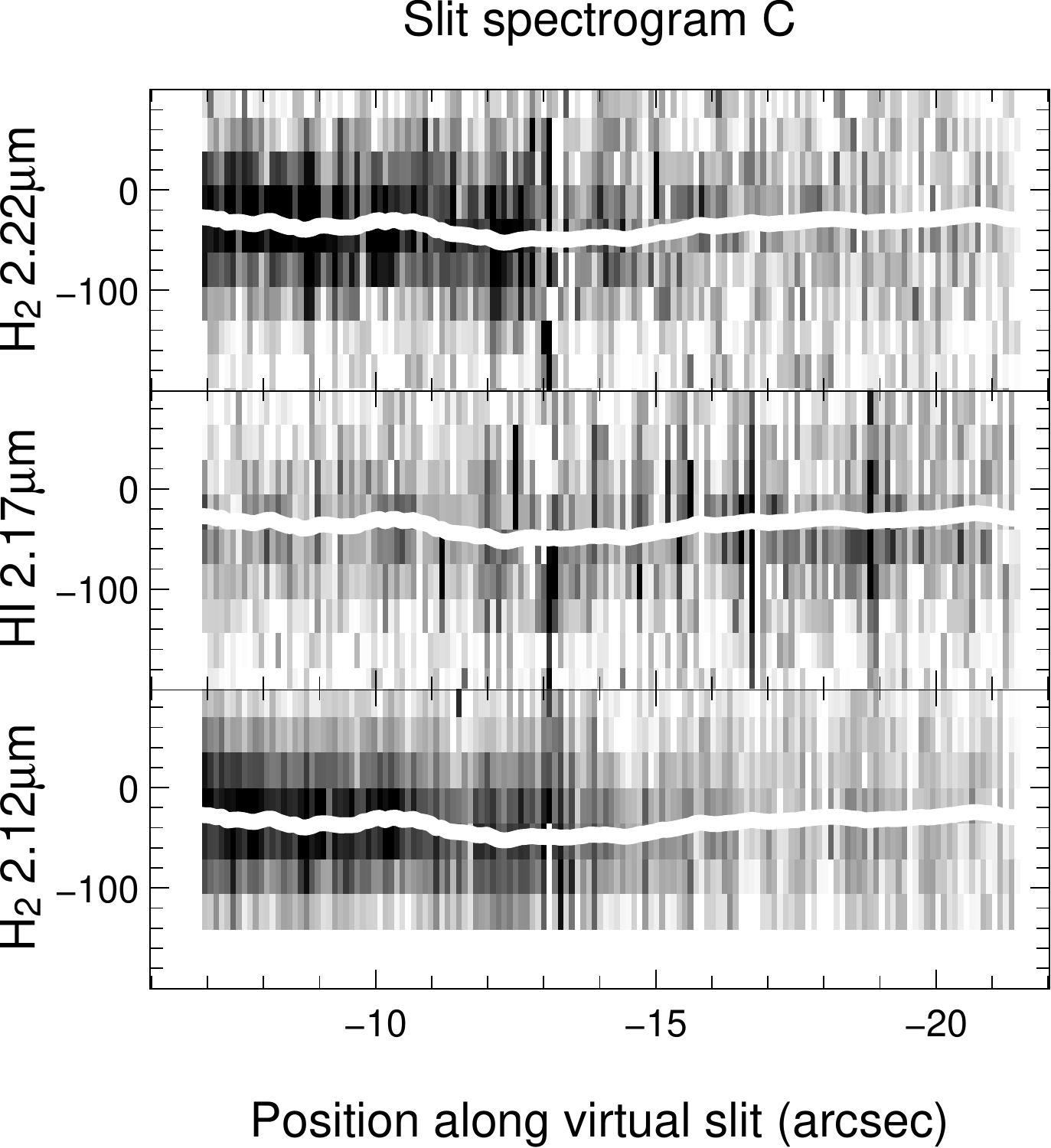}%
  \includegraphics[width=0.25\textwidth]{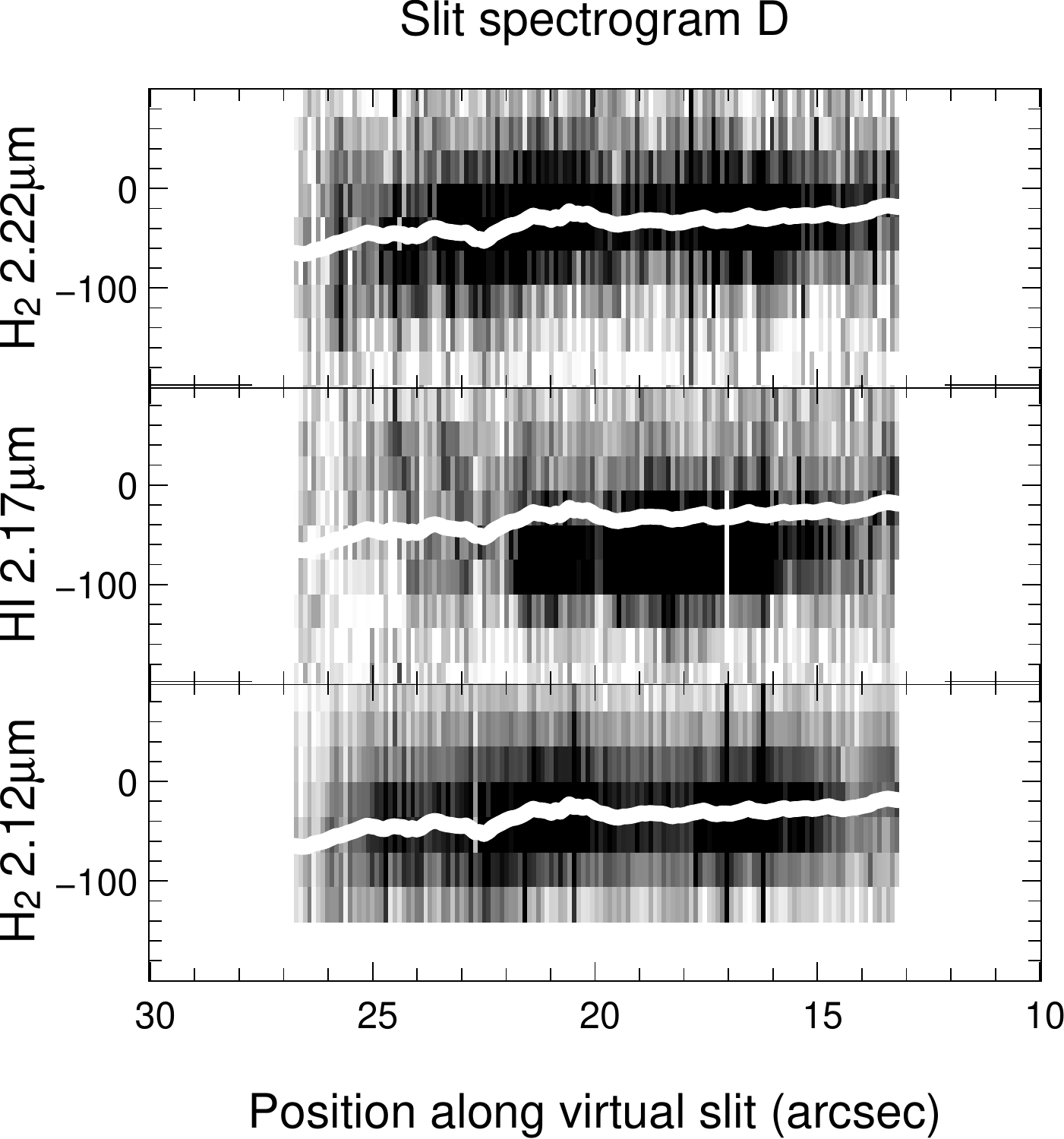}
  \caption{Slit-like spectrograms extracted from the SW mosaic data
    along the lines displayed on Fig.~\ref{fig:SW-maps-H2}. The
    virtual slit has a width of $0.3$ arcsec. Position-velocity images around each of
    the three spectral lines studied in this paper are displayed. The spectral
    axis is expressed as $v_\mathrm{LSR}$ for each line. Overplotted in white
    is the H$_2$ radial velocity derived with CubeFit at the same
    location, the curve is the same in each subpanel. The green dash-dotted lines highlight the two-component solution described in the main text.}
  \label{fig:SW-spectrograms}
\end{figure*}

\begin{figure*}[!ht]
  \centering
  \includegraphics[width=0.25\textwidth]{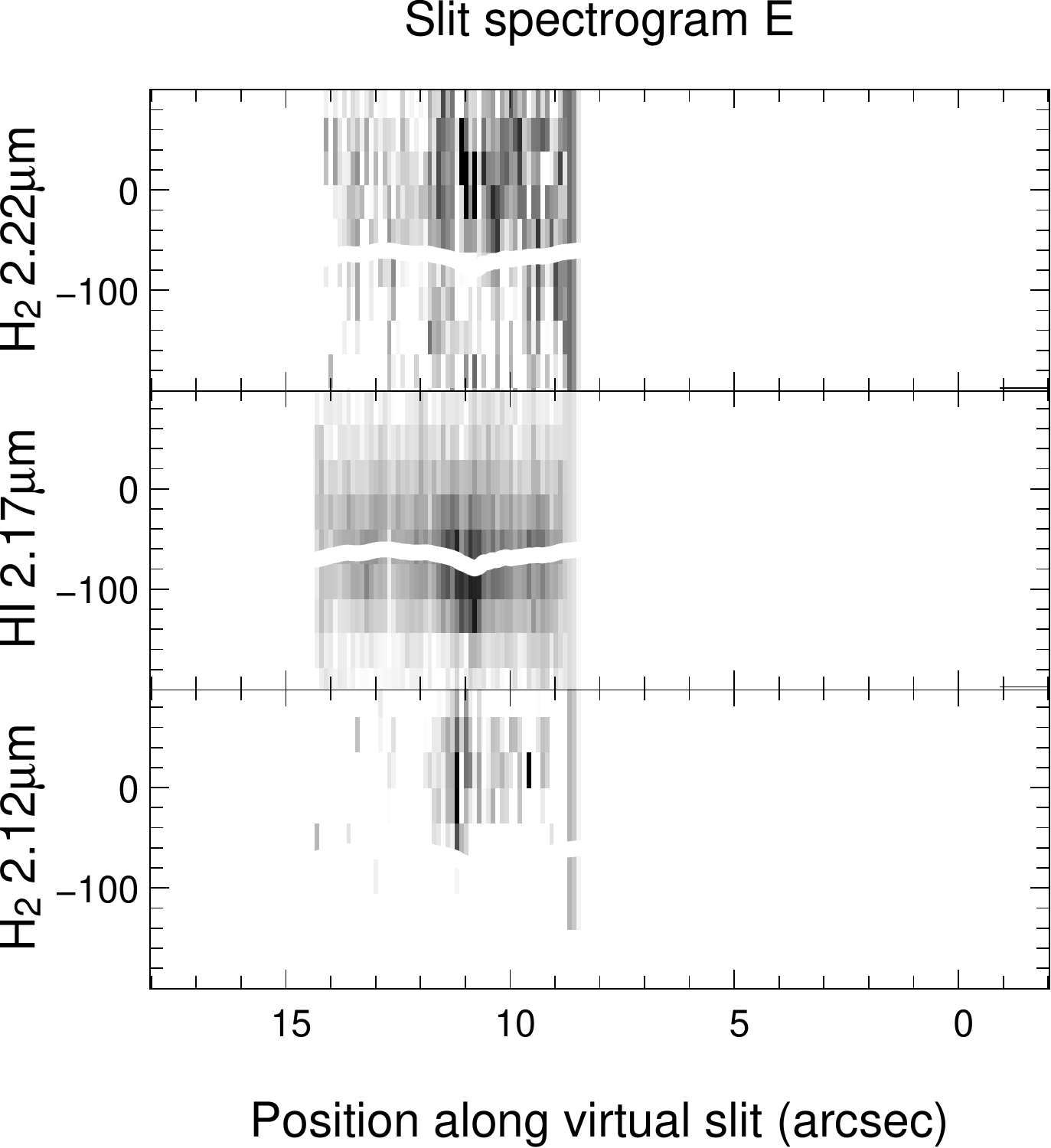}%
  \includegraphics[width=0.25\textwidth]{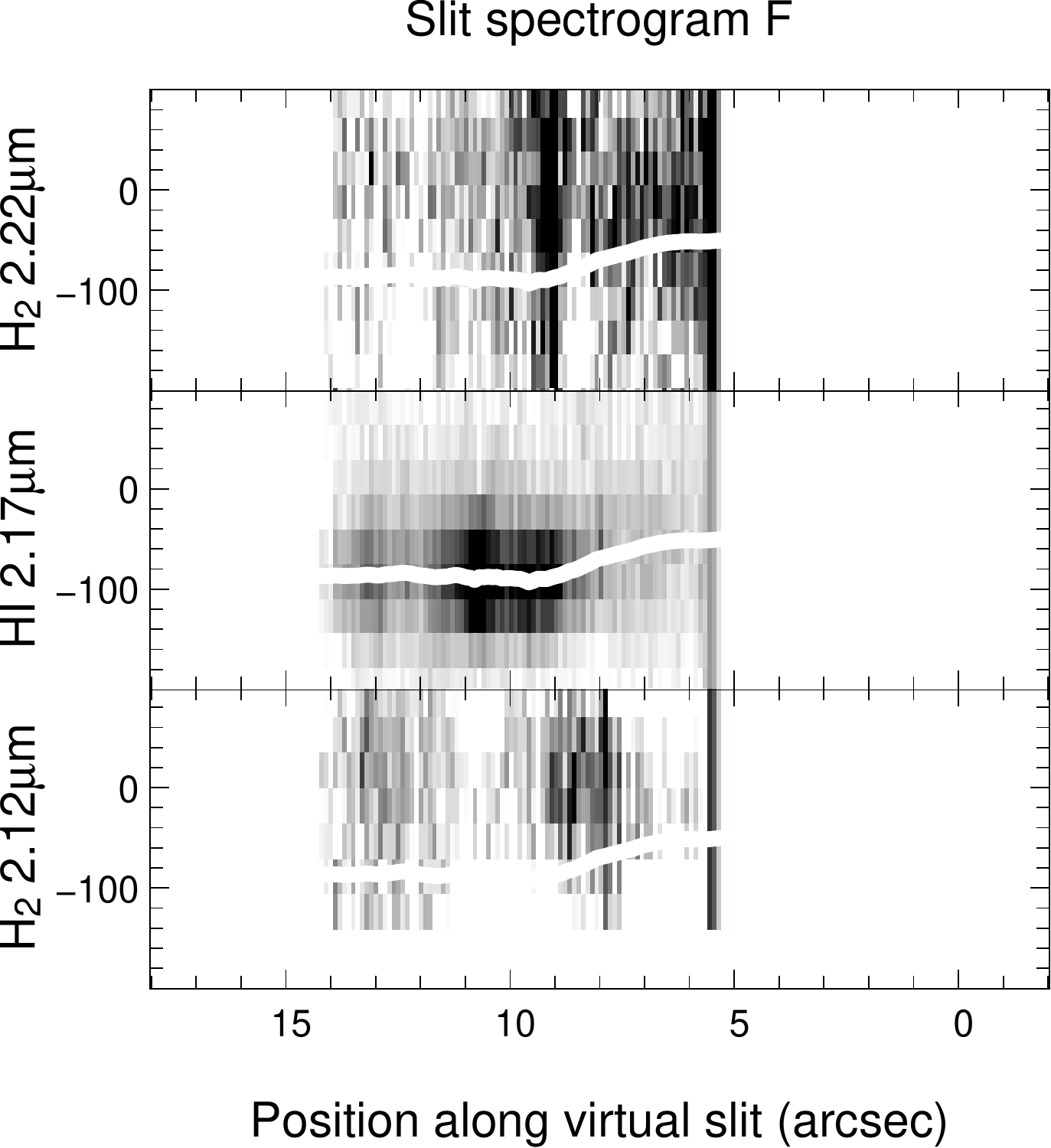}%
  \includegraphics[width=0.25\textwidth]{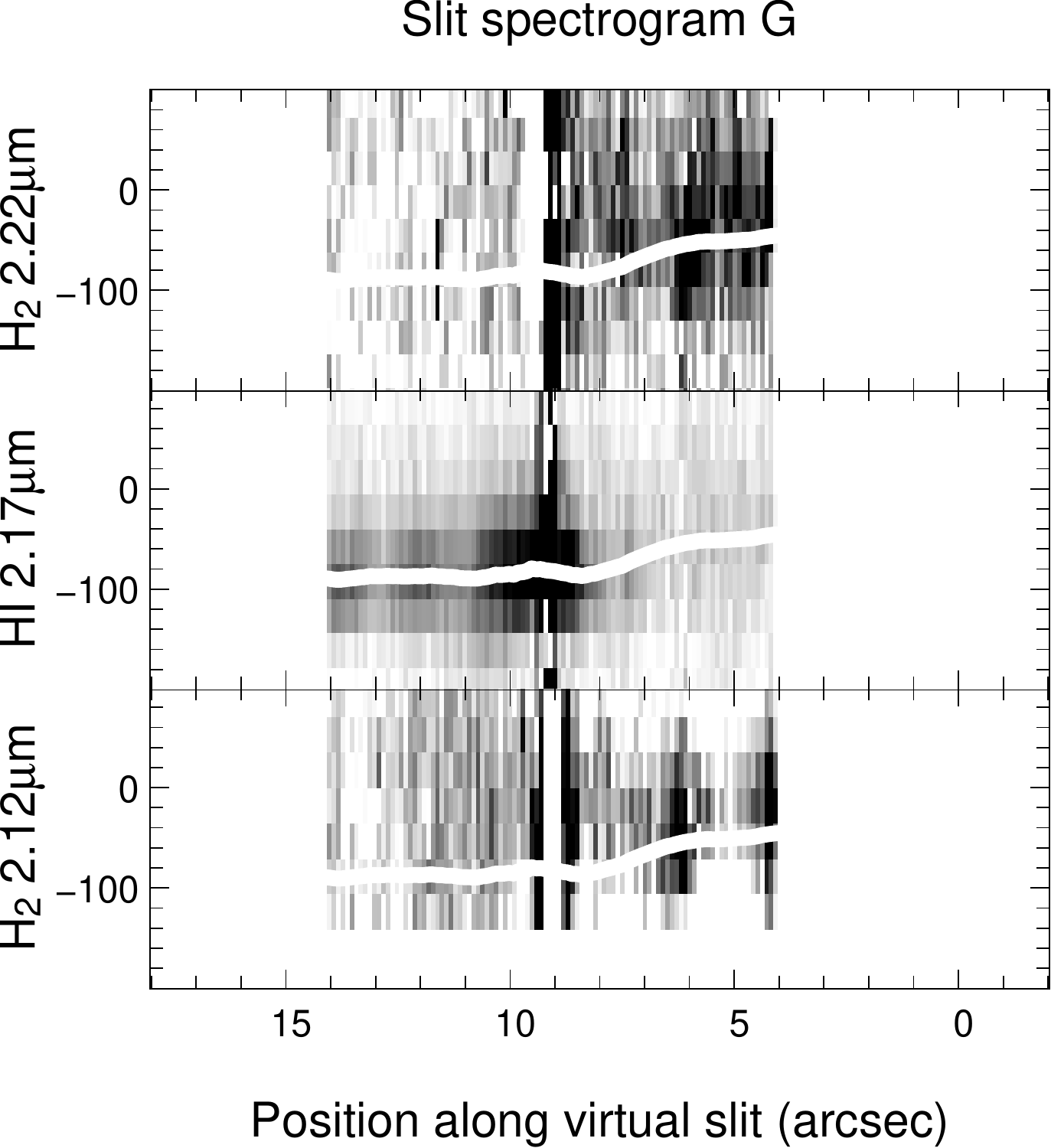}%
  \includegraphics[width=0.25\textwidth]{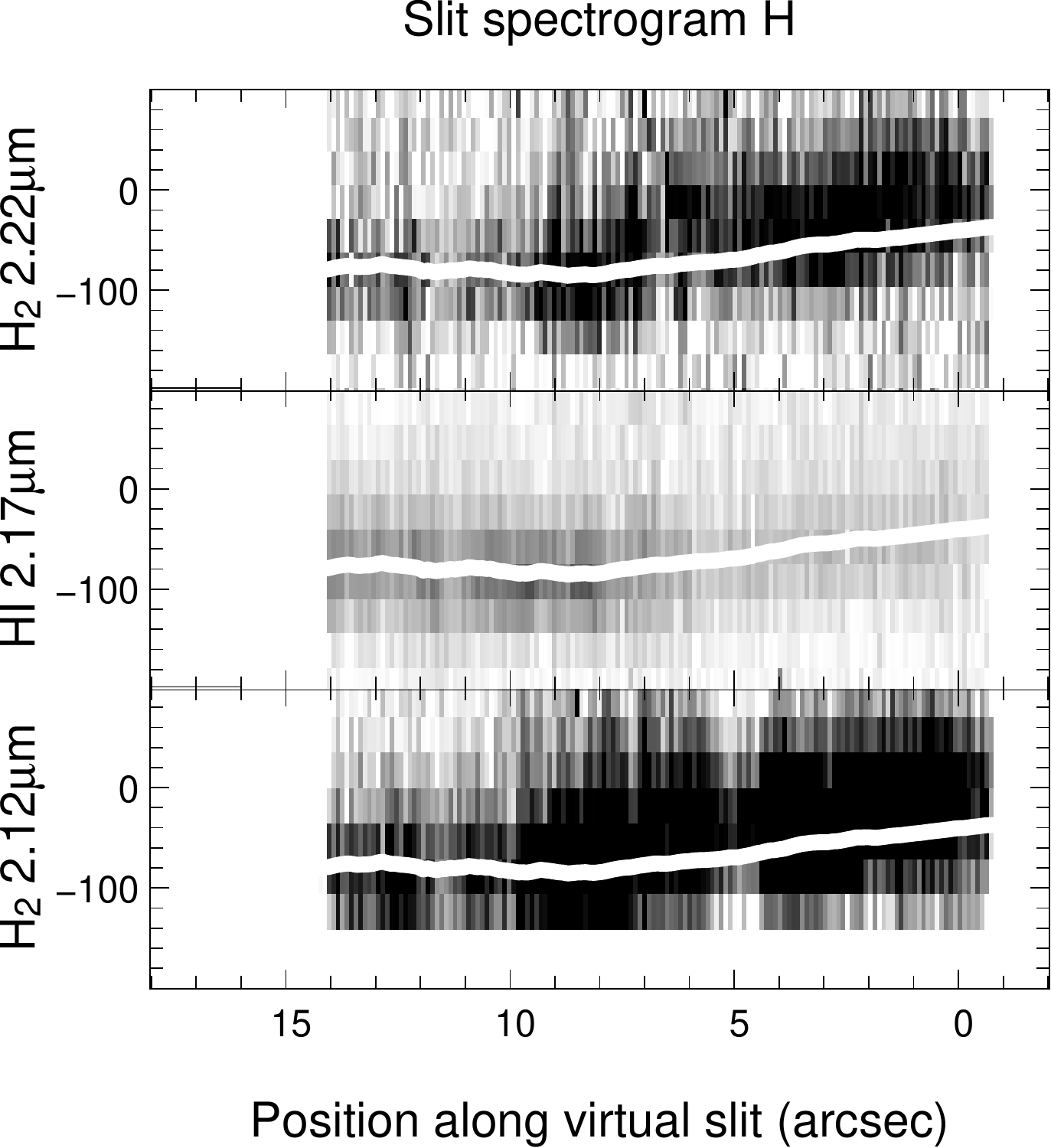}
  \caption{Slit-like spectrograms extracted from the SW mosaic data
    along the lines displayed on Fig.~\ref{fig:SW-maps-HI}. The
    virtual slit has a width of $0.3$ arcsec. Position-velocity images around each of
    the three spectral lines studied in this paper are displayed. The spectral
    axis is expressed as $v_\mathrm{LSR}$ for each line. Overplotted
    is the \ion{H}{ii} radial velocity derived with CubeFit at the
    same location, the curve is the same in each subpanel. }
  \label{fig:SW-spectrograms-b}
\end{figure*}

The results for the SW mosaic are presented in
Figs.~\ref{fig:SW-maps-H2} and \ref{fig:SW-maps-HI} for H$_2$ and
Br$\gamma$, respectively. The corresponding uncertainty maps are in
Figs.~\ref{fig:SW-errmaps-H2} and \ref{fig:SW-errmaps-HI} in
Appendix~\ref{appendix:errmaps}. The median uncertainties are listed
in Table~\ref{table:errs}.

\begin{figure}
  \centering
  \includegraphics[scale=0.4, viewport=0 83 475 380, clip]{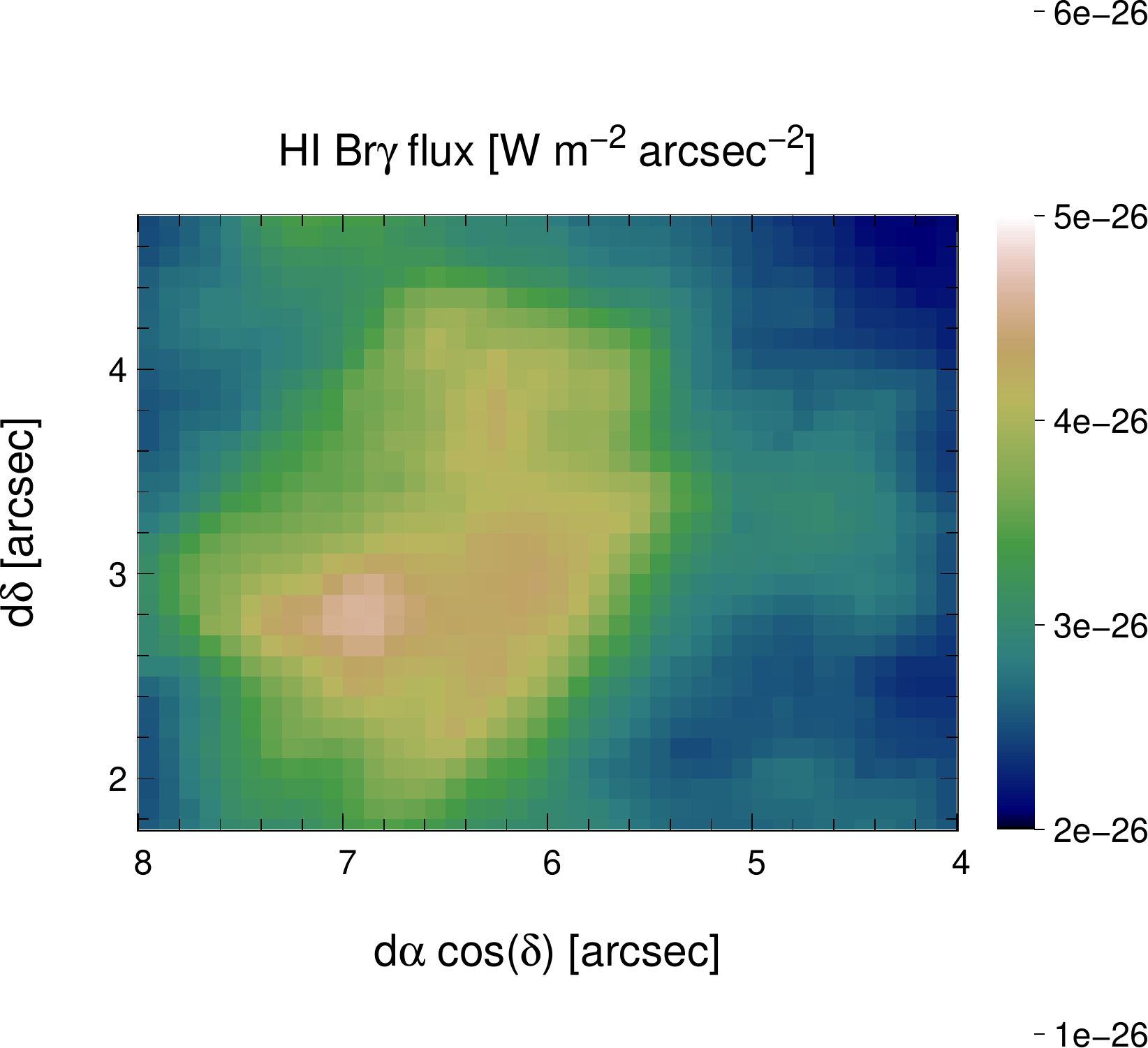}
  \includegraphics[scale=0.4, viewport=0 00 475 380, clip]{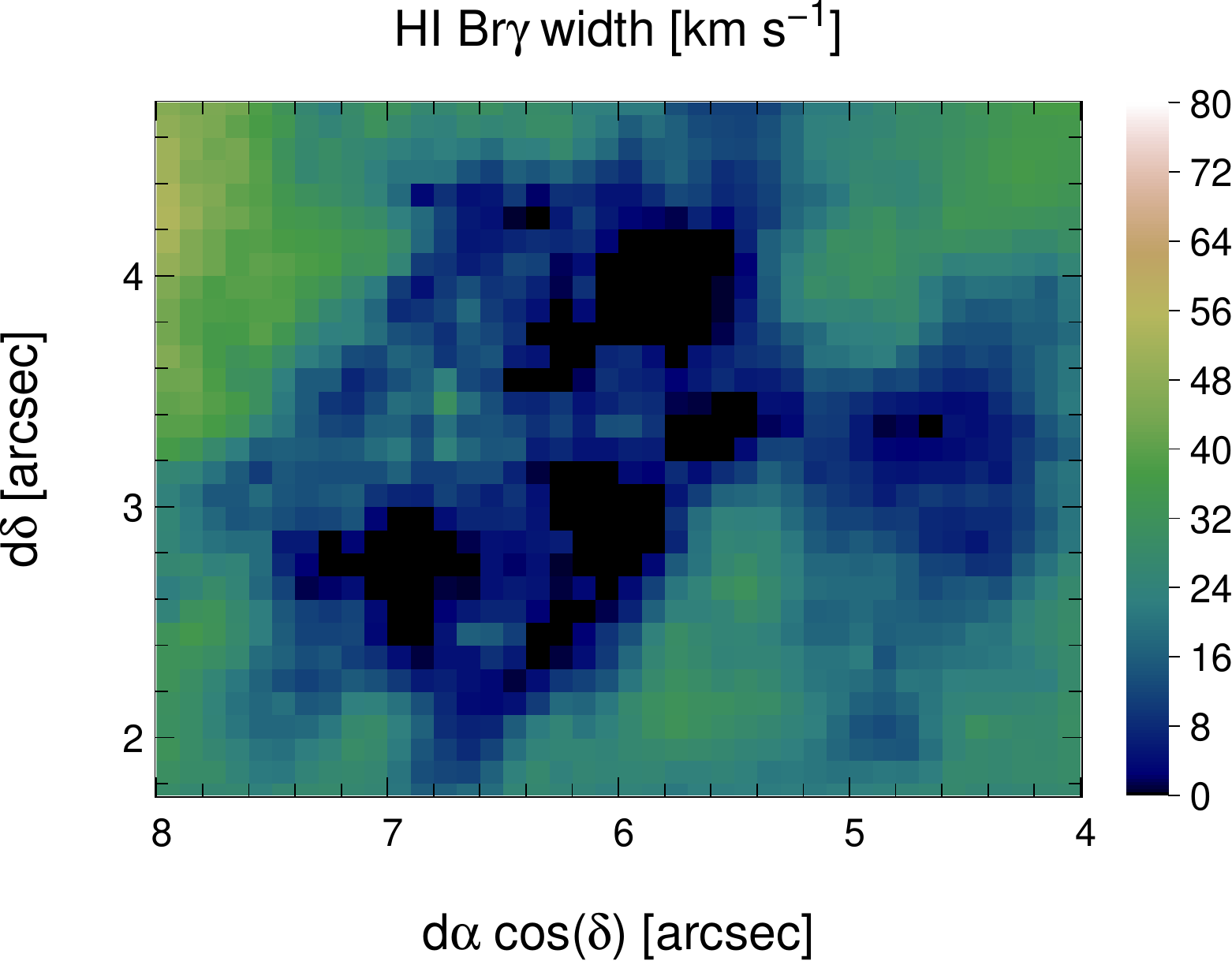}
  \caption{Zoom-in on a feature in the SW mosaic: Br$\gamma$
      line flux and width maps.\label{fig:SWzoom}}
\end{figure}

H$_2$ is reliably detected almost over the entire field, except along
the eastern edge. Br$\gamma$ is also detected almost everywhere,
except along the southern edge.  In contrast to the NE field, the
various maps show coherent structure in the form of elongated features
presumably of filamentary nature. These elongated features can be seen
consistently in all parameter maps (clearly in the linewidth,
flux and velocity maps, less clearly in the flux ratio map).

\begin{figure}
  \includegraphics[width=\columnwidth]{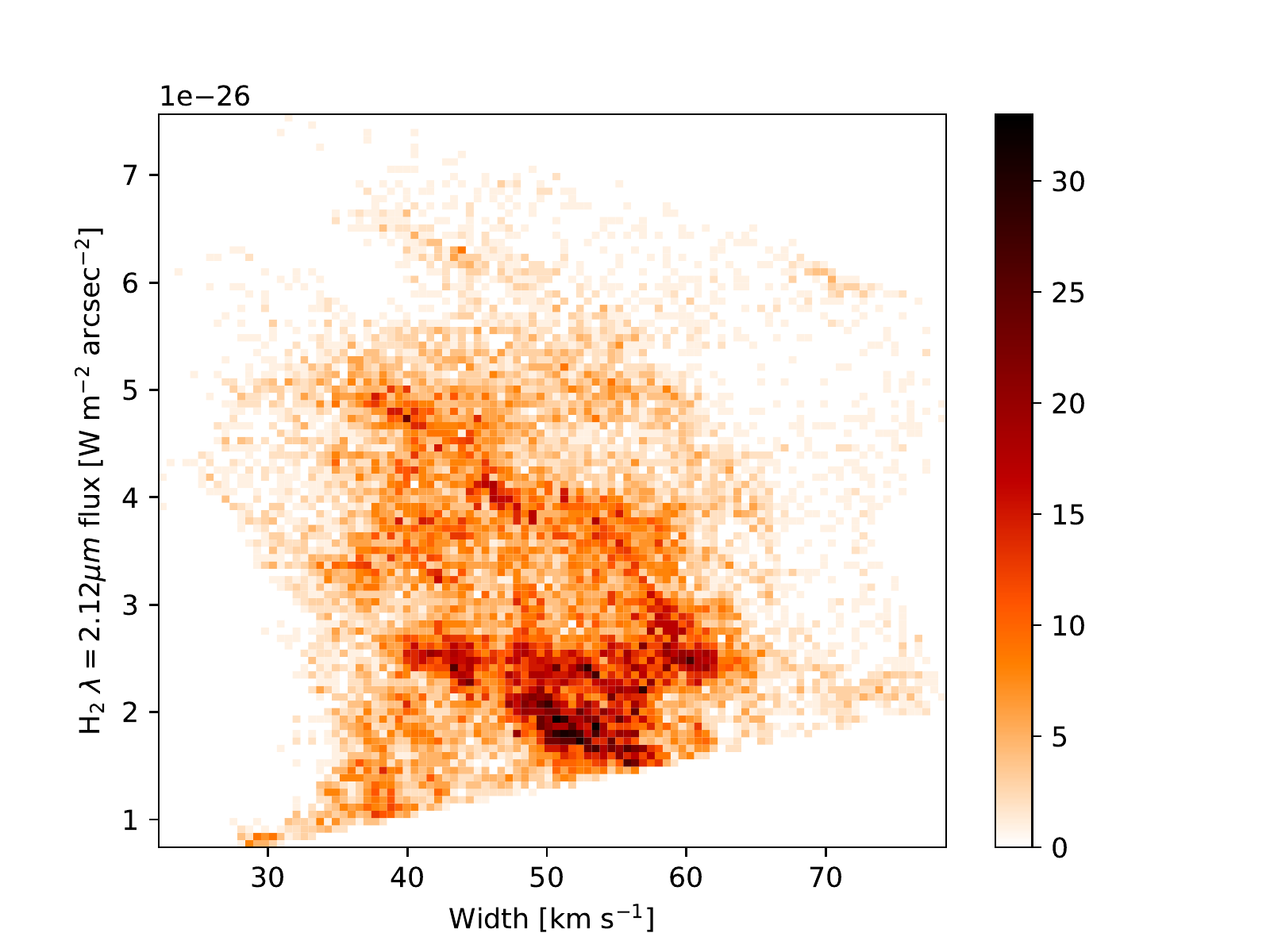}
  \includegraphics[width=\columnwidth]{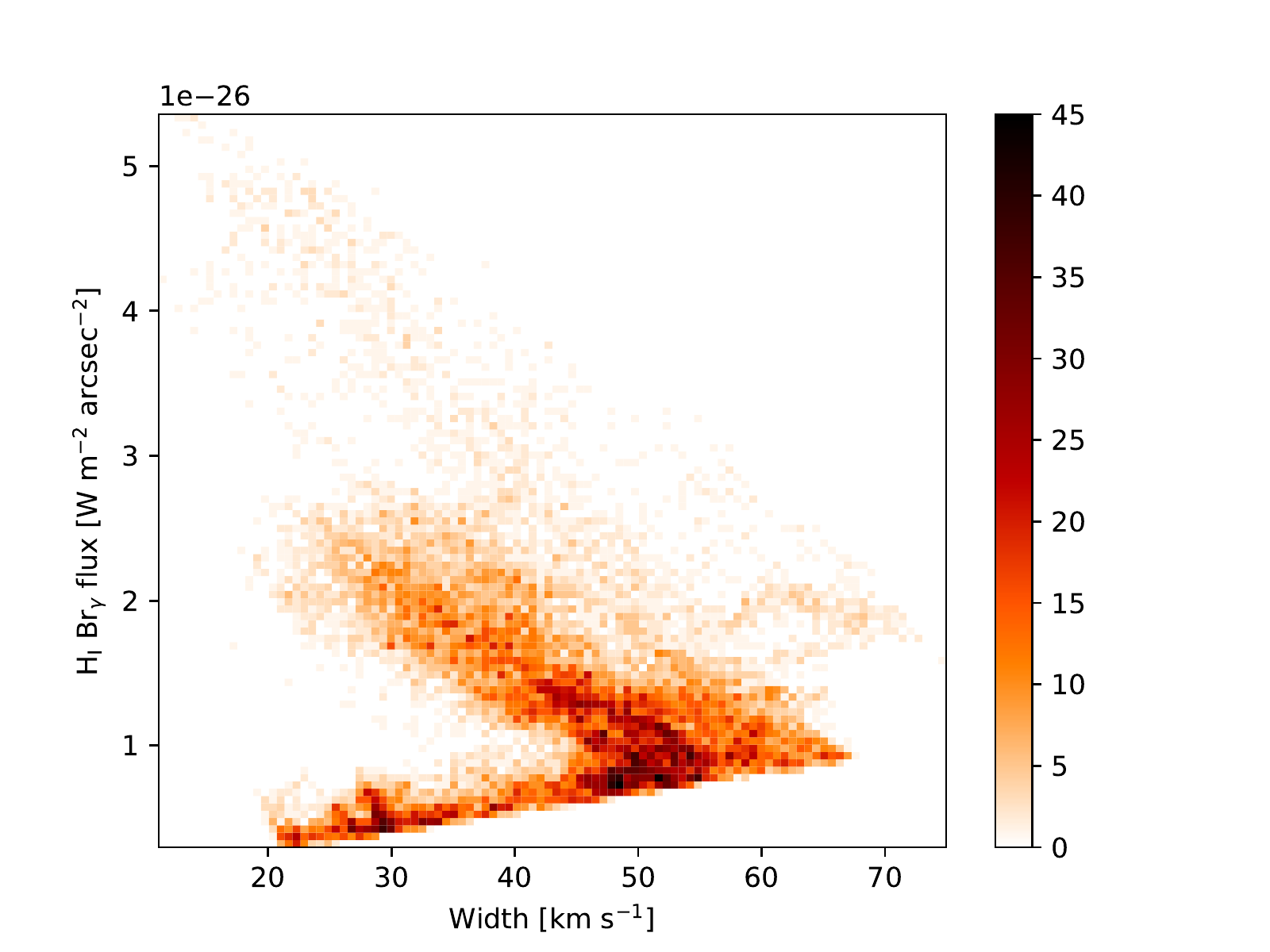}
  \caption{Line flux vs. width density plot (arbitrary units) in the
    SW mosaic (see Fig.~\ref{fig:NEcor}). The anti-correlation between
    line flux and width is quite remarkable on the lower panel
    (Br$\gamma$) where it takes the form of several narrow, densely
    populated streaks on the density plot. In the top panel (H$_2$),
    the anti-correlation is weaker than in the bottom panel, and the
    density distribution is more diffuse, but one can still identify
    linear features, showing that such a correlation exists
    locally.\label{fig:SWcor}}
\end{figure}

As in the NE field, linewidth is anti-correlated with flux, at
  least locally (Fig.~\ref{fig:SWcor}). However, the brightest H$_2$
feature is not associated with a particularly strong local minimum in
the width map. The filaments appear as deep valleys in the width map
and ridges on the flux map. Two of the four such elongated features in
the H$_2$ maps, labeled $A$ and $B$ on Fig.~\ref{fig:SW-maps-H2}, are
oriented in the south-east -- north-west direction and parallel to the
four features labeled $E$ to $H$ on the Br$\gamma$ maps
(Fig.~\ref{fig:SW-maps-HI}). The H$_2$ and \ion{H}{ii} velocity maps
also offer strong similarities: they can both be described as a
plateau near $\boldsymbol v\approx-30$~km~s$^{-1}$ occupying most of
the field with the set of parallel filaments ($A$, $B$ and $E$ to $H$)
at a significantly different velocity, near $-80$~km~s$^{-1}$. We also
note that Br$\gamma$ is brightest along the eastern edge of the field,
precisely where H$_2$ is not robustly detected.

In order to better probe the velocity field in the filaments, we have
extracted spectrograms (Figs.~\ref{fig:SW-spectrograms} and
\ref{fig:SW-spectrograms-b}) along the four most prominent filaments
of each map. The spectrograms extracted from the 3D cubes are similar
to what would be achieved with a $0.3$-arcsec-wide long-slit
spectrometer aligned on each filament. On slits $A$ and $B$, the bulk
of the emission has a velocity displacement of about $-30$~km/s with a
drop to about $-80$~km~s$^{-1}$ on the eastern end (where the
filamentary structures can be seen on the flux, linewidth and radial
velocity maps). The (single-component) 3D fit yields a smooth
transition between those two regions (white curve), especially for
slit $B$. However, for both slits, another interpretation is also
possible: that of two distinct, overlapping components (green,
dash-dotted lines). The single component fit then gives a weighted
average of the two actual components in the transition region. This
interpretation is corroborated by looking at spectrogram $D$:
Br$\gamma$ emission is detected in this slit, but at
$\simeq-80$~km~s$^{-1}$ which is not well matched by the white line
delineating the H$_2$ velocity. This indicates that H$_2$ and
\ion{H}{ii} are separated into two components along the line of
sight. Similarly, H$_2$ emission can be seen at slightly positive
radial velocity on slits $E$ to $G$ (Fig.~\ref{fig:SW-spectrograms-b})
where \ion{H}{ii} is again near $-80$~km~s$^{-1}$. Conversely, in slit
$H$ (which is located about $1''$ north of slit $B$), both species are
detected at compatible velocities ($\simeq-80$~km~s$^{-1}$).

\section{Discussion}
\label{sect:discussion}

\subsection{Comparison with radio maps}

The two Br$\gamma$ flux maps offer a striking resemblance to the Very Large Array (VLA)
6-cm continuum image (Fig.~\ref{fig:fieldmaps}). This is most obvious for the SW
mosaic where the Br$\gamma$ emission occupies most of the field. The
system of parallel elongated features seen in
Fig.~\ref{fig:SW-maps-HI} ($E$ to $H$) are seen very clearly in the
VLA image as a detail in the Western Arc of the Minispiral. The fact
that this system continues at larger distance from Sgr~A* in H$_2$
(features labeled $A$ and $B$ on Fig.~\ref{fig:SW-maps-H2}) is a
confirmation that the Western Arc is the ionized inner edge of the CND
\citep{1999ASPC..186..265V, 2020ApJ...896...68N}. The same similarity
between Br$\gamma$ and 6-cm continuum also exists in the field of the
NE mosaic but is less obvious because both are much less luminous
there than in the SW field. However one can still clearly see that the Br$\gamma$ and 6-cm
emission are concentrated along the southern and western sides of the
NE field (see also Fig.~\ref{fig:both-mosaic-3color}), and a small filament in the 6-cm continuum image in the
south-eastern corner of the NE field can be recognized in the
Br$\gamma$ image.

Likewise, the two H$_2$ maps are very similar to the CS image from
Fig.~\ref{fig:fieldmaps}.
The $-80$~km~s$^{-1}$ filamentary features seen in H$_2$ emission in the SW mosaic
are evident in the CS image. The
main H$_2$ features that can be seen in this mosaic are a thin filament
running from the top center to the south-western corner labeled $C$
in Fig.~\ref{fig:SW-maps-H2}, and a broad north-south ridge labeled
$D$.

The fact that the morphology of many of the structures we identify in H$_2$ and \ion{H}{ii} match those observed in the radio (CS and continuum, respectively) represents a very strong, independent proof of the robustness of CubeFit results, even at small scale.

\subsection{Bright features are compact}

In both fields and in both species, we consistently see an
anti-correlation between linewidth and flux, with large areas of low flux and broad linewidth contrasting with small regions of brighter flux and narrow linewidth. That can be explained in the following way. Hydrogen emission in \ion{H}{ii} regions is known to occur only (for H$_2$) or primarily (for Br$\gamma$) at the surface of clumps because clumps are usually optically thick in the ultra-violet. The fact that we see broad lines means that several clumps are stacked along the line-of-sight, so that one sees the integral of a velocity gradient or a velocity dispersion among clumps. Then, we can interpret the narrow line areas in the field as individual compact clumps that are particularly bright and therefore dominate the integral over the line-of-sight.
Those clumps can be presumed to be brighter as a consequence of a stronger UV field or a higher rate of collisional excitation. The particular H$_2$ lines that we probe in this paper do not allow us to discriminate between the two excitation mechanisms, both of which are possible in this environment \citep{2016A&A...594A.113C}.

In this context, the brightest feature in
the H$_2$ flux map for the SW field requires some attention as it is
not associated with a strong local minimum of linewidth. Actually,
this particular feature is at the intersection of the filaments
labeled $A$ and $D$. It is also at the transition between the
$-30$~km~s$^{-1}$ plateau of the velocity map and the
$-80$~km~s$^{-1}$ filaments. Therefore, this location is special. The
appearance as the brightest flux maximum results from the overlap
of several bright features. The velocity dispersion inside each of
these individual features is small, as demonstrated by the width minima
elsewhere in the $A$ and $D$ filaments, but the velocity dispersion
among those individual features is large, which explains the overall large
linewidth.

\subsection{The filaments are thin clumps}

Recognizing the Western Arc as the ionized inner edge of the CND
raises the question of whether each clump is mostly molecular with an
ionized surface, or whether some clumps are ionized and others
neutral. In Figs.~\ref{fig:SW-maps-H2} and
\ref{fig:SW-maps-HI}, filament $A$ can be seen only in H$_2$, $E$ to $G$ can be
seen only in \ion{H}{ii}, but $B$ and $H$, which are less than $1''$ apart, are essentially detected in both
species.  There are two ways to interpret those two features. It is possible that 
ridges $H$ (seen in \ion{H}{ii}) and $B$ (seen in H$_2$) trace two layers
of different ionization states in a somewhat thick filament. Alternatively, it is also possible that they really are the trace of two distinct thin filaments, one of which is fully ionized and the
other fully neutral. In this case, the fact that we also detect
\ion{H}{ii} in slit $B$ and H$_2$ in slit $H$ could be attributed to
spatial resolution and slit width ($0.3''$). The fact that we see only
one such transition between the two ionization states, and not one in
each filament, supports the interpretation as separate thin
filaments. 
For the filaments presented here, the apparent shape of the emission feature (be it Br$\gamma$ or one of the two H$_2$ lines) is a good
representation of the actual shape of those clumps:
rectilinear, thin ($\lesssim 0.3''$ thickness) and long (reaching
$\gtrsim10''$ length). 
This is in contrast with what has been seen in the central cavity
\citep{2019A&A...621A..65C} where clumps appear to be only partially ionized. 
The observed elongation of the features we observe here is an argument against self-gravitating cloudlets.

\section{Conclusion}
\label{sect:conclusion}

The 3D, regularized fitting method that we propose in
Sect.~\ref{sect:3Dfitting} proves to be a robust way of estimating
maps of physical parameters such as line flux, velocity dispersion,
and radial velocity, recovering those parameters in regions where the
local average per-voxel SNR is as low as $\approx1$. The resulting
maps have the desired properties of being smooth while retaining sharp
features (see e.g. the sharp local minima in the linewidth maps). The
uncertainties, estimated by repeating the same process over four
independent data subsets, are very small, corresponding to those that
could be obtained after smoothing the data with a radius of $\approx5$
pixels ($0.5"$) which would have severely degraded the spatial
resolution of the parameter maps (Fig.~\ref{fig:NE_H2_comp}). We
validated the results of CubeFit and the associated uncertainties by
comparing with a more classical spaxel-by-spaxel 1D fit in
Sect.~\ref{sect:1Duncertainties}. Another validation is provided in
Appendix~\ref{appendix:NE_H2_comp} by comparing the results of CubeFit
with a more classical 1D fit at a few locations. With this comparison,
we also demonstrated that variations of linewidth across the field, as
estimated by the 3D method, are robust. This method has also been
applied to another data set in a companion paper,
\citet{2016A&A...594A.113C}, where we detected H$_2$ throughout the
central cavity using the SPIFFI integral-field spectrometer at ESO
VLT.  Furthermore, in the application presented here we find
small-scale features that corresponds to filaments and clumps observed
in the radio and in other molecules. This provides a strong
independent confirmation of the robustness of our findings with
CubeFit.

We have applied CubeFit to Keck/OSIRIS data of the Galactic Center
CND in Sect.~\ref{sect:application}. The linewidth map is a useful tool
for identifying denser knots in the ISM, which appear as sharp local
minima in the velocity dispersion. The ISM in the SW field is organized into
two components: 1) a diffuse component near $-30$~km~s$^{-1}$ containing only a
few features, including two filaments oriented roughly in the
north--south direction, and 2) a number of thin, compact, tidally sheared filaments, aligned along the
orthoradial direction, with radial velocities near
$-80$~km~s$^{-1}$. These two components contain both \ion{H}{ii} and
H$_2$, but each orthoradial filament contains only one of
the two species, with the most ionized closer to the center of the
nuclear star cluster. Those filaments are aligned with the filaments that are evident in the radio continuum image (Fig.~\ref{fig:fieldmaps}) and with the local direction of the magnetic field (Dowell et al., private communication). In contrast, shear seems to play less of a
role in shaping the emission in the NE field, and the Br$\gamma$ and
H$_2$ emission seem to originate from distinct overlapping
components. Here also, most of the surface is occupied by an optically
thin medium emitting at low flux, with a few bright compact regions having presumably higher density and optical depth.

Our observations therefore reveal a complex and clumpy environment,
where low density, high-filling factor material seems to coexist with
higher-density, low-filling factor clumps which are tidally stretched
and some of which are fully ionized. Those components have distinct
kinematics, with radial velocities separated by
$\simeq50$~km~s$^{-1}$. This picture is very different from the assumption of spherical, self-gravitating clumps, which has sometimes been invoked in past studies.

Observing the rest of the CND in the three lines used here would help determine whether our conclusions hold generally.

\begin{acknowledgements}
  TP thanks Damien Gratadour for sharing the Yoda code
  and fruitful discussions concerning regularization.
AC, MRM, TD and AMG  acknowledge the support provided by the U.S. National Science Foundation (grants AST-1412615, AST-1518273), Jim and Lori Keir, the Gordon and Betty Moore Foundation, the Heising-Simons Foundation, the W. M. Keck Foundation, and Howard and Astrid Preston.
  The authors wish to recognize and acknowledge the very significant cultural role and reverence that the summit of Maunakea has always had within the indigenous Hawaiian community.  We are most fortunate to have the opportunity to conduct observations from this mountain. 
\end{acknowledgements}

\bibliographystyle{aa} 
\bibliography{cnd}{}

\appendix

\section{Comparison with classical 1D method}
\label{appendix:NE_H2_comp}

\begin{figure}[b]
  \centering
  \includegraphics[scale=0.4, viewport=0 50 460 413, clip]{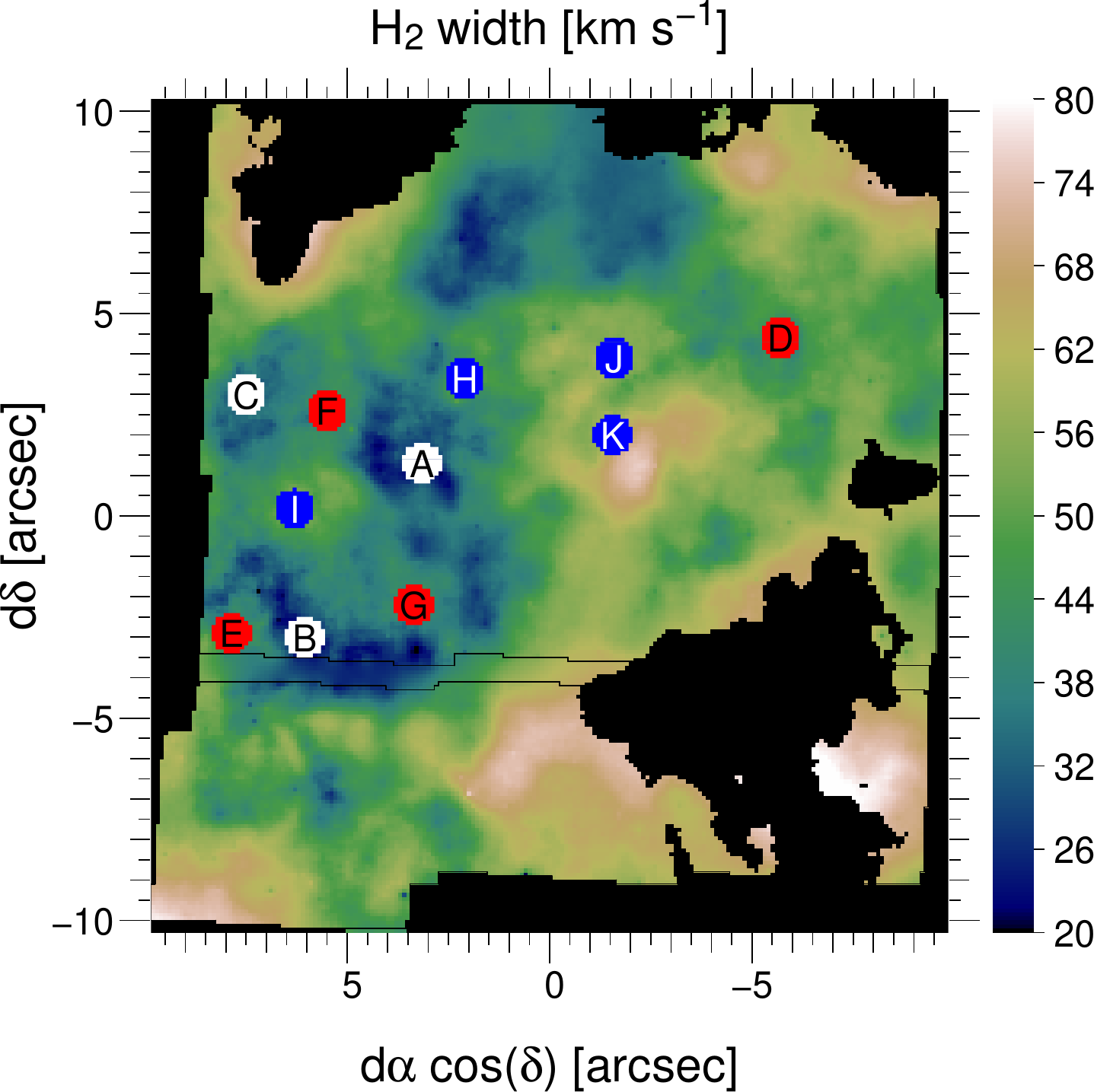}
  \vspace{2mm}\\
  \includegraphics[scale=0.4, viewport=0 0 460 413, clip]{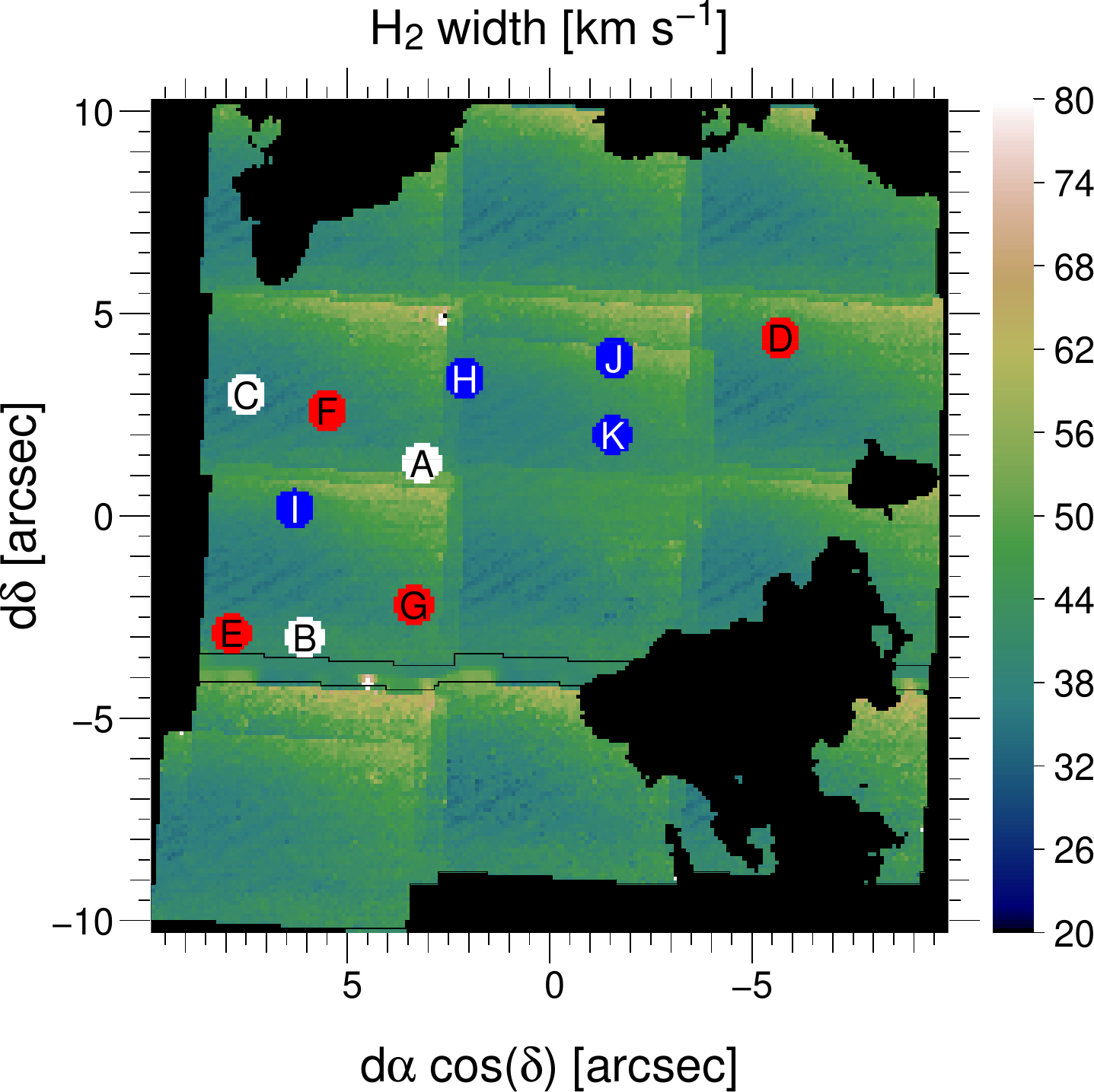}
  \vspace{2mm}\\
  \includegraphics[scale=0.4, viewport=0 -15 361.80 400, clip]{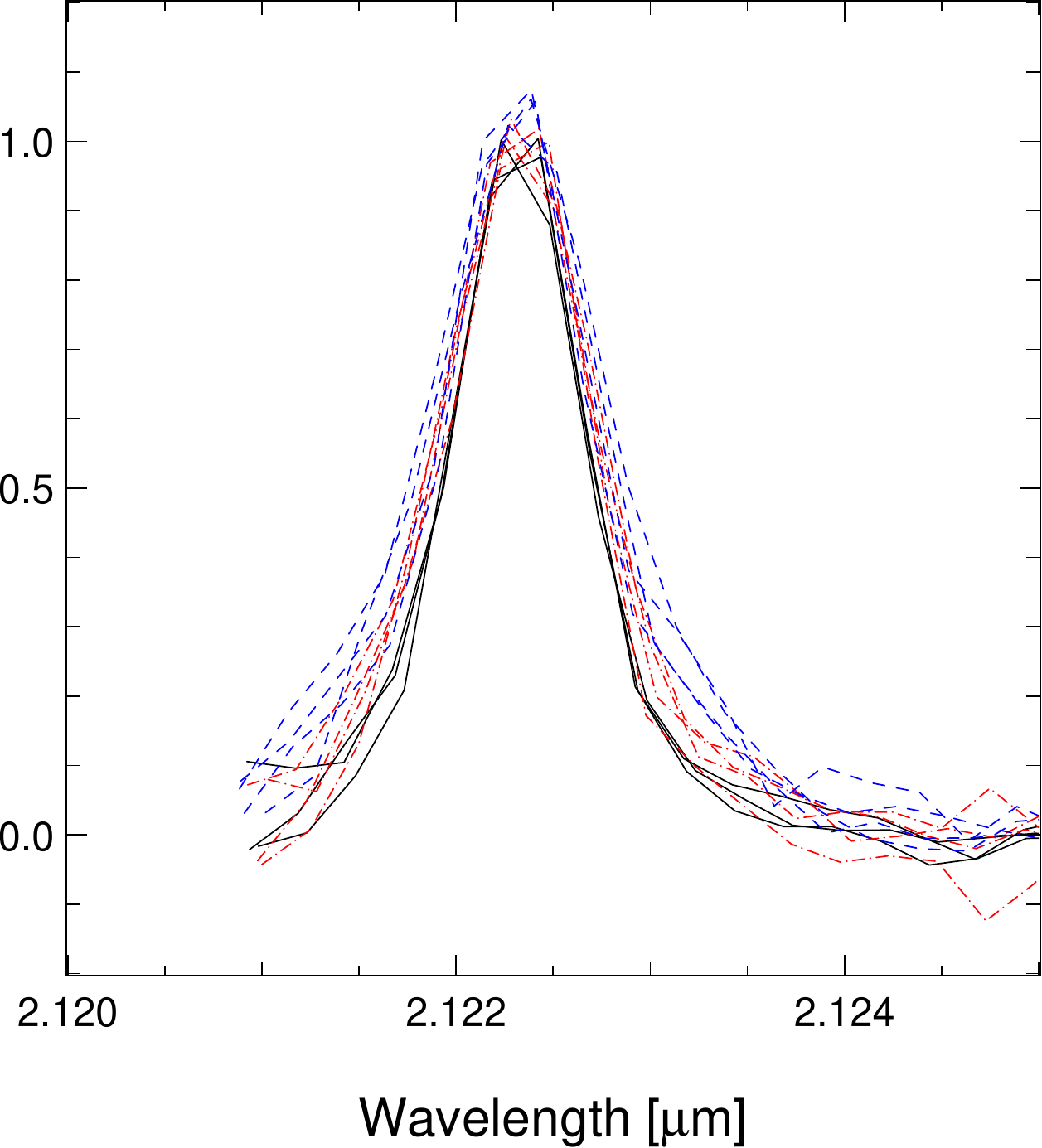}
  \caption{Variation of $\boldsymbol \sigma_{\text{H}_2}$ over the NE
    field. \emph{Top:} H$_2$ linewidth map (as in
    Fig.~\ref{fig:NE-maps-H2}). The colored dots figure the aperture
    over which spectra have been extracted. Three distinct colors
    correspond to three linewidth ranges: $<35$~km~s$^{-1}$ in
    white/black, $35$--$42$~km~s$^{-1}$ in blue, $>42$~km~s$^{-1}$ in
    red. \emph{Middle:} corresponding OH linewidth map.
    \emph{Bottom:} spectra extracted over each aperture, normalized
    according to the 1D Gaussian fit. The black spectra correspond to
    the white apertures while the red and blue spectra correspond to
    the red and blue apertures. The corresponding widths are listed in
    Table~\ref{table:NE_H2_comp}.}
\label{fig:NE_H2_comp}
\end{figure}

\begin{figure*}
  \centering
  \includegraphics[scale=0.4, viewport=0 0 460 334, clip]{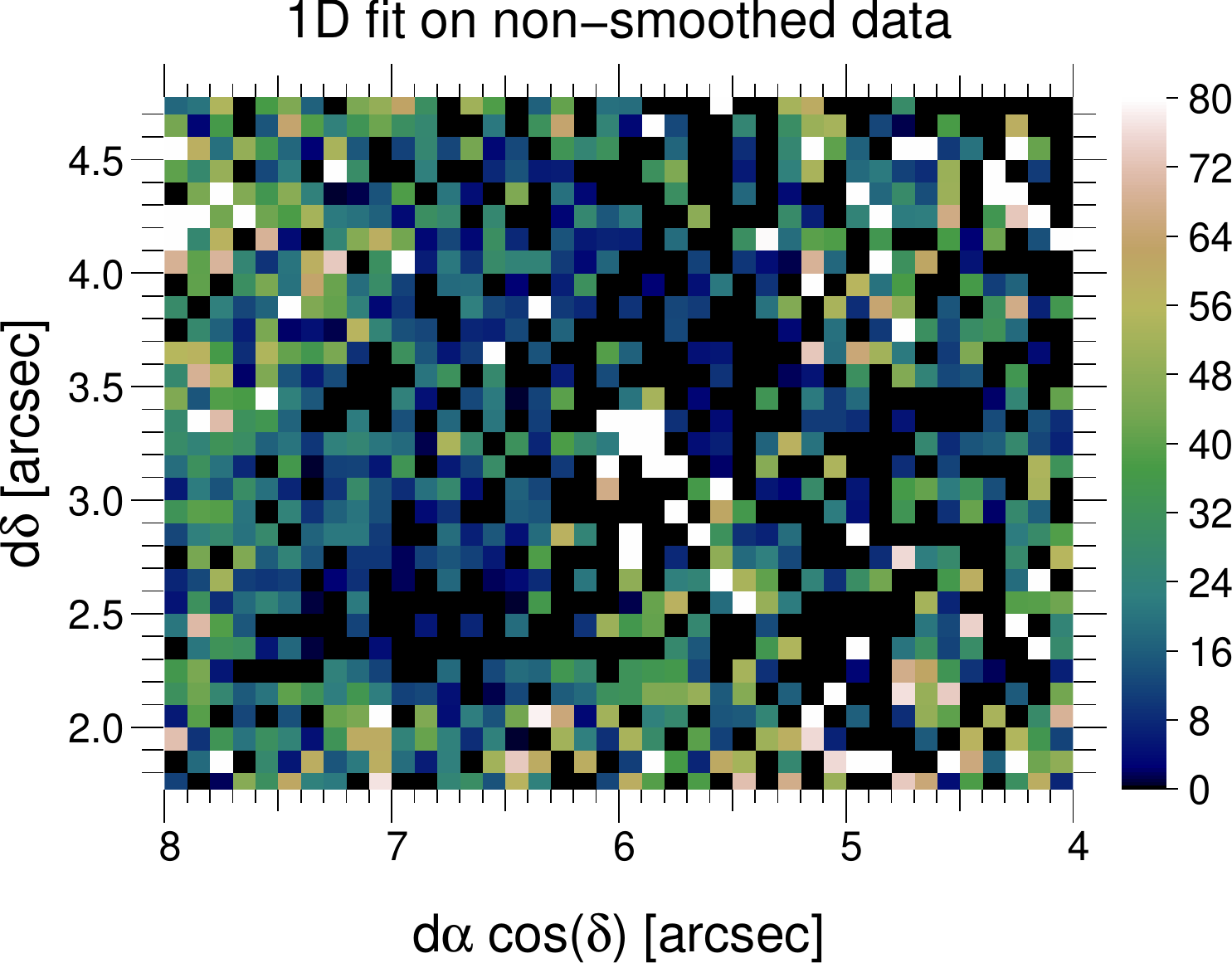}%
  \includegraphics[scale=0.4, viewport=44 0 460 334, clip]{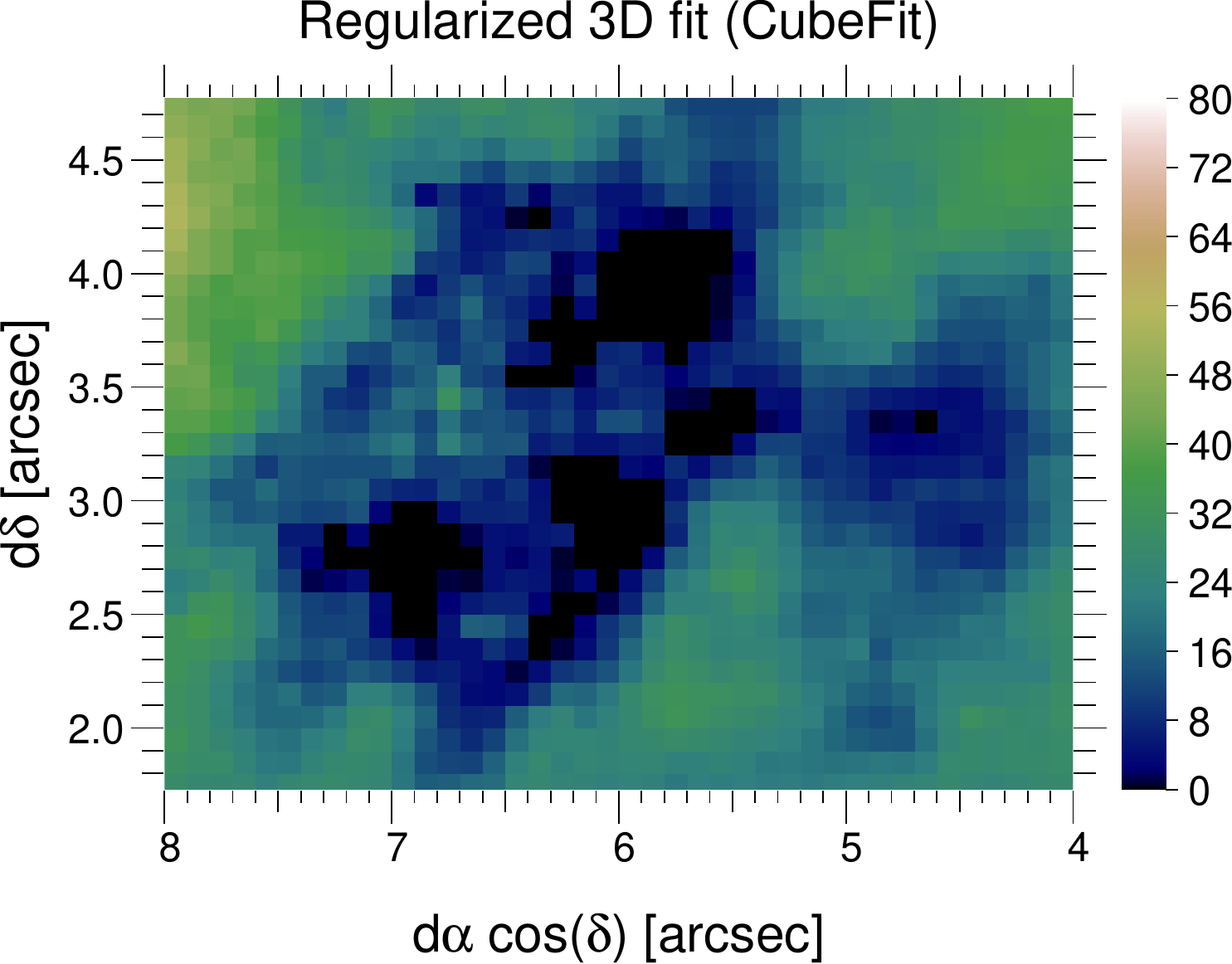}%
  \includegraphics[scale=0.4, viewport=44 0 460 334, clip]{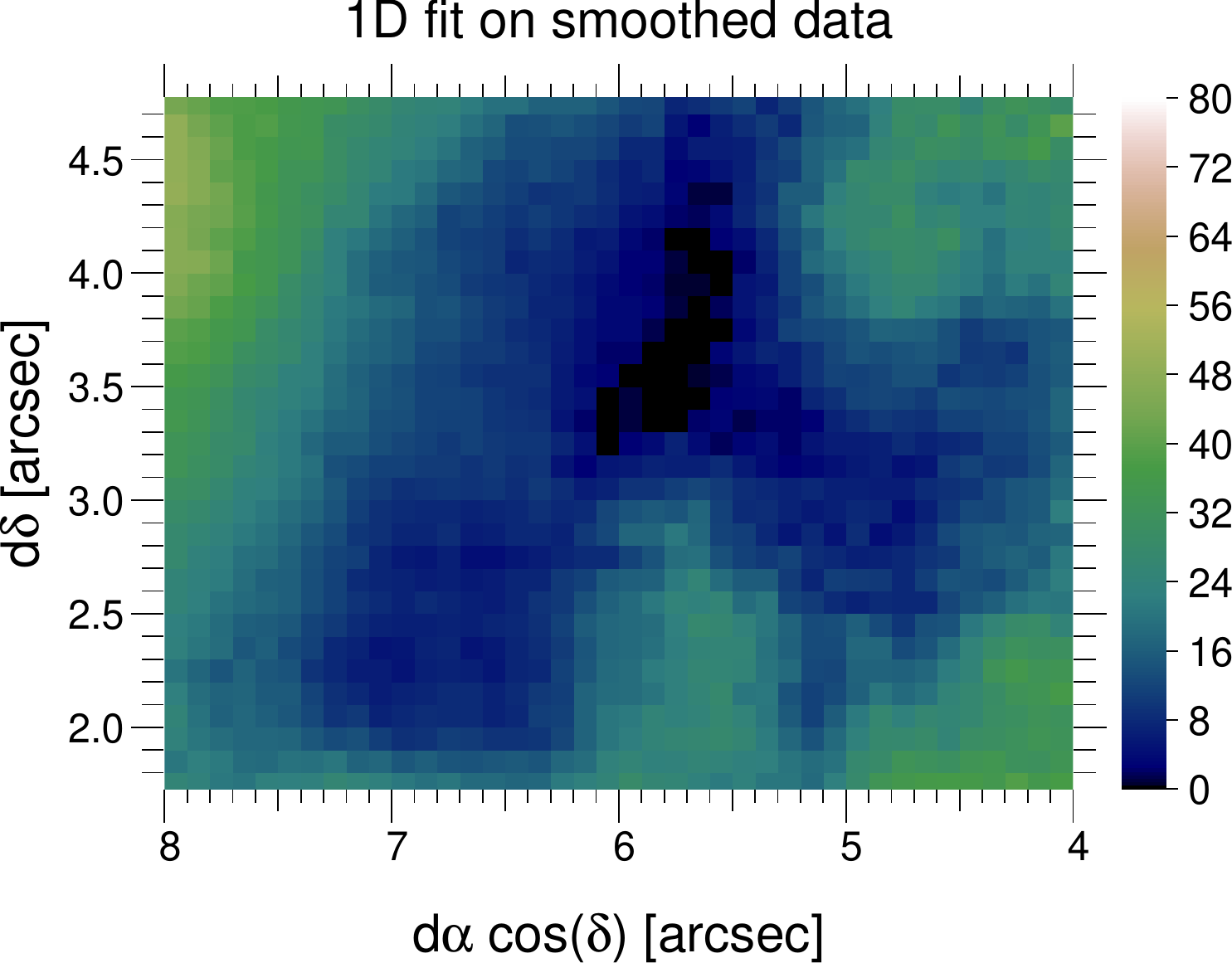}
  \caption{Close-up on three versions of the SW width map. \emph{Left:} 1D fit on original (non-smoothed) cube. \emph{Center:} 3D fit with CubeFit. \emph{Right:} 1D method fit on aperture-smoothed cube.}
\label{fig:SW-widthmap-comp}
\end{figure*}

\begin{table}[b]
  \caption{Intrinsic H$_2$ linewidth (km~s$^{-1}$) at the locations
    represented Fig.~\ref{fig:NE_H2_comp} as estimated by a 1D fit and
    CubeFit.} \centerline{
    \begin{tabular}{c| c @{$\null\pm\null$} c | c @{$\null\pm\null$} c }
      \hline\hline
      Location &
      \multicolumn{2}{c|}{1D fit}  &
      \multicolumn{2}{c}{3D fit} \\
      \hline
A & 23.8 & 2.6 & 24.9  & 1.5 \\
B & 25.9 & 2.6 & 24.8  & 1.9 \\
C & 35.3 & 2.2 & 34.3  & 1.9 \\
D & 46.7 & 2.5 & 39.8  & 1.5 \\
E & 41.9 & 2.4 & 41.2  & 2.3 \\
F & 39.4 & 3.3 & 40.5  & 1.4 \\
G & 41.3 & 2.9 & 40.8  & 1.3 \\
H & 43.6 & 3.6 & 46.7  & 1.7 \\
I & 50.7 & 3.3 & 51.7  & 1.4 \\
J & 56.3 & 6.0 & 54.4  & 1.6 \\
K & 52.5 & 6.4 & 62.8  & 2.0 \\
      \hline\hline
    \end{tabular}
  }
  \label{table:NE_H2_comp}
\end{table}

In order to validate our method, we have extracted a few
spectra at typical places in the H$_2$ field by averaging spaxels over
a $0.5''$ radius (a median on 78 spaxels per aperture) and performed a
classical 1D Gaussian fit on the H$_2$ $\lambda$ 2.12~$\mu$m line in
these spectra. The chosen location and spectra are displayed
Fig.~\ref{fig:NE_H2_comp}. The average OH linewidth (weighted by
$\boldsymbol F_{\text{H}_2}$) over the same aperture is then
subtracted in quadrature and the uncertainty is propagated from the
uncertainty estimated by the fit itself. The corresponding estimate
from the 3D fit and its uncertainty are given by the average of
$\boldsymbol \sigma_{\text{H}_2}$ (Fig.~\ref{fig:NE-maps-H2}) and
$\boldsymbol \sigma_\sigma$ (Fig.~\ref{fig:NE-errmaps-H2}), also weighted by
$\boldsymbol F_{\text{H}_2}$. These 1D and 3D width measurements are
listed in Table~\ref{table:NE_H2_comp} and agree very
well.  The classical 1D-fit approach
corroborates the findings from CubeFit, confirming the intrinsic
width variations over the field and confirming that the order of
magnitude of our estimated uncertainties is correct.

We have also performed the 1D fit over the entire SW cube, once on the original non-smoothed cube and once on a smoothed version of the cube where every spaxel contains the average spectrum over a $0.5''$ radius as above. We have again subtracted the OH linewidth in quadrature. A close-up of the width map derived from these two fits and from CubeFit is displayed Fig.~\ref{fig:SW-widthmap-comp}. Results from the 1D fit on the non-smoothed data contain high-resolution features but are very noisy. Smoothing the cube enhances the signal-to-noise ratio significantly, at the cost of erasing the small-scale features. CubeFit brings the best of the two worlds together, keeping the high resolution features while removing most of the noise, thus enhancing those small-scale features. The sharp astrophysical features that we discuss in Sect.~\ref{sect:discussion}, very clear with our method, are difficult to make out from the non-smoothed 1D fit and are smoothed out in the smoothed 1D fit. Another advantage of our method is its resilience on bad data points. A few bad voxels in the cube result only in individual spikes in our method (some can be seen in the flux ratio map for the NE mosaic, Fig.~\ref{fig:NE-maps-H2}) while smoothing smears them over large areas.

\section{Uncertainty maps}

\begin{figure*}[!ht]
  \centering
  \includegraphics[scale=0.4, viewport=0 50 460 413, clip]{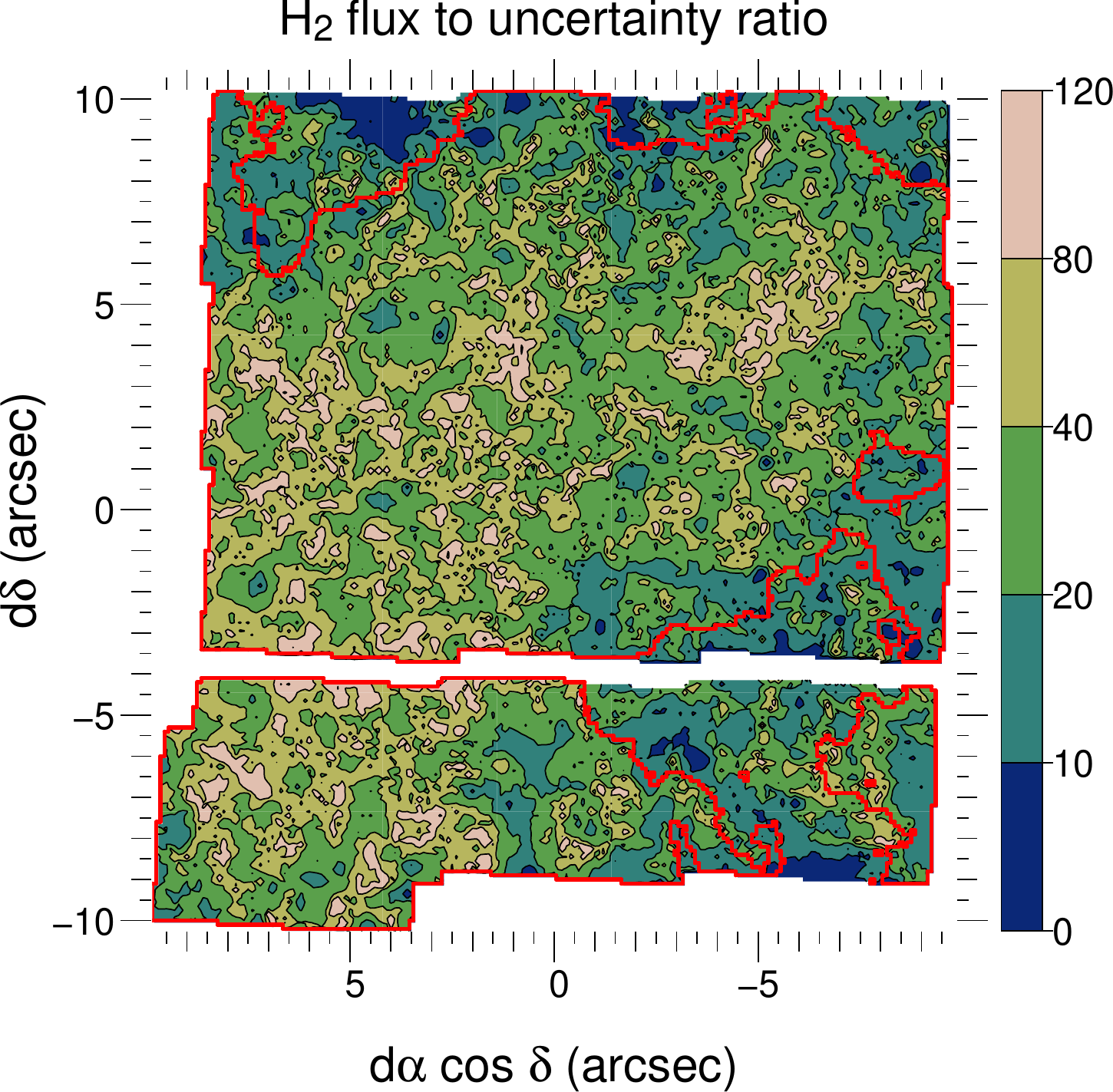}
  \includegraphics[scale=0.4, viewport=44 50 460 413, clip]{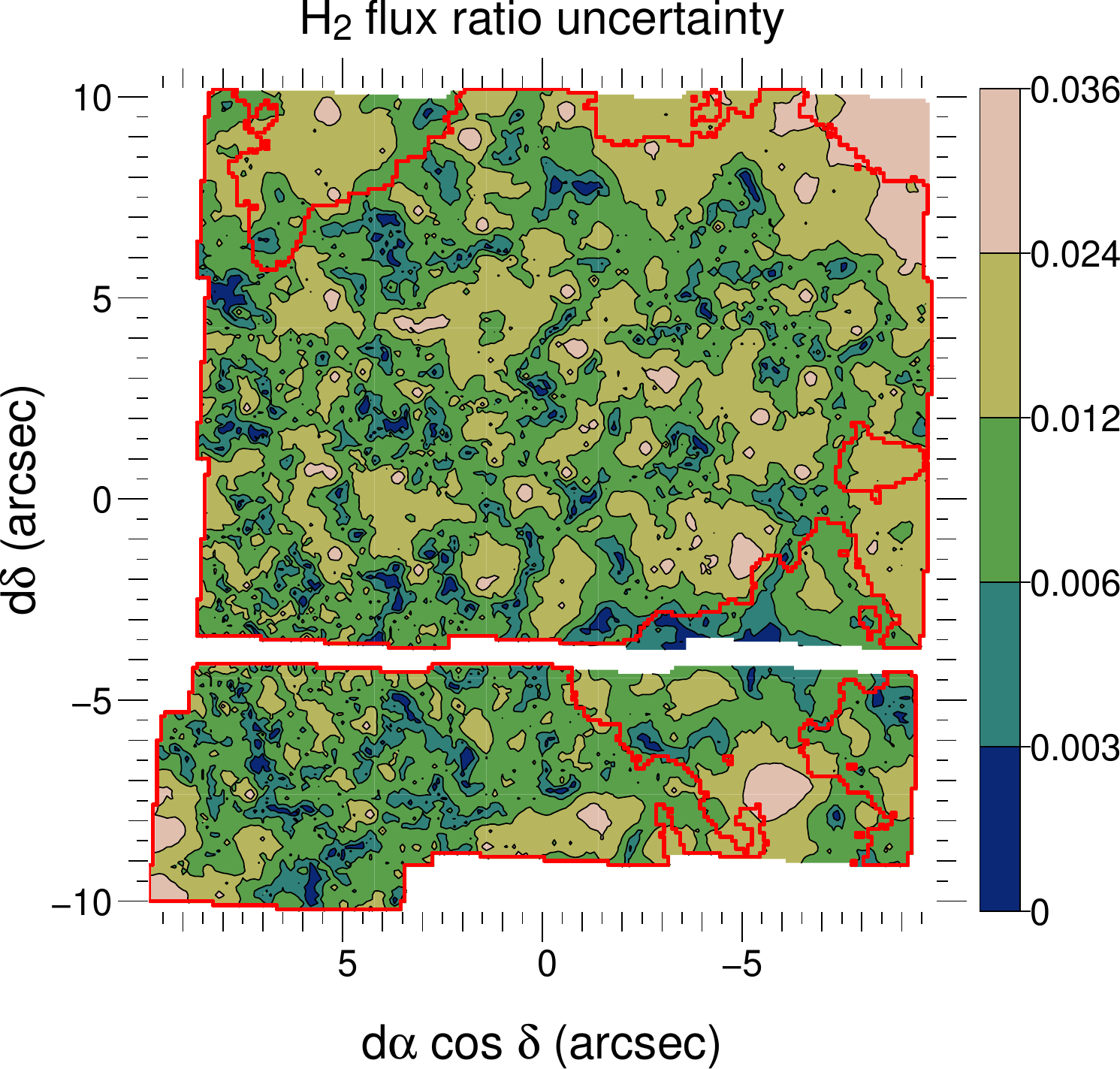}
  \vspace{2mm}\\
  \includegraphics[scale=0.4, viewport=0 0 460 413, clip]{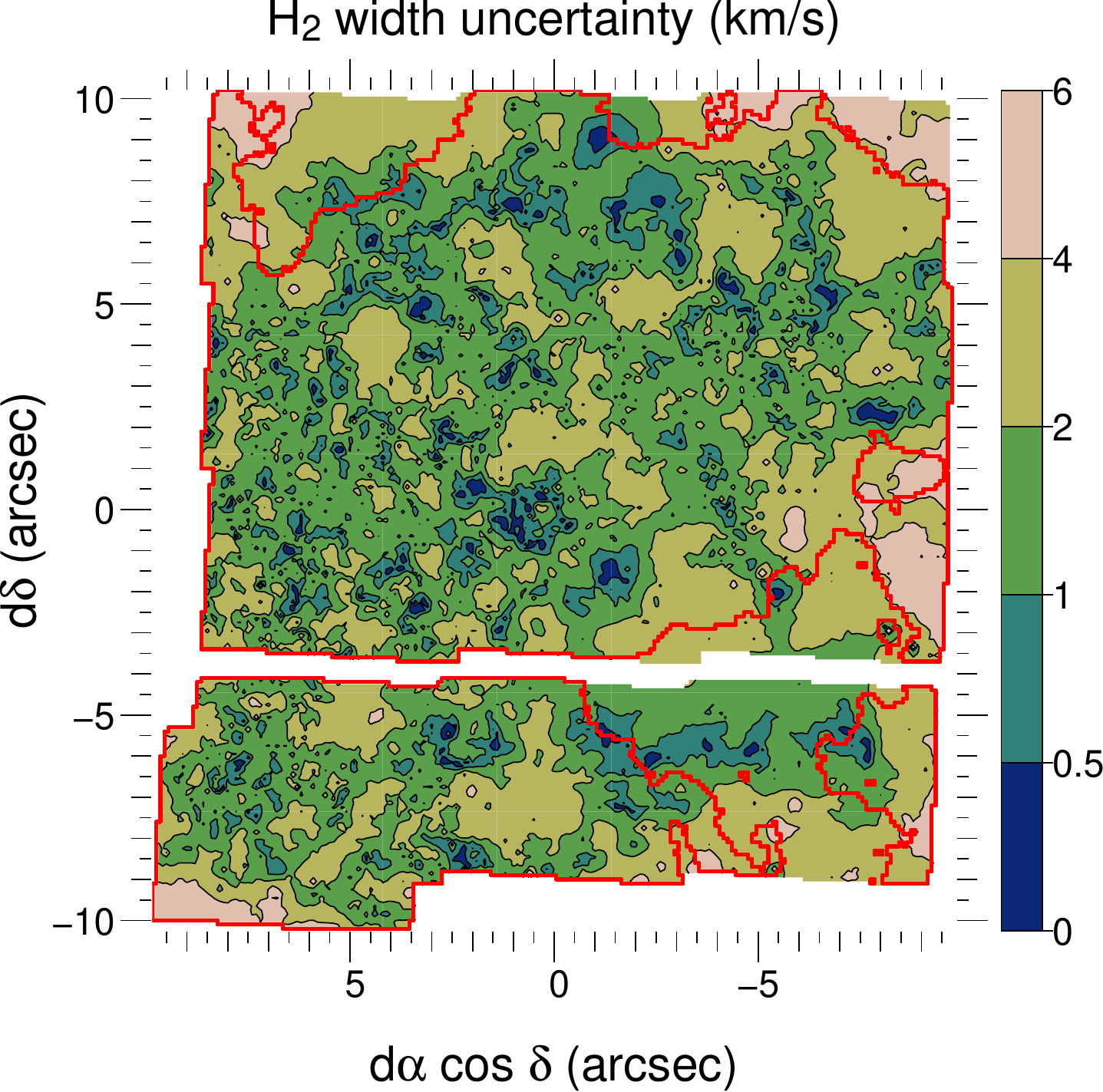}
  \includegraphics[scale=0.4, viewport=44 0 460 413, clip]{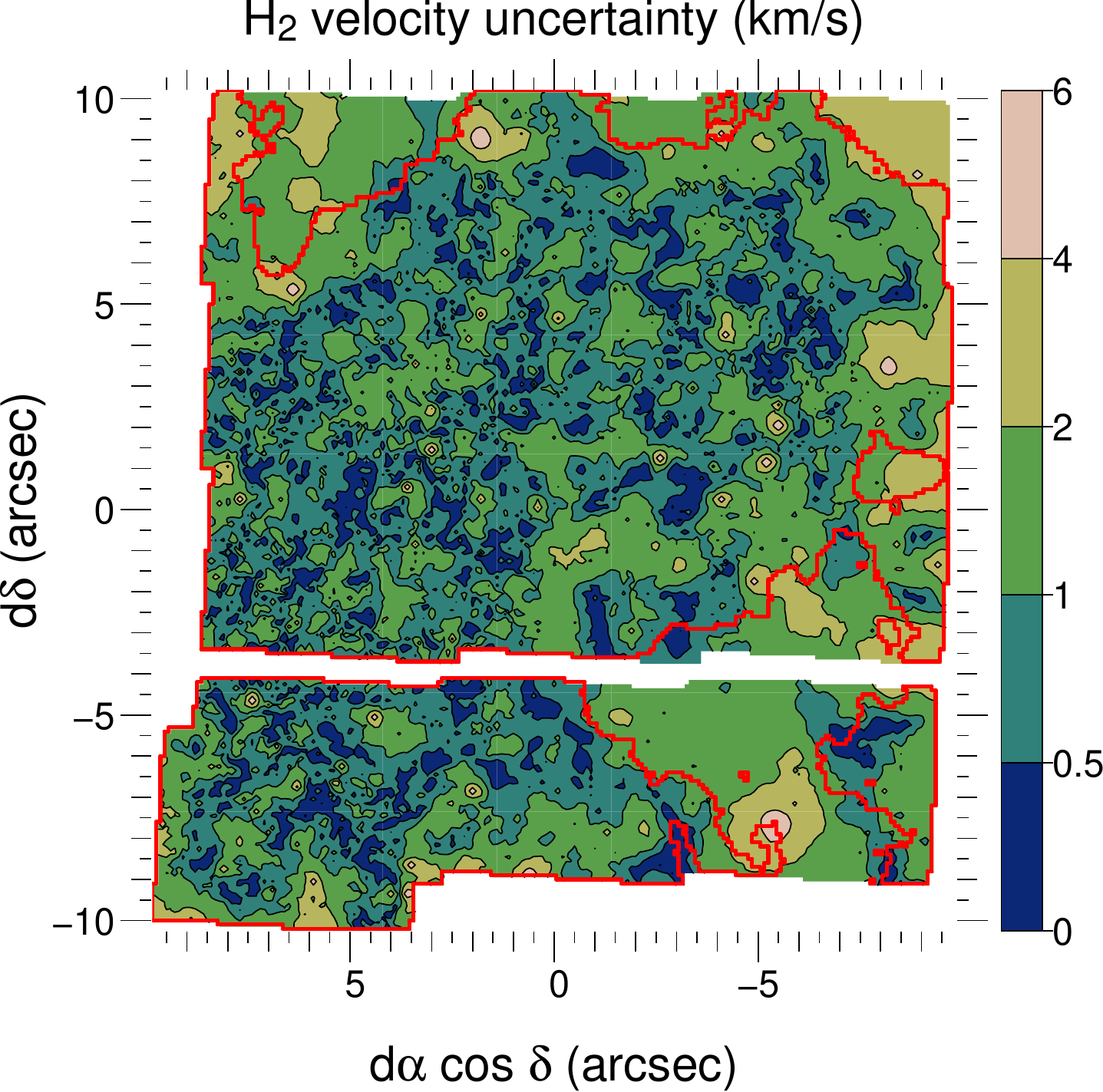}
  \caption{Uncertainty maps corresponding to Fig.~\ref{fig:NE-maps-H2}. A red line
    delineates the same amplitude threshold as on the flux map. 
    We display the flux to
    flux uncertainty ratio rather than the uncertainty itself.}
  \label{fig:NE-errmaps-H2}
\end{figure*}

\begin{figure*}[!ht]
  \centering
  \includegraphics[scale=0.4, viewport=0 0 460 413, clip]{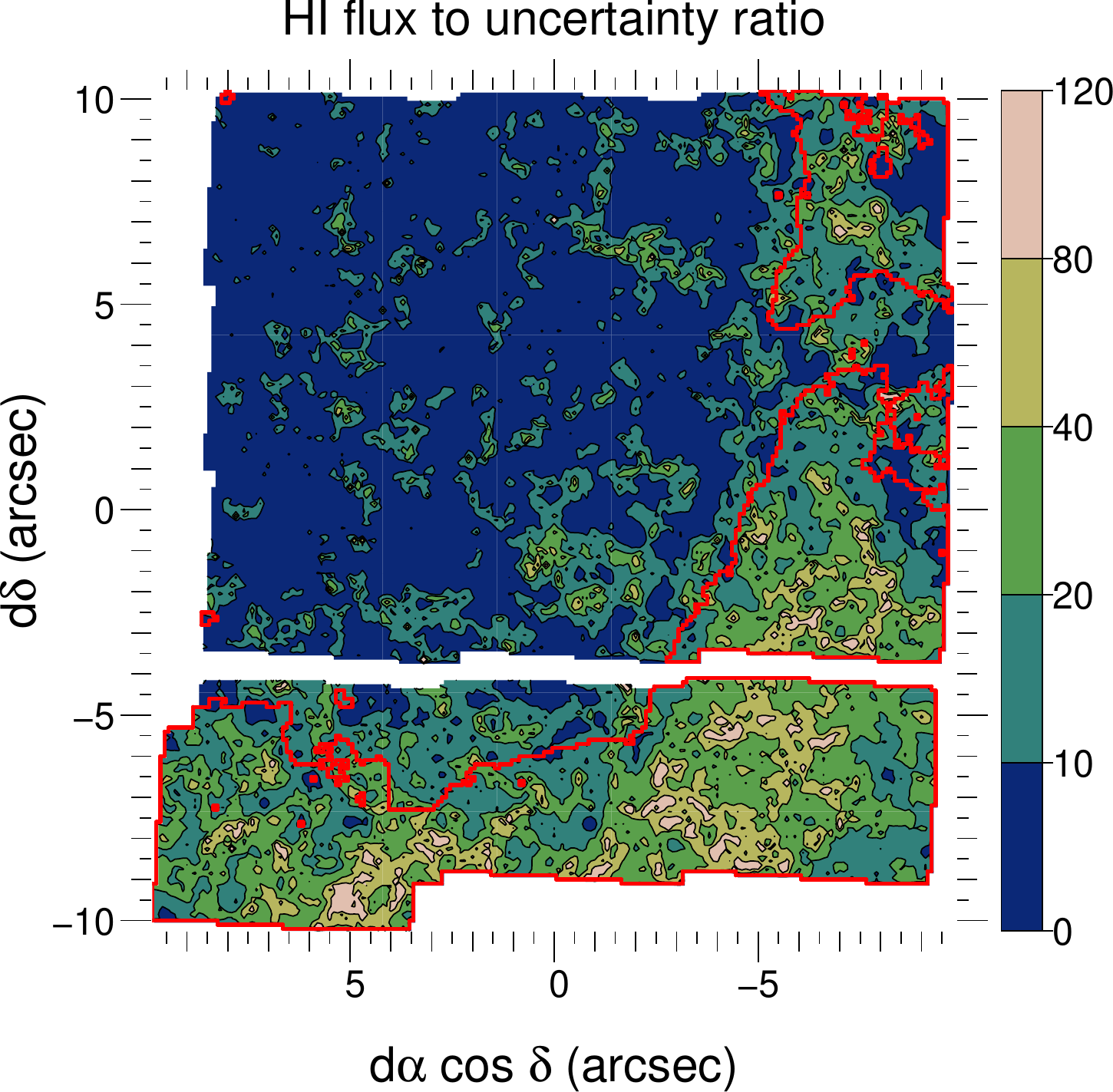}
  \includegraphics[scale=0.4, viewport=44 0 460 413, clip]{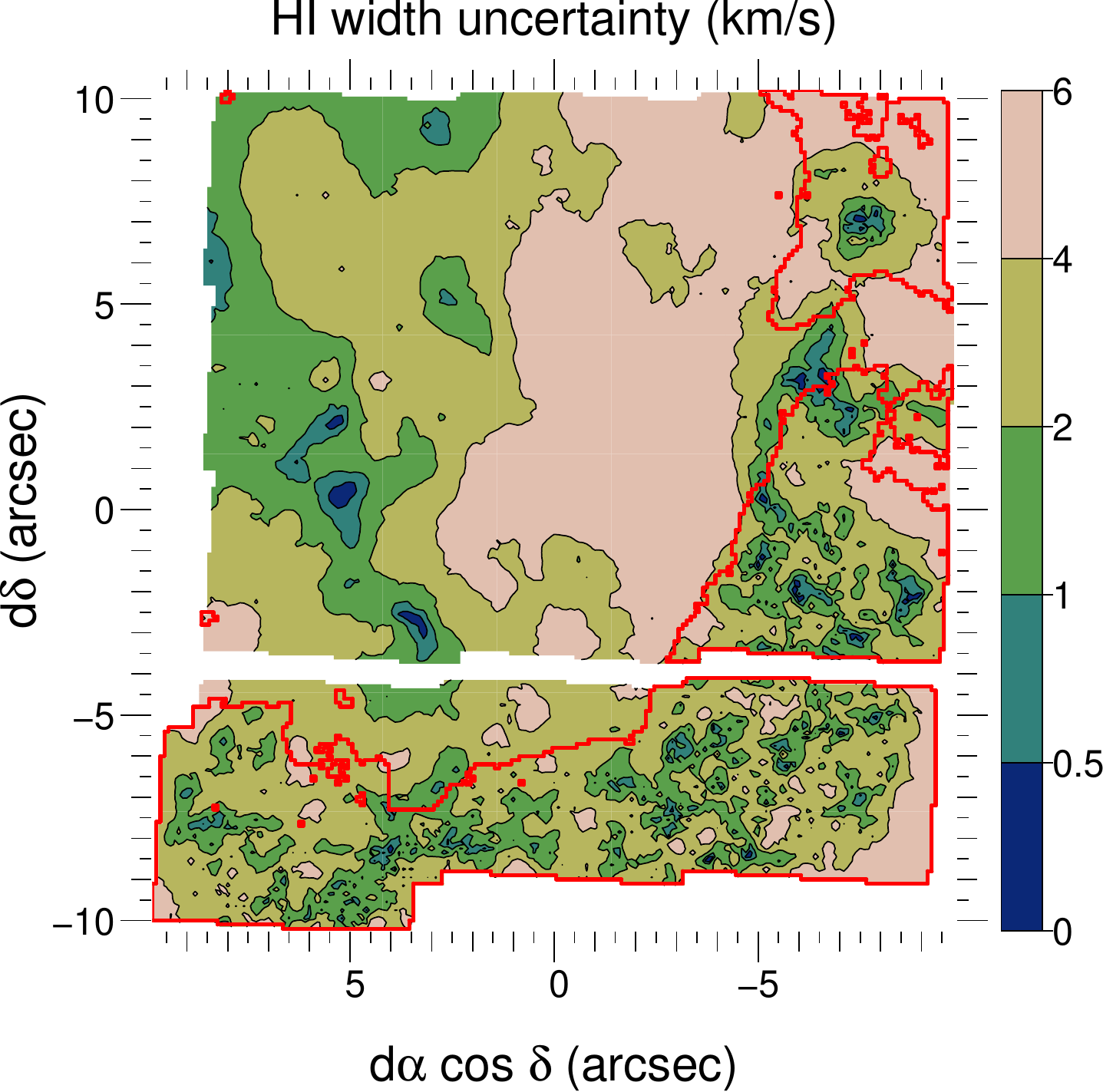}
  \includegraphics[scale=0.4, viewport=44 0 460 413, clip]{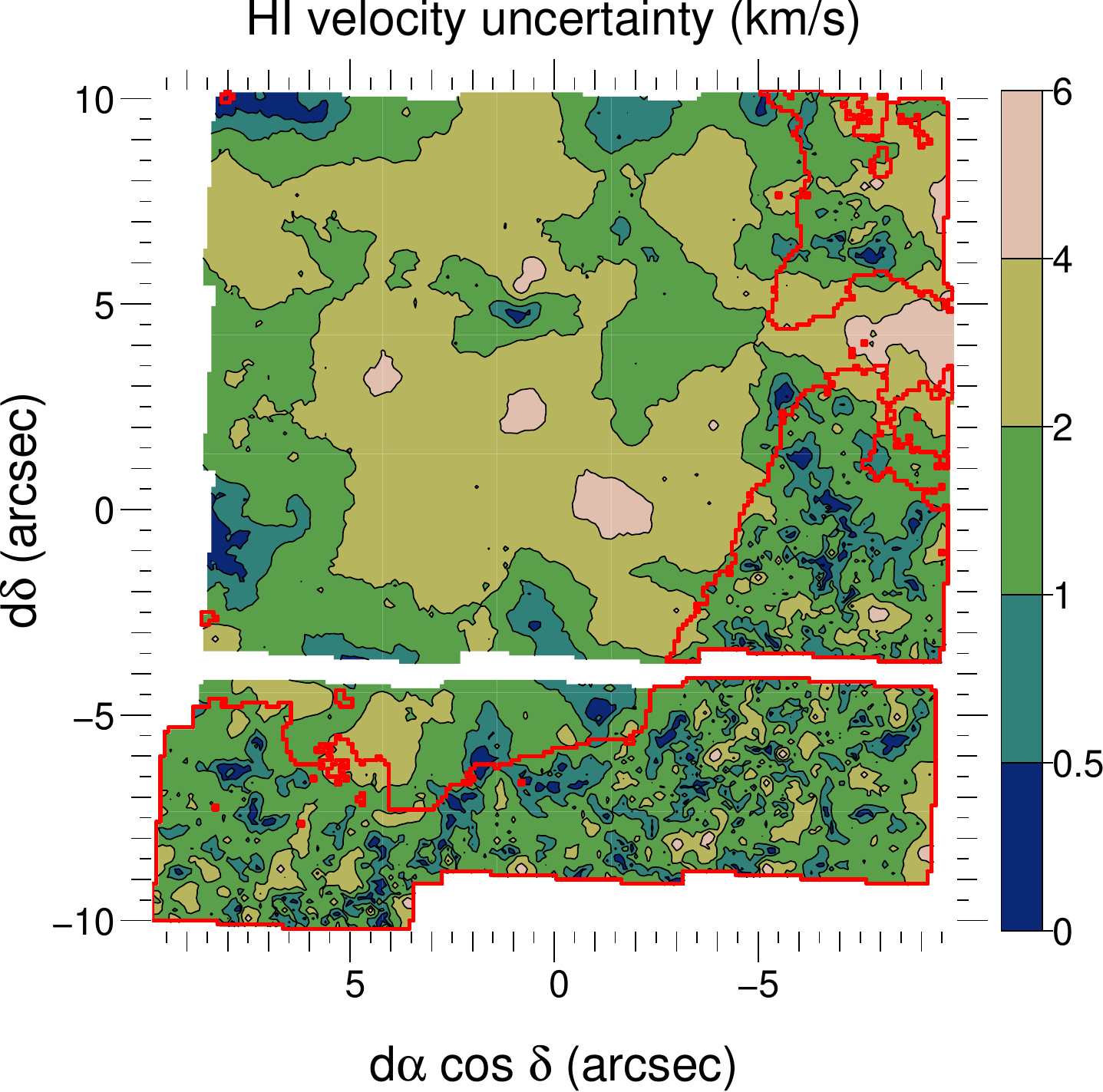}
  \caption{Uncertainty maps corresponding to
      Fig.~\ref{fig:NE-maps-HI}. A red line delineates the same
      amplitude threshold as on the flux map.  We display the flux to
      flux uncertainty ratio rather than the uncertainty itself.}
  \label{fig:NE-errmaps-HI}
\end{figure*}

\begin{figure*}[!ht]
  \centering
  \includegraphics[scale=0.4, viewport=0 50 460 338, clip]{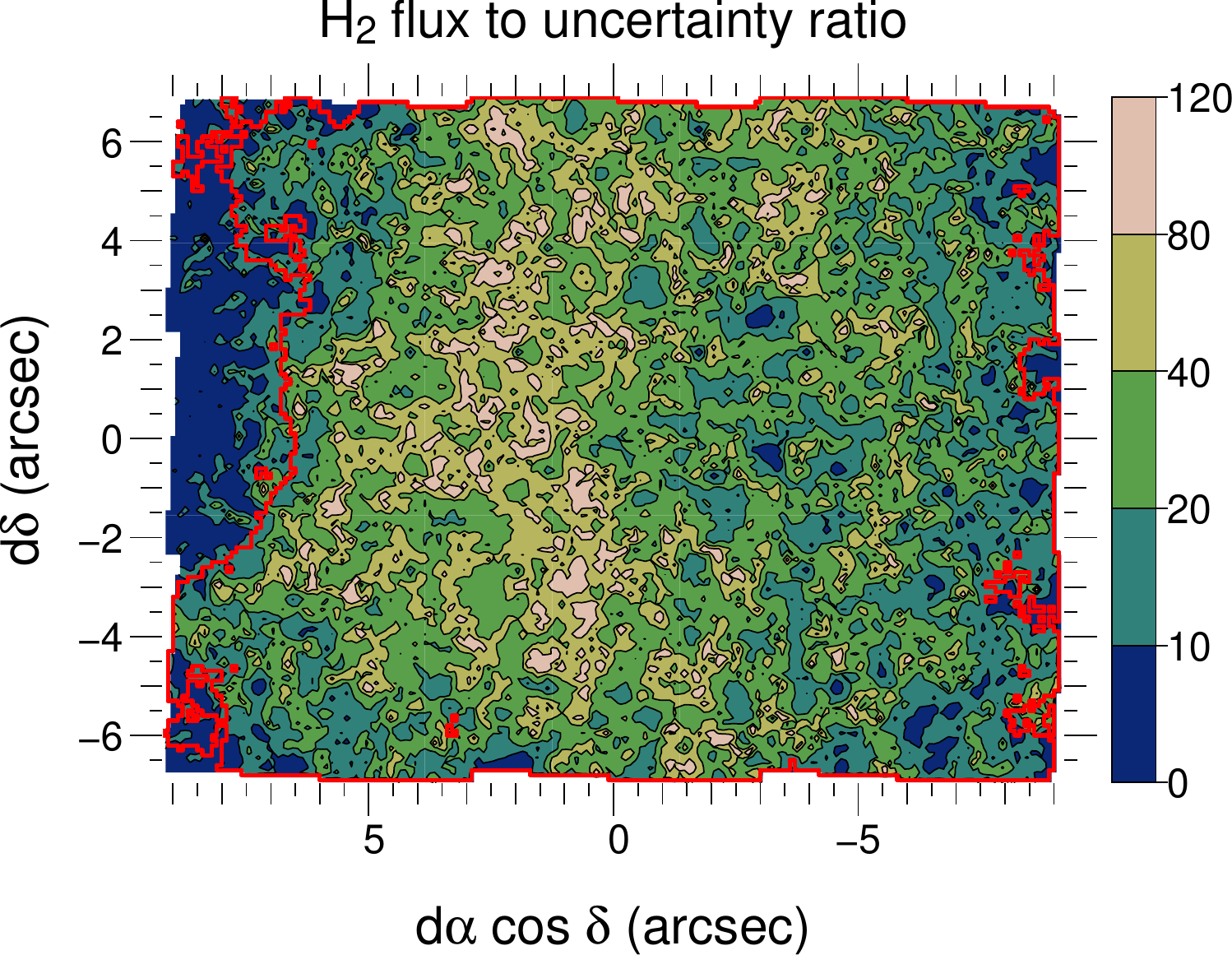}
  \includegraphics[scale=0.4, viewport=44 50 460 338, clip]{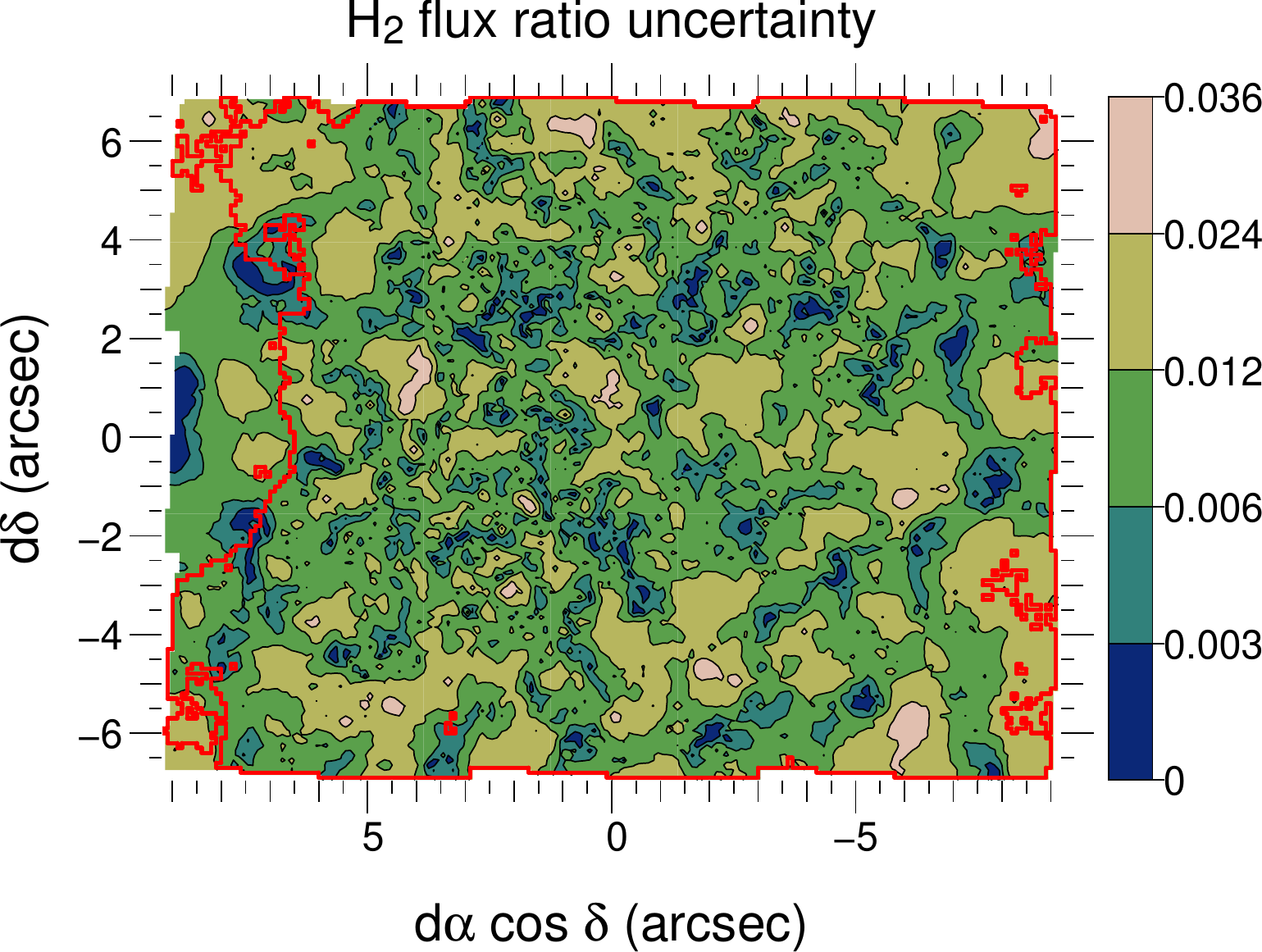}
  \vspace{2mm}\\
  \includegraphics[scale=0.4, viewport=0 0 460 338, clip]{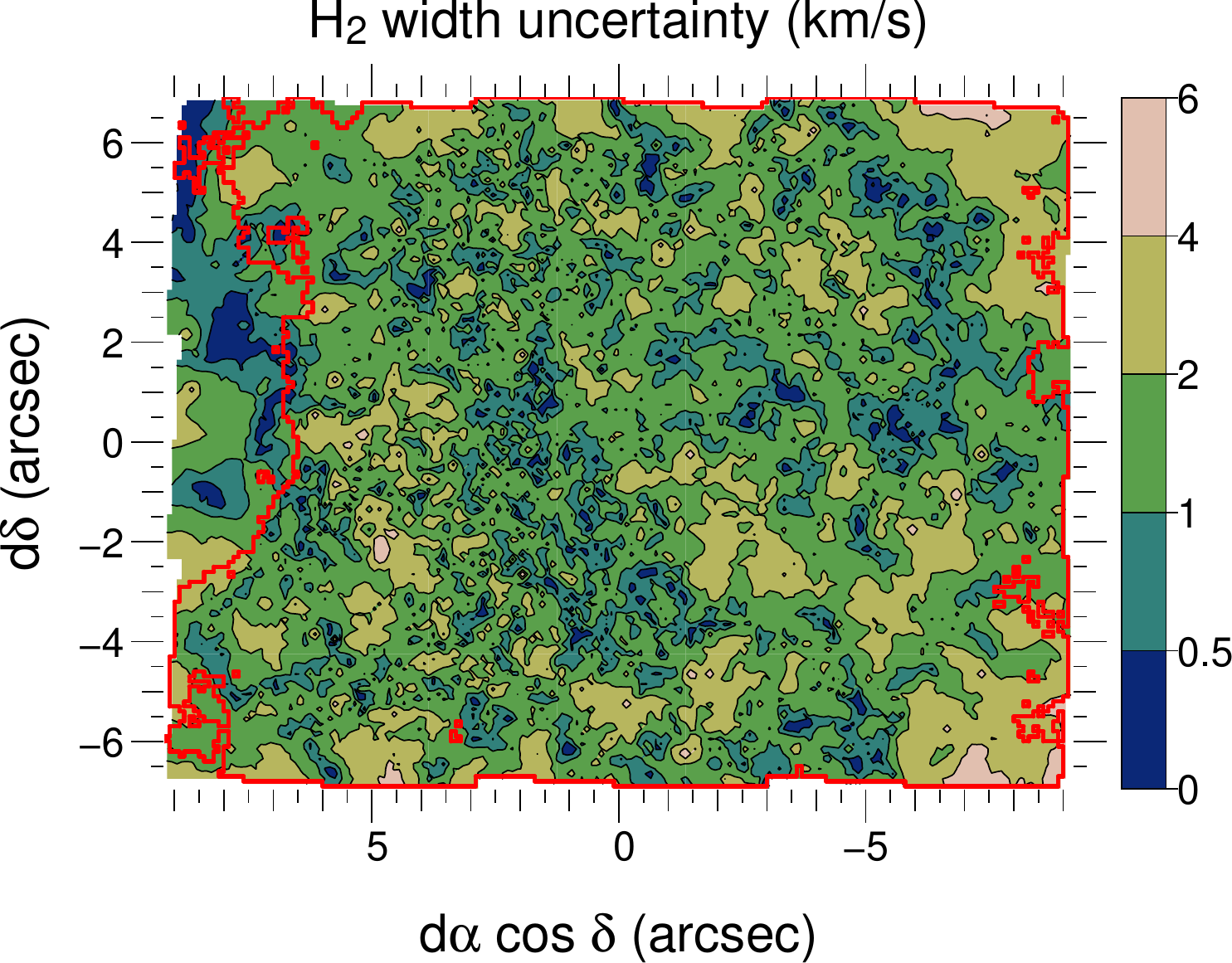}
  \includegraphics[scale=0.4, viewport=44 0 460 338, clip]{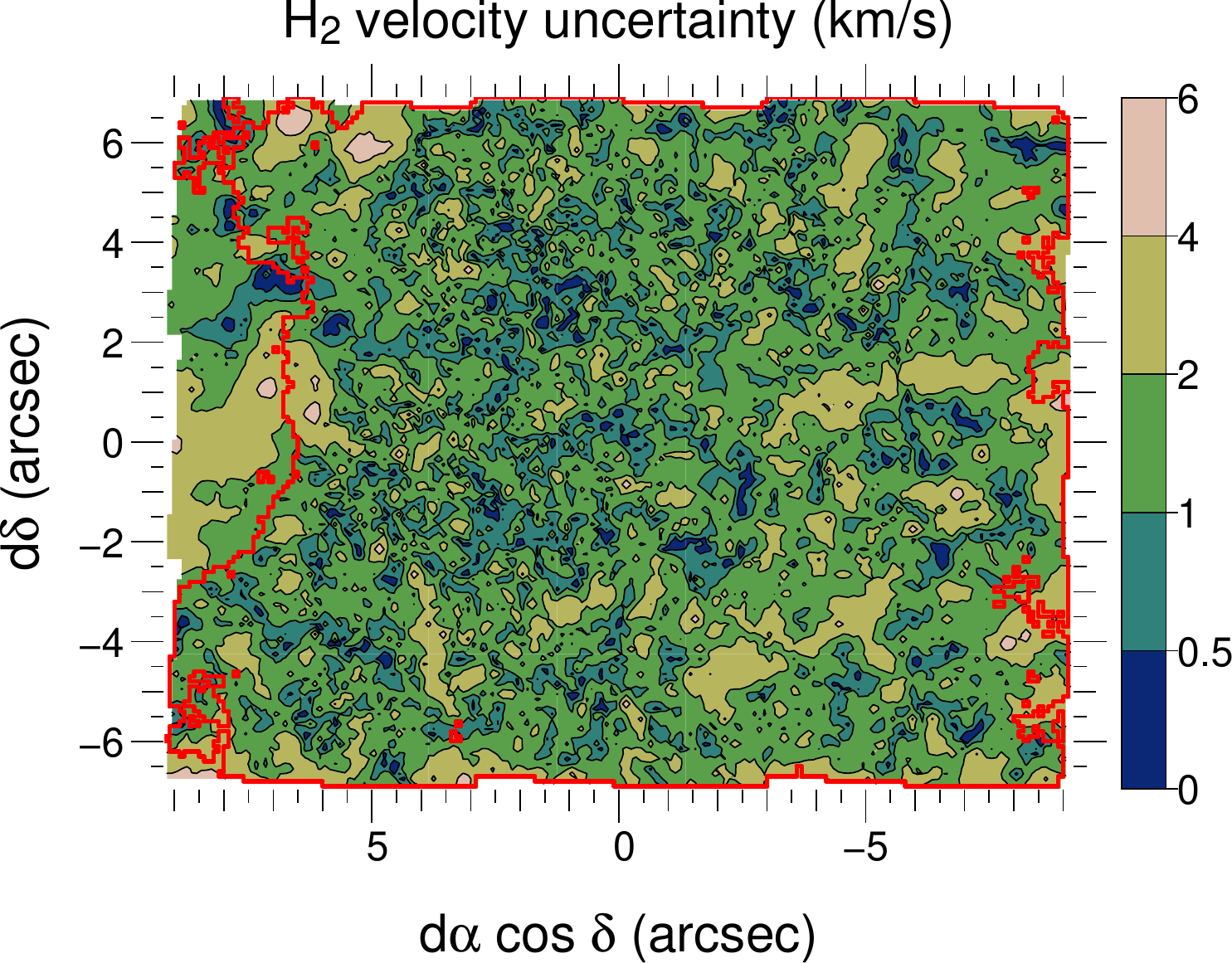}
  \caption{Uncertainty maps corresponding to
      Fig.~\ref{fig:SW-maps-H2}. A red line delineates the same
      amplitude threshold as on the flux map.  We display the flux to
      flux uncertainty ratio rather than the uncertainty itself.}
  \label{fig:SW-errmaps-H2}
\end{figure*}

\begin{figure*}[!ht]
  \centering
  \includegraphics[scale=0.4, viewport=0 0 460 338, clip]{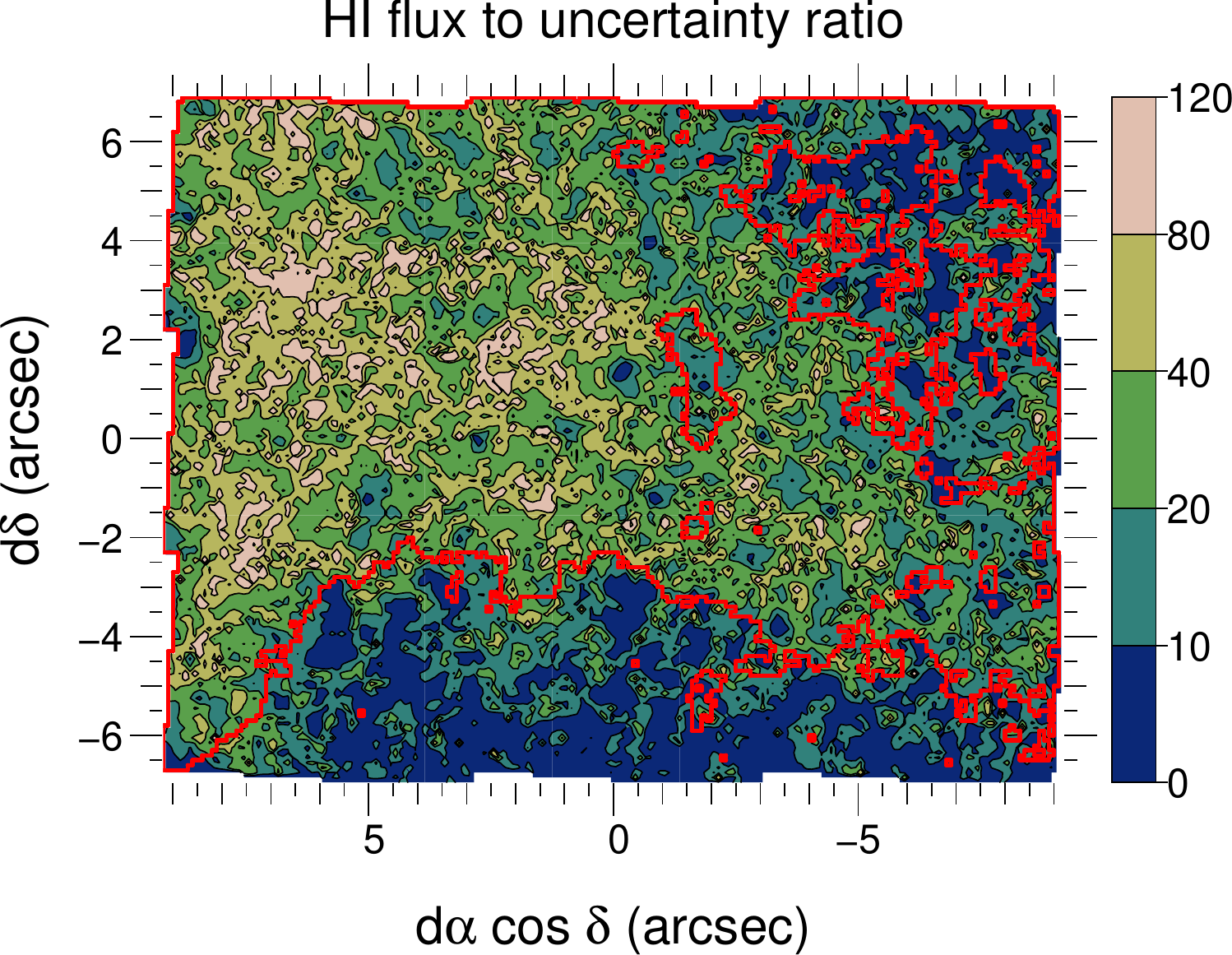}
  \includegraphics[scale=0.4, viewport=44 0 460 338, clip]{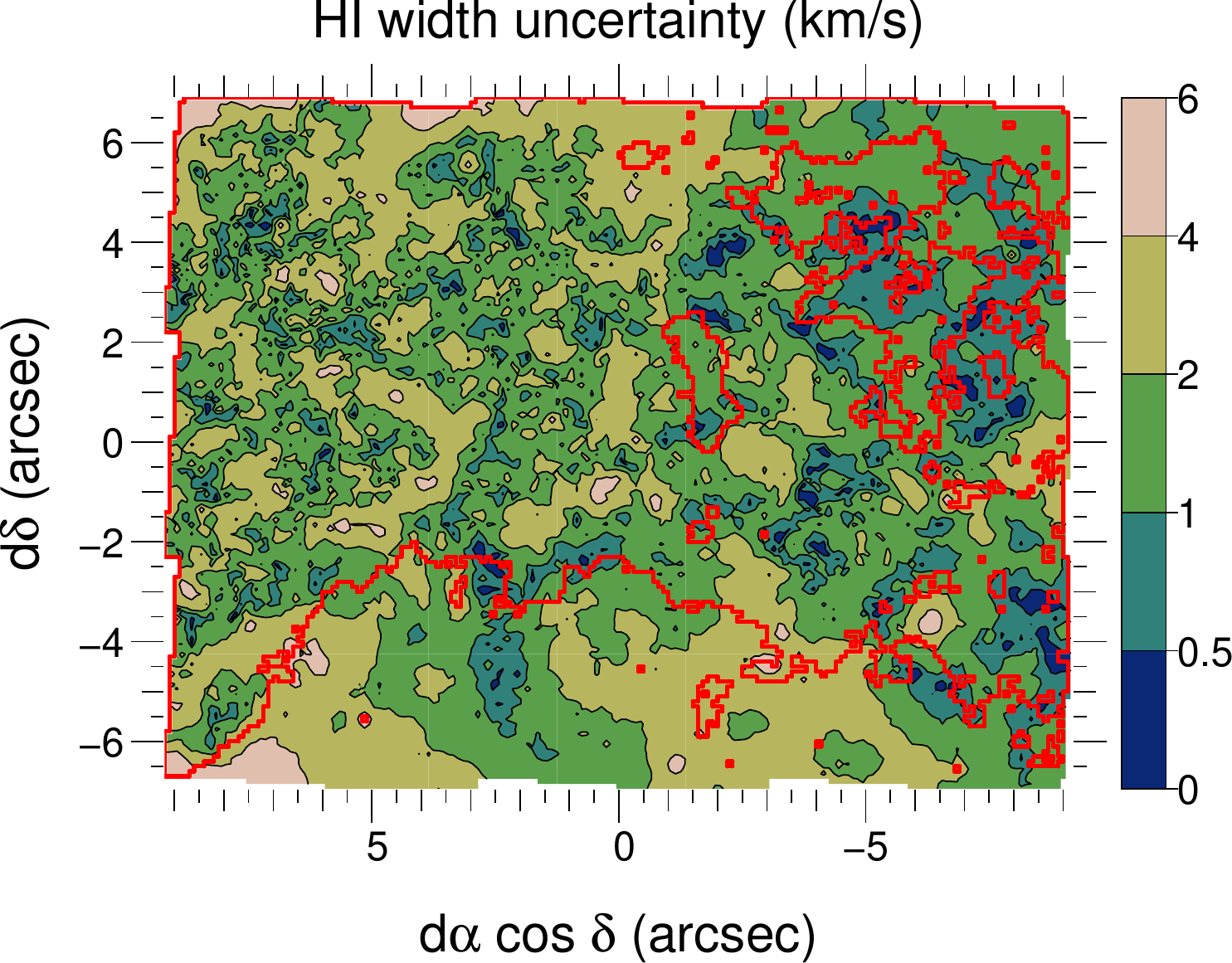}
  \includegraphics[scale=0.4, viewport=44 0 460 338, clip]{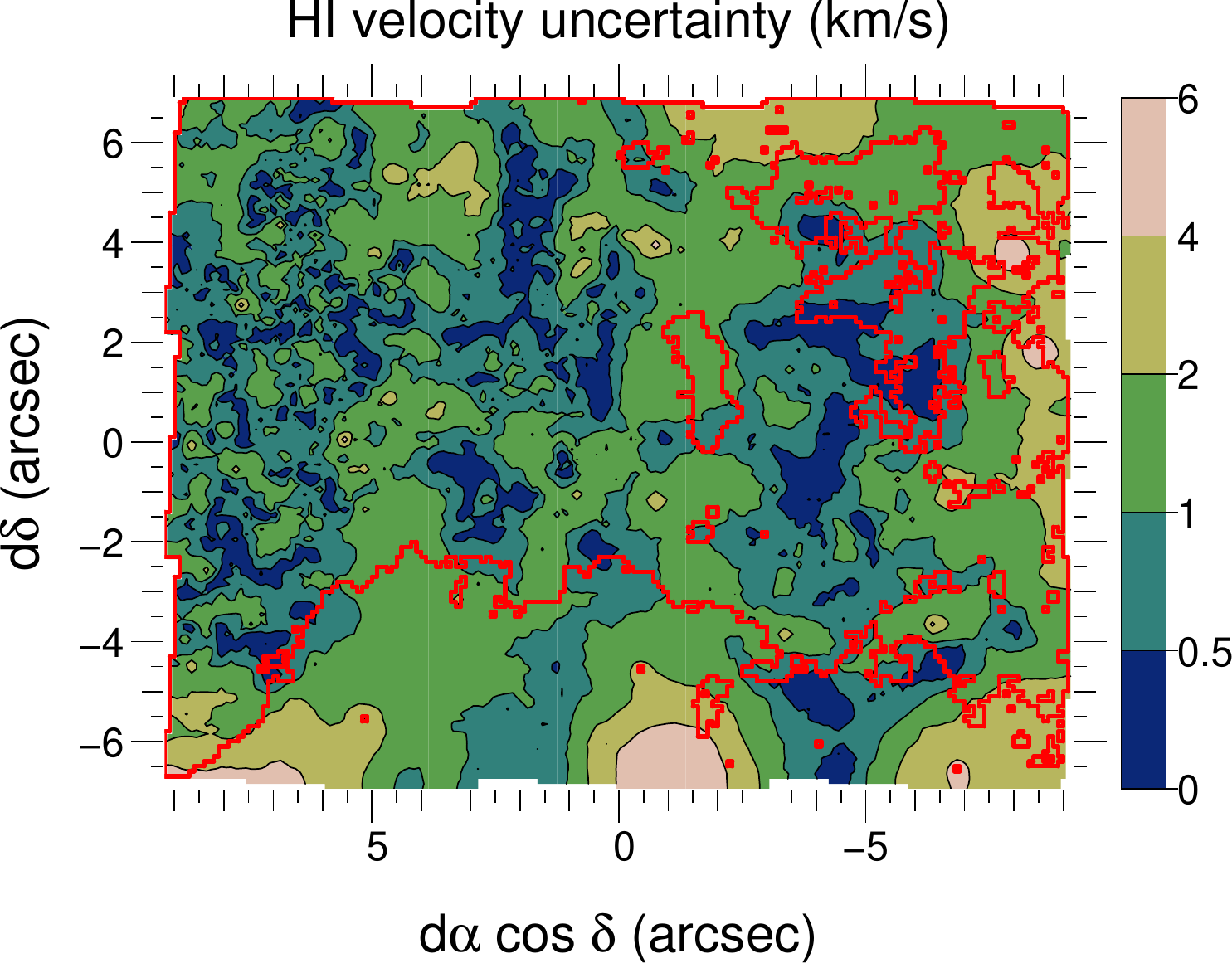}
  \caption{Uncertainty maps corresponding to
      Fig.~\ref{fig:SW-maps-HI}. A red line delineates the same
      amplitude threshold as on the flux map.  We display the flux to
      flux uncertainty ratio rather than the uncertainty itself.}
  \label{fig:SW-errmaps-HI}
\end{figure*}

Figures~\ref{fig:NE-errmaps-H2}, \ref{fig:NE-errmaps-HI},
\ref{fig:SW-errmaps-H2} and \ref{fig:SW-errmaps-HI} show the
uncertainty maps associated with the parameter maps in
Figs.~\ref{fig:NE-maps-H2}, \ref{fig:NE-maps-HI}, \ref{fig:SW-maps-H2}
and \ref{fig:SW-maps-HI}, respectively. Median values over the
considered field-of-view for each mosaic and specie are listed in
Table~\ref{table:errs}.
\label{appendix:errmaps}

\end{document}